%
\expandafter \def \csname CHAPLABELintro\endcsname {1}
\expandafter \def \csname EQLABELintegrals\endcsname {1.1?}
\expandafter \def \csname EQLABELquintic\endcsname {1.2?}
\expandafter \def \csname EQLABELyukone\endcsname {1.3?}
\expandafter \def \csname EQLABELyuktwo\endcsname {1.4?}
\expandafter \def \csname EQLABELmmap\endcsname {1.5?}
\expandafter \def \csname EQLABELPF\endcsname {1.6?}
\expandafter \def \csname EQLABELperiods\endcsname {1.7?}
\expandafter \def \csname EQLABELsemiperiod\endcsname {1.8?}
\expandafter \def \csname EQLABELfirstresult\endcsname {1.9?}
\expandafter \def \csname CHAPLABELprelims\endcsname {2}
\expandafter \def \csname EQLABELfactone\endcsname {2.1?}
\expandafter \def \csname EQLABELfacttwo\endcsname {2.2?}
\expandafter \def \csname EQLABELNseries\endcsname {2.3?}
\expandafter \def \csname FIGLABELdisk\endcsname {2.1?}
\expandafter \def \csname EQLABELp-order\endcsname {2.4?}
\expandafter \def \csname EQLABELsigdef\endcsname {2.5?}
\expandafter \def \csname EQLABELap\endcsname {2.6?}
\expandafter \def \csname EQLABELgammadef\endcsname {2.7?}
\expandafter \def \csname EQLABELnfac\endcsname {2.8?}
\expandafter \def \csname EQLABELgammaformula\endcsname {2.9?}
\expandafter \def \csname EQLABELclassicalref\endcsname {2.10?}
\expandafter \def \csname EQLABELpremult\endcsname {2.11?}
\expandafter \def \csname EQLABELpremultclass\endcsname {2.12?}
\expandafter \def \csname EQLABELmultclass\endcsname {2.13?}
\expandafter \def \csname FIGLABELsquare\endcsname {2.2?}
\expandafter \def \csname FIGLABELcube\endcsname {2.3?}
\expandafter \def \csname EQLABELacongruence\endcsname {2.14?}
\expandafter \def \csname EQLABELFrobzero\endcsname {2.15?}
\expandafter \def \csname CHAPLABELallperiods\endcsname {3}
\expandafter \def \csname EQLABELautomorphisms\endcsname {3.1?}
\expandafter \def \csname EQLABELgenericperiod\endcsname {3.2?}
\expandafter \def \csname FIGLABELloci\endcsname {3.1?}
\expandafter \def \csname EQLABELexact\endcsname {3.3?}
\expandafter \def \csname EQLABELcaseoneid\endcsname {3.4?}
\expandafter \def \csname EQLABELPF\endcsname {3.5?}
\expandafter \def \csname EQLABELFfrob\endcsname {3.6?}
\expandafter \def \csname EQLABELAm\endcsname {3.7?}
\expandafter \def \csname FIGLABELloopandwedge\endcsname {3.2?}
\expandafter \def \csname EQLABELupsilonzero\endcsname {3.8?}
\expandafter \def \csname EQLABELLv\endcsname {3.9?}
\expandafter \def \csname EQLABELgencoeffs\endcsname {3.10?}
\expandafter \def \csname CHAPLABELcalczero\endcsname {4}
\expandafter \def \csname EQLABELfacttwo\endcsname {4.1?}
\expandafter \def \csname EQLABELnmcondition\endcsname {4.2?}
\expandafter \def \csname EQLABELpiofalpha\endcsname {4.3?}
\expandafter \def \csname EQLABELsimplecase\endcsname {4.4?}
\expandafter \def \csname CHAPLABELcalcone\endcsname {5}
\expandafter \def \csname EQLABELnewnu\endcsname {5.1?}
\expandafter \def \csname TABLABELvtableone\endcsname {5.1?}
\expandafter \def \csname EQLABELpoch\endcsname {5.2?}
\expandafter \def \csname EQLABELvcontribs\endcsname {5.3?}
\expandafter \def \csname EQLABELdiscrepant\endcsname {5.4?}
\expandafter \def \csname EQLABELorderone\endcsname {5.5?}
\expandafter \def \csname CHAPLABELhigher\endcsname {6}
\expandafter \def \csname EQLABELAm\endcsname {6.1?}
\expandafter \def \csname EQLABELAmfour\endcsname {6.2?}
\expandafter \def \csname EQLABELfk\endcsname {6.3?}
\expandafter \def \csname EQLABELNpsi\endcsname {6.4?}
\expandafter \def \csname EQLABELhratio\endcsname {6.5?}
\expandafter \def \csname EQLABELaovera\endcsname {6.6?}
\expandafter \def \csname EQLABELpadicN\endcsname {6.7?}
\expandafter \def \csname EQLABELbetam\endcsname {6.8?}
\expandafter \def \csname EQLABELAminfty\endcsname {6.9?}
\expandafter \def \csname EQLABELlimit\endcsname {6.10?}
\expandafter \def \csname EQLABELBoyarski\endcsname {6.11?}
\expandafter \def \csname EQLABELDworkfn\endcsname {6.12?}
\expandafter \def \csname EQLABELDworkcoeffs\endcsname {6.13?}
\expandafter \def \csname EQLABELAcoeffs\endcsname {6.14?}
\expandafter \def \csname EQLABELnumericalbeta\endcsname {6.15?}
\expandafter \def \csname EQLABELends\endcsname {6.16?}
\expandafter \def \csname EQLABELshortbeta\endcsname {6.17?}
\expandafter \def \csname EQLABELforcalc\endcsname {6.18?}
\expandafter \def \csname EQLABELfinalN\endcsname {6.19?}
\expandafter \def \csname EQLABELnus\endcsname {6.20?}
\expandafter \def \csname EQLABELbetas\endcsname {6.21?}
\expandafter \def \csname CHAPLABELzerodim\endcsname {7}
\expandafter \def \csname EQLABELreclaw\endcsname {7.1?}
\expandafter \def \csname EQLABELperiod\endcsname {7.2?}
\expandafter \def \csname EQLABELprenpsi\endcsname {7.3?}
\expandafter \def \csname EQLABELnpsi\endcsname {7.4?}
\expandafter \def \csname EQLABELmpsi\endcsname {7.5?}
\expandafter \def \csname CHAPLABELcubics\endcsname {8}
\expandafter \def \csname EQLABELAcubic\endcsname {8.1?}
\expandafter \def \csname EQLABEL\endcsname {8.2?}
\expandafter \def \csname CHAPLABELgauss\endcsname {9}
\expandafter \def \csname EQLABELseries\endcsname {9.1?}
\expandafter \def \csname EQLABELgausssums\endcsname {9.2?}
\expandafter \def \csname EQLABELinversion\endcsname {9.3?}
\expandafter \def \csname EQLABELGKformula\endcsname {9.4?}
\expandafter \def \csname EQLABELyPeqn\endcsname {9.5?}
\expandafter \def \csname EQLABELcharcases\endcsname {9.6?}
\expandafter \def \csname EQLABELintermediategauss\endcsname {9.7?}
\expandafter \def \csname EQLABELnuexpr\endcsname {9.8?}
\expandafter \def \csname EQLABELyg\endcsname {9.9?}
\expandafter \def \csname EQLABELssum\endcsname {9.10?}
\expandafter \def \csname TABLABELvtabletwo\endcsname {9.1?}
\expandafter \def \csname EQLABELlsum\endcsname {9.11?}
\expandafter \def \csname EQLABELnuY\endcsname {9.12?}
\expandafter \def \csname EQLABELnewnewnu\endcsname {9.13?}
\expandafter \def \csname EQLABELcoeffrels\endcsname {9.14?}
\expandafter \def \csname EQLABELprecnf\endcsname {9.15?}
\expandafter \def \csname EQLABELgammamult\endcsname {9.16?}
\expandafter \def \csname EQLABELnewcoefrels\endcsname {9.17?}
\expandafter \def \csname EQLABELgaussqsums\endcsname {9.18?}
\expandafter \def \csname EQLABELnewnewnuq\endcsname {9.19?}
%
\magnification=1200

\font\eightrm=cmr8 at 8pt
\font\fourteenrm=cmr12 at 14pt
\font\seventeenrm=cmr17 at 17pt
\font\twentyonerm=cmr17 at 21pt

\font\ss=cmss10

\font\csc=cmcsc10

\font\twelvecal=cmsy10 at 12pt

\font\twelvemath=cmmi12

\font\fourteenbold=cmbx12 at 14pt
\font\seventeenbold=cmbx7 at 17pt

\font\fively=lasy5
\font\sevenly=lasy7
\font\tenly=lasy10

\textfont10=\tenly
\scriptfont10=\sevenly    
\scriptscriptfont10=\fively
\parskip=10pt
\parindent=20pt
\def\today{\ifcase\month\or January\or February\or March\or April\or May\or June
       \or July\or August\or September\or October\or November\or December\fi
       \space\number\day, \number\year}

\def\title#1{\footline={\ifnum\pageno<2\hfil
       \else\hss\tenrm\folio\hss\fi}\vskip1truein\centerline{{#1}   
       \footnote{\raise1ex\hbox{*}}{\eightrm Supported in part
       by the Robert A. Welch Foundation and N.S.F. Grants 
       PHY-880637 and\break PHY-8605978.}}}

\def\newpage{\vfill\eject}
\def\abstract#1{\centerline{\bf ABSTRACT}\vskip.2truein{\narrower\noindent#1
       \smallskip}}

\def\runninghead#1#2{\voffset=2\baselineskip\nopagenumbers
       \headline={\ifodd\pageno\rightheadline\else \leftheadline\fi}
       \def\rightheadline{{\sl#1}\hfill{\rm\folio}}
       \def\leftheadline{{\rm\folio}\hfill{\sl#2}}}
\def\SS{\mathhexbox278}

\newcount\footnoteno
\def\Footnote#1{\advance\footnoteno by 1
                \let\SF=\empty 
                \ifhmode\edef\SF{\spacefactor=\the\spacefactor}\/\fi
                $^{\the\footnoteno}$\ignorespaces
                \SF\vfootnote{$^{\the\footnoteno}$}{#1}}

\def\figbox#1#2#3{\vbox{\vskip15pt
                   \vbox{\hrule
                    \hbox{\vrule
                     \vbox{\vskip12truept\centerline #1 \vskip6truept
                          {\hskip.4truein\vbox{\hsize=5truein\noindent
                          {\bf Figure\hskip5truept#2:}\hskip5truept#3}}
                     \vskip18truept}
                    \vrule}
                   \hrule}}}
\def\place#1#2#3{\vbox to0pt{\kern-\parskip\kern-7pt
                             \kern-#2truein\hbox{\kern#1truein #3}
                             \vss}\nointerlineskip}
\def\figurecaption#1#2{\kern.75truein\vbox{\hsize=5truein\noindent{\bf Figure
    \figlabel{#1}:} #2}}
\def\tablecaption#1#2{\kern.75truein\lower12truept\hbox{\vbox{\hsize=5truein
    \noindent{\bf Table\hskip5truept\tablabel{#1}:} #2}}}
\def\boxed#1{\lower3pt\hbox{
                       \vbox{\hrule\hbox{\vrule
                         \vbox{\kern2pt\hbox{\kern3pt#1\kern3pt}\kern3pt}\vrule}
                         \hrule}}}
\def\a{\alpha}
\def\b{\beta}
\def\g{\gamma}\def\G{\Gamma}
\def\d{\delta}\def\D{\Delta}
\def\e{\epsilon}\def\ve{\varepsilon}
\def\z{\zeta}

\def\th{\theta}\def\Th{\Theta}\def\vth{\vartheta}

\def\l{\lambda}
\def\m{\mu}
\def\n{\nu}
\def\x{\xi}

\def\p{\pi}\def\vp{\varpi}
\def\r{\rho}
\def\s{\sigma}

\def\U{\Upsilon}
\def\ph{\phi}\def\Ph{\Phi}\def\vph{\varphi}

\def\ps{\psi}\def\Ps{\Psi}
\def\O{\Omega}

\def\ca#1{\relax\ifmmode {{\cal #1}}\else $\cal #1$\fi}

\def\calb{{\cal B}}

\def\calm{{\cal M}}

\def\inbar{\vrule height1.5ex width.4pt depth0pt}
\def\IB{\relax{\rm I\kern-.18em B}}
\def\IC{\relax\hbox{\kern.25em$\inbar\kern-.3em{\rm C}$}}
\def\ID{\relax{\rm I\kern-.18em D}}
\def\IE{\relax{\rm I\kern-.18em E}}
\def\IF{\relax{\rm I\kern-.18em F}}
\def\IG{\relax\hbox{\kern.25em$\inbar\kern-.3em{\rm G}$}}
\def\IH{\relax{\rm I\kern-.18em H}}
\def\II{\relax{\rm I\kern-.18em I}}
\def\IK{\relax{\rm I\kern-.18em K}}
\def\IL{\relax{\rm I\kern-.18em L}}
\def\IM{\relax{\rm I\kern-.18em M}}
\def\IN{\relax{\rm I\kern-.18em N}}
\def\IO{\relax\hbox{\kern.25em$\inbar\kern-.3em{\rm O}$}}
\def\IP{\relax{\rm I\kern-.18em P}}
\def\IQ{\relax\hbox{\kern.25em$\inbar\kern-.3em{\rm Q}$}}
\def\IR{\relax{\rm I\kern-.18em R}}
\def\IZ{\relax\ifmmode\hbox{\ss Z\kern-.4em Z}\else{\ss Z\kern-.4em Z}\fi}
\def\IGa{\relax{\rm I}\kern-.18em\Gamma}
\def\IPi{\relax{\rm I}\kern-.18em\Pi}
\def\ITh{\relax\hbox{\kern.25em$\inbar\kern-.3em\Theta$}}
\def\IOm{\relax\thinspace\inbar\kern1.95pt\inbar\kern-5.525pt\Omega}


\def\ie{{\it i.e.,\ \/}}

\def\noblackboxes{\overfullrule=0pt}
\def\define{\buildrel\rm def\over =}

\def\cy{Calabi--Yau} 
\def\cym{Calabi--Yau manifold}
\def\cys{Calabi--Yau manifolds}

\def\K{K\"ahler}

\def\H#1#2{\relax\ifmmode {H^{#1#2}}\else $H^{#1 #2}$\fi}
\def\M{\relax\ifmmode{\calm}\else $\calm$\fi}

\def\Bigcheck{\lower3.8pt\hbox{\smash{\hbox{{\twentyonerm \v{}}}}}}
\def\bigboldcheck{\smash{\hbox{{\seventeenbold\v{}}}}}

\def\Bighat{\lower3.8pt\hbox{\smash{\hbox{{\twentyonerm \^{}}}}}}

\def\Msharp{\relax\ifmmode{\calm^\sharp}\else $\smash{\calm^\sharp}$\fi}
\def\Mflat{\relax\ifmmode{\calm^\flat}\else $\smash{\calm^\flat}$\fi}
\def\preMcheck{\kern2pt\hbox{\Bigcheck\kern-12pt{$\cal M$}}}
\def\Mcheck{\relax\ifmmode\preMcheck\else $\preMcheck$\fi}
\def\preMhat{\kern2pt\hbox{\Bighat\kern-12pt{$\cal M$}}}
\def\Mhat{\relax\ifmmode\preMhat\else $\preMhat$\fi}

\def\Bsharp{\relax\ifmmode{\calb^\sharp}\else $\calb^\sharp$\fi}
\def\Bflat{\relax\ifmmode{\calb^\flat}\else $\calb^\flat$ \fi}
\def\preBcheck{\hbox{\Bigcheck\kern-9pt{$\cal B$}}}
\def\Bcheck{\relax\ifmmode\preBcheck\else $\preBcheck$\fi}
\def\preBhat{\hbox{\Bighat\kern-9pt{$\cal B$}}}
\def\Bhat{\relax\ifmmode\preBhat\else $\preBhat$\fi}

\def\figBcheck{\kern3pt\hbox{\raise1pt\hbox{\bigboldcheck}\kern-11pt
    {\twelvecal B}}}
\def\figBsharp{{\twelvecal B}\raise5pt\hbox{$\twelvemath\sharp$}}
\def\figBflat{{\twelvecal B}\raise5pt\hbox{$\twelvemath\flat$}}

\def\gcheck{\hbox{\lower2.5pt\hbox{\Bigcheck}\kern-8pt$\g$}}
\def\lhat{\hbox{\raise.5pt\hbox{\Bighat}\kern-8pt$\l$}}

\def\Fcheck{\kern2pt\hbox{\raise1pt\hbox{\Bigcheck}\kern-10pt{$\cal F$}}}
\def\Fhat{\kern2pt\hbox{\raise1pt\hbox{\Bighat}\kern-10pt{$\cal F$}}}
 
\def\cp#1{\relax\ifmmode {\IP\kern-2pt{}_{#1}}\else $\IP\kern-2pt{}_{#1}$\fi}
\def\h#1#2{\relax\ifmmode {b_{#1#2}}\else $b_{#1#2}$\fi}

\def\half{{1\over 2}}
\def\tr{{\rm tr}}
\def\frac#1#2{{#1\over #2}}

\def\cone{\relax\thinspace\hbox{$<\kern-.8em{)}$}}
\mathchardef\mho"0A30

\def\-{\hphantom{-}}


\def\npb#1{Nucl.\ Phys.\ {\bf B#1}}

\def\cmp#1{Commun. Math. Phys. {\bf #1}}


\def\picture #1 by #2 (#3){\vbox to #2{\hrule width #1 height 0pt depth 0pt
                                       \vfill\special{picture #3}}}
\def\scaledpicture #1 by #2 (#3 scaled #4){{\dimen0=#1 \dimen1=#2
           \divide\dimen0 by 1000 \multiply\dimen0 by #4
            \divide\dimen1 by 1000 \multiply\dimen1 by #4
            \picture \dimen0 by \dimen1 (#3 scaled #4)}}
\def\illustration #1 by #2 (#3){\vbox to #2{\hrule width #1 height 0pt depth 0pt
                                       \vfill\special{illustration #3}}}
\def\scaledillustration #1 by #2 (#3 scaled #4){{\dimen0=#1 \dimen1=#2
           \divide\dimen0 by 1000 \multiply\dimen0 by #4
            \divide\dimen1 by 1000 \multiply\dimen1 by #4
            \illustration \dimen0 by \dimen1 (#3 scaled #4)}}


\def\delaOssa{\nobreak\vskip1truein\hbox to\hsize
       {\hskip 4truein Xenia de la Ossa\hfill}}

\def\hoy{\number\day\space de \ifcase\month\or enero\or febrero\or marzo\or
       abril\or mayo\or junio\or julio\or agosto\or septiembre\or octubre\or
       noviembre\or diciembre\fi\space de \number\year}

\def\cropen#1{\crcr\noalign{\vskip #1}}

\newif\ifproofmode
\proofmodefalse

\newif\ifforwardreference
\forwardreferencefalse

\newif\ifchapternumbers
\chapternumbersfalse

\newif\ifcontinuousnumbering
\continuousnumberingfalse

\newif\iffigurechapternumbers
\figurechapternumbersfalse

\newif\ifcontinuousfigurenumbering
\continuousfigurenumberingfalse

\newif\iftablechapternumbers
\tablechapternumbersfalse

\newif\ifcontinuoustablenumbering
\continuoustablenumberingfalse

\font\eqsixrm=cmr6

\def\marginstyle{\eqsixrm}

\newtoks\chapletter
\newcount\chapno
\newcount\eqlabelno
\newcount\figureno
\newcount\tableno

\chapno=0
\eqlabelno=0
\figureno=0
\tableno=0

\def\chapfolio{\ifnum\chapno>0 \the\chapno\else\the\chapletter\fi}

\def\bumpchapno{\ifnum\chapno>-1 \global\advance\chapno by 1
\else\global\advance\chapno by -1 \setletter\chapno\fi
\ifcontinuousnumbering\else\global\eqlabelno=0 \fi
\ifcontinuousfigurenumbering\else\global\figureno=0 \fi
\ifcontinuoustablenumbering\else\global\tableno=0 \fi}

\def\setletter#1{\ifcase-#1{}\or{}%
\or\global\chapletter={A}%
\or\global\chapletter={B}%
\or\global\chapletter={C}%
\or\global\chapletter={D}%
\or\global\chapletter={E}%
\or\global\chapletter={F}%
\or\global\chapletter={G}%
\or\global\chapletter={H}%
\or\global\chapletter={I}%
\or\global\chapletter={J}%
\or\global\chapletter={K}%
\or\global\chapletter={L}%
\or\global\chapletter={M}%
\or\global\chapletter={N}%
\or\global\chapletter={O}%
\or\global\chapletter={P}%
\or\global\chapletter={Q}%
\or\global\chapletter={R}%
\or\global\chapletter={S}%
\or\global\chapletter={T}%
\or\global\chapletter={U}%
\or\global\chapletter={V}%
\or\global\chapletter={W}%
\or\global\chapletter={X}%
\or\global\chapletter={Y}%
\or\global\chapletter={Z}\fi}

\def\tempsetletter#1{\ifcase-#1{}\or{}%
\or\global\chapletter={A}%
\or\global\chapletter={B}%
\or\global\chapletter={C}%
\or\global\chapletter={D}%
\or\global\chapletter={E}%
\or\global\chapletter={F}%
\or\global\chapletter={G}%
\or\global\chapletter={H}%
\or\global\chapletter={I}%
\or\global\chapletter={J}%
\or\global\chapletter={K}%
\or\global\chapletter={L}%
\or\global\chapletter={M}%
\or\global\chapletter={N}%
\or\global\chapletter={O}%
\or\global\chapletter={P}%
\or\global\chapletter={Q}%
\or\global\chapletter={R}%
\or\global\chapletter={S}%
\or\global\chapletter={T}%
\or\global\chapletter={U}%
\or\global\chapletter={V}%
\or\global\chapletter={W}%
\or\global\chapletter={X}%
\or\global\chapletter={Y}%
\or\global\chapletter={Z}\fi}

\def\chapshow#1{\ifnum#1>0 \relax#1%
\else{\tempsetletter{\number#1}\chapno=#1\chapfolio}\fi}

\def\chaplabel#1{\bumpchapno\ifproofmode\ifforwardreference
\immediate\write1{\noexpand\expandafter\noexpand\def
\noexpand\csname CHAPLABEL#1\endcsname{\the\chapno}}\fi\fi
\global\expandafter\edef\csname CHAPLABEL#1\endcsname
{\the\chapno}\ifproofmode\llap{\hbox{\marginstyle #1\ }}\fi\chapfolio}

\def\chapref#1{\ifundefined{CHAPLABEL#1}??\ifproofmode\ifforwardreference%
\else\write16{ ***Undefined Chapter Reference #1*** }\fi
\else\write16{ ***Undefined Chapter Reference #1*** }\fi
\else\edef\LABxx{\getlabel{CHAPLABEL#1}}\chapshow\LABxx\fi
\ifproofmode\write2{Chapter #1}\fi}

\def\eqnum{\global\advance\eqlabelno by 1
\eqno(\ifchapternumbers\chapfolio.\fi\the\eqlabelno)}

\def\eqlabel#1{\global\advance\eqlabelno by 1 \ifproofmode\ifforwardreference
\immediate\write1{\noexpand\expandafter\noexpand\def
\noexpand\csname EQLABEL#1\endcsname{\the\chapno.\the\eqlabelno?}}\fi\fi
\global\expandafter\edef\csname EQLABEL#1\endcsname
{\the\chapno.\the\eqlabelno?}\eqno(\ifchapternumbers\chapfolio.\fi
\the\eqlabelno)\ifproofmode\rlap{\hbox{\marginstyle #1}}\fi}

\def\eqalignnum{\global\advance\eqlabelno by 1
&(\ifchapternumbers\chapfolio.\fi\the\eqlabelno)}

\def\eqalignlabel#1{\global\advance\eqlabelno by 1 \ifproofmode 
\ifforwardreference\immediate\write1{\noexpand\expandafter\noexpand\def
\noexpand\csname EQLABEL#1\endcsname{\the\chapno.\the\eqlabelno?}}\fi\fi
\global\expandafter\edef\csname EQLABEL#1\endcsname
{\the\chapno.\the\eqlabelno?}&(\ifchapternumbers\chapfolio.\fi
\the\eqlabelno)\ifproofmode\rlap{\hbox{\marginstyle #1}}\fi}

\def\eqref#1{\hbox{(\ifundefined{EQLABEL#1}***)\ifproofmode\ifforwardreference%
\else\write16{ ***Undefined Equation Reference #1*** }\fi
\else\write16{ ***Undefined Equation Reference #1*** }\fi
\else\edef\LABxx{\getlabel{EQLABEL#1}}%
\def\LAByy{\expandafter\stripchap\LABxx}\ifchapternumbers%
\chapshow{\LAByy}.\expandafter\stripeq\LABxx%
\else\ifnum\number\LAByy=\chapno\relax\expandafter\stripeq\LABxx%
\else\chapshow{\LAByy}.\expandafter\stripeq\LABxx\fi\fi)\fi}%
\ifproofmode\write2{Equation #1}\fi}

\def\fignum{\global\advance\figureno by 1
\relax\iffigurechapternumbers\chapfolio.\fi\the\figureno}

\def\figlabel#1{\global\advance\figureno by 1
\relax\ifproofmode\ifforwardreference
\immediate\write1{\noexpand\expandafter\noexpand\def
\noexpand\csname FIGLABEL#1\endcsname{\the\chapno.\the\figureno?}}\fi\fi
\global\expandafter\edef\csname FIGLABEL#1\endcsname
{\the\chapno.\the\figureno?}\iffigurechapternumbers\chapfolio.\fi
\ifproofmode\llap{\hbox{\marginstyle#1
\kern1.2truein}}\relax\fi\the\figureno}

\def\figref#1{\hbox{\ifundefined{FIGLABEL#1}!!!!\ifproofmode\ifforwardreference%
\else\write16{ ***Undefined Figure Reference #1*** }\fi
\else\write16{ ***Undefined Figure Reference #1*** }\fi
\else\edef\LABxx{\getlabel{FIGLABEL#1}}%
\def\LAByy{\expandafter\stripchap\LABxx}\iffigurechapternumbers%
\chapshow{\LAByy}.\expandafter\stripeq\LABxx%
\else\ifnum \number\LAByy=\chapno\relax\expandafter\stripeq\LABxx%
\else\chapshow{\LAByy}.\expandafter\stripeq\LABxx\fi\fi\fi}%
\ifproofmode\write2{Figure #1}\fi}

\def\tabnum{\global\advance\tableno by 1
\relax\iftablechapternumbers\chapfolio.\fi\the\tableno}

\def\tablabel#1{\global\advance\tableno by 1
\relax\ifproofmode\ifforwardreference
\immediate\write1{\noexpand\expandafter\noexpand\def
\noexpand\csname TABLABEL#1\endcsname{\the\chapno.\the\tableno?}}\fi\fi
\global\expandafter\edef\csname TABLABEL#1\endcsname
{\the\chapno.\the\tableno?}\iftablechapternumbers\chapfolio.\fi
\ifproofmode\llap{\hbox{\marginstyle#1
\kern1.2truein}}\relax\fi\the\tableno}

\def\tabref#1{\hbox{\ifundefined{TABLABEL#1}!!!!\ifproofmode\ifforwardreference%
\else\write16{ ***Undefined Table Reference #1*** }\fi
\else\write16{ ***Undefined Table Reference #1*** }\fi
\else\edef\LABtt{\getlabel{TABLABEL#1}}%
\def\LABTT{\expandafter\stripchap\LABtt}\iftablechapternumbers%
\chapshow{\LABTT}.\expandafter\stripeq\LABtt%
\else\ifnum\number\LABTT=\chapno\relax\expandafter\stripeq\LABtt%
\else\chapshow{\LABTT}.\expandafter\stripeq\LABtt\fi\fi\fi}%
\ifproofmode\write2{Table#1}\fi}

\def\eq{Eq.~}

\newdimen\sectionskip     \sectionskip=20truept
\newcount\sectno
\def\section#1#2{\sectno=0 \null\vskip\sectionskip
    \centerline{\chaplabel{#1}.~~{\bf#2}}\nobreak\vskip.2truein
    \noindent\ignorespaces}

\def\advancesectno{\global\advance\sectno by 1}
\def\sectfolio{\number\sectno}
\def\subsection#1{\goodbreak\advancesectno\null\vskip10pt
                  \noindent\chapfolio.~\sectfolio.~{\bf #1}
                  \nobreak\vskip.05truein\noindent\ignorespaces}

\def\uttg#1{\null\vskip.1truein
    \ifproofmode \line{\hfill{\bf Draft}:
    UTTG--{#1}--\number\year}\line{\hfill\today}
    \else \line{\hfill UTTG--{#1}--\number\year}
    \line{\hfill\ifcase\month\or January\or February\or March\or April\or May\or June
    \or July\or August\or September\or October\or November\or December\fi
    \space\number\year}\fi}

\def\contents{\noindent
   {\bf Contents\Z}\nobreak\vskip.05truein\noindent\ignorespaces}

\def\getlabel#1{\csname#1\endcsname}
\def\ifundefined#1{\expandafter\ifx\csname#1\endcsname\relax}
\def\stripchap#1.#2?{#1}
\def\stripeq#1.#2?{#2}

%
\catcode`@=11 
\def\space@ver#1{\let\@sf=\empty\ifmmode#1\else\ifhmode%
\edef\@sf{\spacefactor=\the\spacefactor}\unskip${}#1$\relax\fi\fi}
\newcount\referencecount     \referencecount=0
\newif\ifreferenceopen       \newwrite\referencewrite
\newtoks\rw@toks
\def\refmark#1{\relax[#1]}
\def\refend{\refmark{\number\referencecount}}
\newcount\lastrefsbegincount \lastrefsbegincount=0
\def\refsend{\refmark{\count255=\referencecount%
\advance\count255 by -\lastrefsbegincount%
\ifcase\count255 \number\referencecount%
\or\number\lastrefsbegincount,\number\referencecount%
\else\number\lastrefsbegincount-\number\referencecount\fi}}
\def\refch@ck{\chardef\rw@write=\referencewrite
\ifreferenceopen\else\referenceopentrue
\immediate\openout\referencewrite=referenc.texauxil \fi}
%
{\catcode`\^^M=\active 
  \gdef\obeyendofline{\catcode`\^^M\active \let^^M\ }}%
%
{\catcode`\^^M=\active 
  \gdef\ignoreendofline{\catcode`\^^M=5}}
{\obeyendofline\gdef\rw@start#1{\def\t@st{#1}\ifx\t@st\blankend%
\endgroup\@sf\relax\else\ifx\t@st\bl@nkend\endgroup\@sf\relax%
\else\rw@begin#1
\backtotext
\fi\fi}}
{\obeyendofline\gdef\rw@begin#1
{\def\n@xt{#1}\rw@toks={#1}\relax%
\rw@next}}
\def\blankend{}
{\obeylines\gdef\bl@nkend{
}}
\newif\iffirstrefline  \firstreflinetrue
\def\rwr@teswitch{\ifx\n@xt\blankend\let\n@xt=\rw@begin%
\else\iffirstrefline\global\firstreflinefalse%
\immediate\write\rw@write{\noexpand\obeyendofline\the\rw@toks}%
\let\n@xt=\rw@begin%
\else\ifx\n@xt\rw@@d \def\n@xt{\immediate\write\rw@write{%
\noexpand\ignoreendofline}\endgroup\@sf}%
\else\immediate\write\rw@write{\the\rw@toks}%
\let\n@xt=\rw@begin\fi\fi\fi}
\def\rw@next{\rwr@teswitch\n@xt}
\def\rw@@d{\backtotext} \let\rw@end=\relax
\let\backtotext=\relax

\newdimen\refindent     \refindent=30pt
\def\Textindent#1{\noindent\llap{#1\enspace}\ignorespaces}
\def\refitem#1{\par\hangafter=0 \hangindent=\refindent\Textindent{#1}}
\def\REFNUM#1{\space@ver{}\refch@ck\firstreflinetrue%
\global\advance\referencecount by 1 \xdef#1{\the\referencecount}}
\def\refnum#1{\space@ver{}\refch@ck\firstreflinetrue%
\global\advance\referencecount by 1\xdef#1{\the\referencecount}\refend}

\def\REF#1{\REFNUM#1%
\immediate\write\referencewrite{%
\noexpand\refitem{#1.}}%
\begingroup\obeyendofline\rw@start}
\def\ref{\refnum\?%
\immediate\write\referencewrite{\noexpand\refitem{\?.}}%
\begingroup\obeyendofline\rw@start}
\def\Ref#1{\refnum#1%
\immediate\write\referencewrite{\noexpand\refitem{#1.}}%
\begingroup\obeyendofline\rw@start}
\def\REFS#1{\REFNUM#1\global\lastrefsbegincount=\referencecount%
\immediate\write\referencewrite{\noexpand\refitem{#1.}}%
\begingroup\obeyendofline\rw@start}

\def\REFSCON#1{\REF#1}

\def\cite#1{\refmark#1}
\catcode`@=12 
%
%
\input epsf.tex
%
%
\proofmodefalse
\baselineskip=15pt plus 1pt minus 1pt
\parskip=5pt
\chapternumberstrue
\forwardreferencefalse
\figurechapternumberstrue
\tablechapternumberstrue
\ifproofmode
\immediate\openout2=allcrossreferfile \fi
\ifforwardreference\input labelfile
\ifproofmode\immediate\openout1=labelfile \fi\fi
\noblackboxes
\hfuzz=1pt
\vfuzz=2pt
%
%
\def\hourandminute{\count255=\time\divide\count255 by 60
\xdef\hour{\number\count255}
\multiply\count255 by -60\advance\count255 by\time
\hour:\ifnum\count255<10 0\fi\the\count255}

\def\chaplabel#1{\bumpchapno\ifproofmode\ifforwardreference
\immediate\write1{\noexpand\expandafter\noexpand\def
\noexpand\csname CHAPLABEL#1\endcsname{\the\chapno}}\fi\fi
\global\expandafter\edef\csname CHAPLABEL#1\endcsname
{\the\chapno}\ifproofmode
\llap{\hbox{\marginstyle #1\ifnum\chapno > -1\ \else
\hskip1.3truein\fi}}\fi\chapfolio}

\def\section#1#2{\sectno=0 \null\vskip\sectionskip
    \ifnum\chapno > -1 
    \centerline{\fourteenrm\chaplabel{#1}.~~\fourteenbold#2}
    \else
    \centerline{\fourteenbold Appendix\ \chaplabel{#1}: {#2}}\fi
    \nobreak\vskip.2truein
    \noindent\ignorespaces}
\def\subsection#1{\goodbreak\advancesectno\null\vskip10pt
                  \noindent{\it \chapfolio.\sectfolio.~#1}
                  \nobreak\vskip.05truein\noindent\ignorespaces}
\def\subsubsection#1{\goodbreak
                  \noindent$\underline{\hbox{#1}}$
                  \nobreak\vskip-5pt\noindent\ignorespaces}
\def\cite#1{\refmark{#1}}
\def\\{\hfill\break}
\def\cropen#1{\crcr\noalign{\vskip #1}}
\def\contents{\line{{\fourteenbold Contents}\hfill}\nobreak\vskip.05truein\noindent%
              \ignorespaces}

\def\titlebox#1#2{\lower7pt\hbox{%
\hsize=.75in\vbox{\vskip5pt\centerline{#1}\vskip5pt\centerline{#2}}}}

\font\mathbb msbm7 at 10pt
\font\sevenmathbb msbm7
\font\frak eufm10

\def\ie{{\it i.e.\ \/}}

\def\Fp{\hbox{\mathbb F}_{\kern-2pt p}}
\def\sevenFp{\hbox{\sevenmathbb F}_p}
\def\sevenFpstar{\hbox{\sevenmathbb F}^*_p}
\def\Fpstar{\Fp^*}
\def\F{\Fp^5}
\def\sevenF{\sevenFp^5}
\def\Fstar{(\Fp^*)^5}
\def\sevenFstar{(\sevenFp^*)^5}

\def\Fq{\hbox{\mathbb F}_q}
\def\sevenFq{\hbox{\sevenmathbb F}_q}
\def\sevenFqstar{\hbox{\sevenmathbb F}^*_q}
\def\Fqstar{\Fq^*}

\def\goth{\hbox{\frak G}}

\font\bigcmmib=cmmib7 at 14pt
\font\bigcmbx=cmbx10 at 14pt
\def\bignupsi{\hbox{\bigcmmib\char'027\bigcmbx(\bigcmmib\char'040\bigcmbx)}}

\def\bb#1{\hbox{\mathbb #1}}
\def\bm{{\bf m}}
\def\bone{{\bf 1}}

\def\notdiv{\hbox{$\not|$\kern3pt}}
\def\vptrunc#1{\,{}^{(#1)}\vp_0}
\def\ftrunc#1#2{\,\,{}^{#2}f_{#1}}
\def\modp{\hbox{(mod $p$)}}
\def\ord#1{\ca{O}\kern-2pt\left(#1\right)}
\def\ordp#1{\hbox{ord}_p\left(#1\right)}
\def\teich{\hbox{Teich}}
\def\cnf{\hbox{cnf}}
\def\poch#1#2{\left(#1\right)_{#2}}
\def\smallfrac#1#2{\textstyle{\scriptstyle #1\over \scriptstyle #2}}
\def\pts{\hbox{pts}}
\def\smallpts{\hbox{\sevenrm pts}}
%
%
\nopagenumbers\pageno=0
\null
\vbox{\baselineskip=12pt
\rightline{\eightrm hep-th/0012233}\vskip-3pt
\rightline{\eightrm 24 December 2000}
\vskip0.7truein
\centerline{\seventeenrm Calabi-Yau Manifolds}
\vskip.3truein
\centerline{\seventeenrm Over}
\vskip.3truein
\centerline{\seventeenrm Finite Fields, I}
\vskip0.6truein
\centerline{%
      {\csc Philip~Candelas}$^1$,\qquad
      {\csc Xenia~de~la~Ossa}$^1$}}
\vskip.1truein
\centerline{\csc and}
\vskip.1truein
\centerline{\csc \hphantom{$^2$}{\csc Fernando Rodriguez Villegas}$^2$}
\vskip.4truein\bigskip
\centerline{
\vtop{\hsize = 3.0truein
\centerline{$^1$\it Mathematical Institute}
\centerline{\it Oxford University}
\centerline{\it 24-29 St.\ Giles'}
\centerline{\it Oxford OX1 3LB, England}}
\vtop{\hsize = 3.0truein
\centerline{$^2$\it Department of Mathematics}
\centerline{\it University of Texas}
\centerline{\it Austin}
\centerline{\it TX 78712, USA}}}
\vskip0.4in\bigskip
\centerline{\bf ABSTRACT}
\vskip.1truein 
\noindent We study \cys\ defined over finite fields. These manifolds have 
parameters, which now also take values in the field and we compute the number
of rational points of the manifold as a function of the parameters. The
intriguing result is that it is possible to give explicit expressions
for the number of rational points in terms of the periods of the holomorphic
three-form. We show also, for a one parameter family of quintic threefolds, that the
number of rational points of the manifold is closely related to as the number of rational
points of the mirror manifold. Our interest is primarily with \cy\ threefolds however we
consider also the interesting case of elliptic curves and even the case of a quadric in
$\bb{P}_1$ which   is a zero dimensional \cym. This zero dimensional manifold has trivial
dependence on the parameter over $\bb{C}$ but a not trivial arithmetic~structure.  
%
%
\newpage
\vbox{\baselineskip=5pt
\contents
\vskip2pt
\item{1.~}Introduction
\vskip3pt
\item{2.~}Preliminaries
\itemitem{2.1~}{\it Field theory for physicists}
\itemitem{2.2~}{\it p-adic numbers}
\itemitem{2.3~}{\it Some useful congruences}
\itemitem{2.4~}{\it The p-adic $\G$-Function}
\itemitem{2.5~}{\it The Teichm\"uller representative}
\itemitem{2.6~}{\it The coefficients of the fundamental period}
\vskip3pt
\item{3.~}The Periods and Semiperiods of the Quintic
\itemitem{3.1~}{\it The periods}
\itemitem{3.2~}{\it The semiperiods}
\vskip3pt
\item{4.~}$\nu(\ps)$ in Zeroth Order
\itemitem{4.1~}{\it Generalities}
\itemitem{4.2~}{\it The calculation to zeroth order}
\itemitem{4.3~}{\it The zeroth order expression for a general CY hypersurface in a toric
variety}
\itemitem{4.4~}{\it A congruence involving the Frobenius map}
\vskip3pt
\item{5.~}$\nu(\ps)$ in First Order
\itemitem{5.1~}{\it The first order calculation}
\itemitem{5.2~}{\it The sum over monomials}
\vskip3pt
\item{6.~}$\nu(\ps)$ in Higher Order and for Finer Fields
\itemitem{6.1~}{\it An ansatz for the case $5\notdiv p-1$}
\itemitem{6.2~}{\it The method of Frobenius}
\itemitem{6.3~}{\it The number of rational points over \smash{$\bb{F}_{p^s}^5$} when
$5\notdiv p^s-1$}
\vskip3pt
\item{7.~}\cy\ Manifolds of Zero Dimension
\itemitem{7.1~}{\it The rational points of a quadric}
\itemitem{7.2~}{\it Solution in terms of periods}
\vskip3pt 
\item{8.~}CY Manifolds of Dimension One: Elliptic Curves
\vskip3pt
\item{9.~}The Relation to Gauss Sums
\itemitem{9.1~}{\it Dwork's character}
\itemitem{9.2~}{\it Evaluation of $\nu(\ps)$ in terms of Gauss sums}
\itemitem{9.3~}{\it The case \smash{$5|p-1$}}
\itemitem{9.4~}{\it Relation to the periods}
\itemitem{9.5~}{\it The calculation for finer fields}
\itemitem{9.6~}{\it A CY hypersurface in a toric variety}
}
\newpage
 %
 %
\headline={\ifproofmode\hfil\eightrm draft:\ \today\
\hourandminute\else\hfil\fi}
\footline={\rm\hfil\folio\hfil}
\pageno=1
\section{intro}{Introduction}
\cys\ owe many remarkable properties to the special role that they play in
relation to supersymmetry and to String Theory. It is a fundamental fact that
these manifolds depend holomorphically on parameters that determine the
complex structure and \K-class. The variation of the complex structure of a
\cym, $M$, is naturally studied through it periods, that is the periods
 $$
\vp_j~=~\int_{\G^j}\O \eqlabel{integrals}$$
of the (unique) holomorphic (3,0)-form over a basis of three dimensional
homology cycles. The periods describe how $\O$ moves inside $H^3(M,\bb{Z})$
and play a fundamental role in the geometry of the parameter space~
\REFS{\SpecialGeomS}{A. Strominger, ``Special Geometry'', \cmp{133} 163 (1990).}
\REFSCON{\SpecialGeomCD}{P. Candelas and X. de la Ossa,\\ ``Moduli
Space of Calabi-Yau Manifolds'', \npb{355} 455 (1991).}
\refsend
~and hence also in mirror symmetry. Closely related periods play a crucial role
also in Seiberg-Witten theory~
\Ref{\SeibergWitten}{N. Seiberg and E. Witten,  ``Electric-Magnetic Duality,
Monopole Condensation, and Confinement in N=2 Supersymmetric Yang-Mills
Theory'',\\ \npb{426} 19 (1994), Erratum-ibid. {\bf B430} 485 (1994), 
hep-th/9407087.}.

It has been known for some time that the periods encode also arithmetic properties
of the underlying manifold~ 
\REFS{\Moore}{G. Moore, ``Attractors and Arithmetic'', hep-th/9807056,\\
``Arithmetic and Attractors, hep-th/9807087'',\\
S. D. Miller and G. Moore, ``Landau-Siegel Zeroes and Black Hole Entropy'',\\ 
hep-th/9903267. } 
\REFSCON{\LianYau}{B. H. Lian and S.-T. Yau, ``Arithmetic Properties of Mirror Map and
Quantum Coupling'',  \cmp{176} 163 (1996), hep-th/9411234.}
\refsend.
Consider, for example, the one parameter family of quintic threefolds, $M_\ps$, defined by
the vanishing of the polynomials
 $$
P(x,\ps)~=~\sum_{i=1}^5 x_i^5 - 5\ps\, x_1 x_2 x_3 x_4 x_5 \eqlabel{quintic} $$
in $\bb{P}^4$. It is a consequence of mirror symmetry that the Yukawa coupling
may be expanded in the form
 $$
y_{ttt}~=~ 
5 + \sum_{k=1}^\infty {n_k k^3 q^k\over 1-q^k} ~~, ~~~q = e^{2\pi i t}
\eqlabel{yukone}$$
where the coordinate $t$ and the Yukawa coupling can both be expressed in terms of
certain periods $\vp_0$ and $\vp_1$
 $$
y_{ttt}~=~\left({5\over 2\p i}\right)^3 {5\,\ps^2\over \vp_0(\ps)^2 \,(1-\ps^5)}\,
\left({d\ps\over dt}\right)^3 ~~,~~~~ 
t~=~{1\over 2\p i}\,{\vp_1(\ps)\over\vp_0(\ps)}~.\eqlabel{yuktwo}$$
The essential point is that \eqref{yuktwo} is a combination
of periods while \eqref{yukone} is a series with integer coefficients $n_k$
that count the numbers of rational curves of $M_\ps$. It has been pointed out by
Lian and Yau \cite{\LianYau} that integral series enter at an earlier stage.  For example,
in order to  expand the Yukawa coupling as a $q$-series we have to invert the relation
between $\ps$ and~$t$. Doing this we find the integral series
 $$\eqalign{
{1\over (5\ps)^5} = q + & 154\,q^2 + 179139\,q^3 + 313195944\,q^4 
+ 657313805125\,q^5 + 1531113959577750\,q^6\cr 
&+ 3815672803541261385\,q^7+ 9970002717955633142112\,q^8+ \ldots~.\cr} \eqlabel{mmap}
  $$
It is shown in \cite{\LianYau} that the coefficients in this and similar series are
indeed all integers and in some examples (though not the series for the quintic) that these
series are Thompson series related to the Monster Group. The fact that the
coefficients are integers in the series \eqref{mmap} is the more remarkable since
the coefficients that appear in the period $\vp_1$ are {\sl not\/} all~integers.
The fact that the construction of \cys\ involves combinatoric if not number theoretic
procedures is evident as soon as one attempts any reasonably general construction and is
especially evident in the toric constructions and the expressions that Batyrev~
\Ref{\Batyrev}{V. V. Batyrev, Duke math J. 69 (1993) 31.}\ 
(see also~
\Ref{\AGM}{P. S. Aspinwall, B. R. Green and D. R. Morrison,\\ Int. Math. Res. Notices
(1993) 319.})  
gives for the Hodge numbers in terms of the toric~data.

We turn now to a statement of our main result which expresses the number of rational
points of the quintic \eqref{quintic} over $\bb{F}_p$ in terms of the periods of $\O$. The
derivation of this expression will be explained in the following sections. To write
out the expression we have to recall some facts pertaining to the periods. The
periods satisfy a system of differential equations, the Picard-Fuchs equations, with
respect to the parameters. Specifically, for the family \eqref{quintic} there are 204
periods, but among them there are four periods which we shall denote by $\vp_0$, $\vp_1$,
$\vp_2$ and $\vp_3$, which coincide with the periods of the mirror manifold and satisfy 
the equation
 $$
\ca{L}\,\vp_j~=~0~~~\hbox{with}~~~
\ca{L}~=~\vth^4 - 5\l\,\prod_{i=1}^4(5\vth+i)\ , \eqlabel{PF}$$
where here and in the following it is convenient to take the parameter to be
 $$
\l~=~{1\over (5\ps)^5}\ ,$$
and $\vth$ denotes the logarithmic derivative $\l{d\over d\l}$. Consider now the
behaviour of the periods in the neighborhood of $\l=0$. The Picard-Fuchs
equation \eqref{PF} has all four of its indices equal to zero. Thus the solutions
are asymptotically like $1$, $\log\l$, $\log^2\l$ and $\log^3\l$. We denote by
$f_0$ the solution that is analytic at $\l=0$. In fact this solution is given by
the series
 $$
f_0(\l)~=~\sum_{m=0}^\infty {(5m)!\over (m!)^5} \l^m~.$$
We can choose the four periods to be of the form
 $$\eqalign{
\vp_0(\l)&~=~f_0(\l)\cr
\vp_1(\l)&~=~f_0(\l)\,\log\l + f_1(\l)\cr
\vp_2(\l)&~=~f_0(\l)\,\log^2\l + 2f_1(\l)\,\log\l + f_2(\l)\cr
\vp_3(\l)&~=~f_0(\l)\,\log^3\l + 3f_1(\l)\,\log^2\l + 3f_2(\l)\,\log\l + f_3(\l)\cr}
\eqlabel{periods}$$ 
where the $f_j(\l)$ are power series in $\l$. 

Now these periods correspond to certain cycles $\G^j$ in \eqref{integrals} and so
can be computed, in principle, by direct integration. The
Picard-Fuchs equation \eqref{PF} is then the differential equation satisfied by the
four $\vp_j(\l)$. For the case of \cys\ that can be realised as hypersurfaces in
toric varieties, which is a wide class with the quintic as the simplest example,
the manifold can be associated with the Newton polyhedron, $\D$, of the
monomials that appear in the polynomial that defines the hypersurface. In these
cases, given $\D$, there is a purely combinatoric way of finding a differential
system that the periods satisfy; this yields the GKZ system~
\Ref{\GKZ}{I. Gelfand, M. Kapranov and A. Zelevinsky,
``Generalized Euler Integrals and A-hypergeometric Functions'', Adv. in Math. {\bf 84}
(1990) 255.}. 
A fact which is not fully understood is that the GKZ system is often of higher
order than the Picard-Fuchs system. So while it is true that the periods satisfy the
differential system that one deduces from $\D$ there are also often additional solutions
of this system that are not periods. These extra solutions are called semiperiods and their
appearance is somewhat mysterious. It turns out that for the quintic there is a
semiperiod and it plays a role in our expressions. 

For the quintic the GKZ operator, $\ca{L}^\D$, is related to the Picard-Fuchs
operator by
 $$
\ca{L}^\D~=~\vth\,\ca{L}~=~\vth^5 - \l\,\prod_{i=1}^5(5\vth+i)~.$$
The first equality shows that the periods $\vp_j(\l)$ satisfy the new equation and
the second equality shows that the new operator has all five of the indices
corresponding to $\l=0$ equal to zero. The semiperiod is thus of the form
 $$
\vp_4(\l)~=~f_0(\l)\,\log^4\l + 4\,f_1(\l)\,\log^3\l + 6\,f_2(\l)\,\log^2\l  
+ 4\,f_3(\l)\,\log\l + f_4(\l)\ , \eqlabel{semiperiod}$$
with the power series $f_0, f_1, f_2, f_3$ as in \eqref{periods}. We denote by
$\Fp$ the finite field with $p$ elements and for $\ps\in\Fp$, $p\neq 5$, we denote by
$\nu(\ps)$ the number of rational points of $M$ 
 $$
\nu(\ps)~=~\#\{x\in\Fp^5 ~|~ P(x,\ps) = 0~~\}~.$$
We denote also by $\ftrunc{j}{n}$ the truncation of the series $f_j$ to $n+1$
terms. Thus for example
 $$
\ftrunc{0}{n}(\l)~=~\sum_{m=0}^n\, {(5m)!\over (m!)^5}\,\l^m~.$$
With these conventions we can state our result most simply for the case that 5 does
not divide $p-1$
$$\eqalign{
\n(\ps)~=~&\ftrunc{0}{(p-1)}(\l^{p^4}) +
\left({p\over 1-p}\right)\ftrunc{1}{(p-1)}'(\l^{p^4}) 
+ {1\over 2!}\left({p\over 1-p}\right)^2\ftrunc{2}{(p-1)}''(\l^{p^4})\cropen{5pt}
&+ {1\over 3!}\left({p\over 1-p}\right)^3\ftrunc{3}{(p-1)}'''(\l^{p^4}) 
+ {1\over 4!}\left({p\over 1-p}\right)^4\ftrunc{4}{(p-1)}''''(\l^{p^4})\hskip20pt
(\hbox{mod}\,p^5)\ ,\cr}\eqlabel{firstresult}$$
where the coefficients ${1\over j!}\left({p\over 1-p}\right)^j$ are understood to be
expanded p-adically. We extend this result in three directions: by writing an exact p-adic
expression, by extending the result to finer fields with $q=p^s$ elements and by extending
the result to cover the interesting case that $5|q-1$. If $5\notdiv q-1$ then the
generalization is immediate. We are simply to replace $p$ by $q$, extend the sum to
infinity with a certain natural definition of the new terms $f_j$ for $j>4$ and replace
$\l$ by its Teichm\"uller representative. The case that
$5|q-1$ is of practical importance since for every $p\neq 5$ there is a least integer $\r$
such that $5|p^\r - 1$ and, depending on $p$, $\r=1,2~\hbox{or}~4$. Moreover this case
corresponds to the existence of nontrivial fifth roots of unity and is of additional
interest in virtue of mirror symmetry. Precisely when $5|q-1$ there are additional
contributions to $\n(\ps)$ arising from the other 200 periods. We compute these
contributions which, apart from corresponding to these other periods, have a structure
analogous to the terms we have exhibited.

There is an intuitive reason that $\nu(\ps)$ should be related to the periods in virtue
of the action of the Frobenius map on the cohomology of $M$. We may regard the
quintic as a polynomial with variables $x_i$ that take values in some sufficiently
large field $K\supset \Fp$ but with coefficients in $\Fp$. In this case the
Frobenius map 
 $$
(x_1,x_2,x_3,x_4,x_5)\mapsto (x_1^p,x_2^p,x_3^p,x_4^p,x_5^p) $$
acts as an automorphism on $M$ since
 $$
P(x,\ps)\equiv 0~~\modp~~\Leftrightarrow~~0\equiv P(x,\ps)^p\equiv P(x^p,\ps)~~\modp $$ 
this is an automorphism since if $x^p\equiv y^p$ then $x^p-y^p\equiv 0$ so
$(x-y)^p\equiv 0$ and $x-y\equiv 0$. The fixed points of the Frobenius map are
the points for which $x^p\equiv x$ which is precisely the condition that characterises
$\Fp\subset K$. Thus the rational points of $M$ are precisely the fixed points of
the Frobenius map. Their number is then the Euler number of the fixed point set and
the Lefshetz theorem relates this to a trace over the Frobenius action on the
cohomology of $M$. In effect what we do in this paper is to evaluate this trace though we
do so somewhat indirectly.

This paper is principally concerned with the calculation of the number of rational points
of the quintic over $\Fq$. We have attempted to make the article reasonably
self-contained and so we include in \SS2 a number of preliminaries relating to finite
fields and p-adic analysis which, though standard, are not widely known to physicists. In
\SS3 we give an account of the periods and a class of semiperiods which we define and
which prove to be of importance for the computation of $\n(\ps)$. In \SS4 we present the
zeroth order calculation, that is the calculation of $\n(\ps)$ mod $p$, and we derive the
result 
 $$
\n(\ps)~\equiv~\ftrunc{0}{\left[p/5\right]}(\l)~~\bmod{p}$$
where $\left[p/5\right]$ denotes the integer part of $p/5$. This generalizes in a
straightforward way a result that is well known for elliptic curves. It was this result
that aroused our interest in this calculation since one inevitably asks why $\n(\ps)$
should be related to a period.

It is perhaps natural to expand the number of points in base $p$
 $$
\n(\ps)~=~\n_0(\ps) + p\,\n_1(\ps) + p^2\,\n_2(\ps) + p^3\,\n_3(\ps) + p^4\,\n_4(\ps), $$
with $0\leq \n_i(\ps)\leq p-1$, and to seek to evaluate the coefficients successively
by performing perturbation theory in powers of $p$. This leads inevitably to p-adic
analysis since it is in this sense that the successive terms are smaller and smaller.
Whereas the zeroth order calculation is elementary the first order calculation,
neglecting terms of $\ord{p^2}$, requires more serious computation at least
computation that is more serious for the untrained physicist. The answer at first
order is
 $$\eqalign{
\n(\ps)&=\ftrunc{0}{(p-1)}(\l^p) + p\,\ftrunc{1}{(p-1)}'(\l^p)\cropen{3pt}  
&\hskip20pt -\d_p\, p\,\sum_{\bf v}{\g_{\bf v}\over \prod_{i=1}^5 (v_ik)!}\kern5pt
{}^{(p-1)}\kern-5pt{}^{}_2F_1^{}\kern-3pt\left(a_{\bf v},b_{\bf v};
c_{\bf v};\,\ps^{-5}\right) +\ord{p^2}\cr} $$
where $\d_p=1$ if $5|p-1$ and $\d_p=0$ otherwise, $k=(p-1)/5$ and the $\bf v$-sum is
over quintic monomials $x^{\bf v}=x_1^{v_1}\ldots x_5^{v_5}$ with certain coefficients
$\g_{\bf v}$ that account for multiplicity. The ${}_2F_1$'s are the other 200 periods
and it is at this order that we first see their contribution. Since the first order
contribution was difficult to calculate and the result seems to show a pattern it is
natural to proceed by making an ansatz and checking this by counting the number of
solutions to $P(x,\ps)=0$ with a computer for low values of $p$. This process leads to
the expression \eqref{firstresult} given above. In \SS6 we take this expression and
extend it to an exact p-adic expression. In effect we extend the the sum in
\eqref{firstresult} to infinity and drop the proviso mod $p^5$. We are then able to
perform the sum over $j$ to obtain the expression in the form
 $$
\n(\ps)~=~\sum_{m=0}^{q-1}\b_{s,m}\teich^m(\l) $$
in terms of the Teichm\"uller representative of $\l$ and coefficients $\b_{s,m}$ for
which we give exact p-adic expressions in terms of the p-adic $\G$-function. It seems
that these expressions are interesting also for \cys\ of lower dimension and we discuss
this in
\SS7 for a \cym\ of dimension zero, a quadric in $\bb{P}_1$,
 $$
x_1^2 + x_2^2 -2\ps\, x_1 x_2 ~=~ 0~. $$
Over $\bb{C}$ there are of course two solutions for $\ps^2\neq 1$ and one solution when
$\ps^2=1$. Over $\bb{P}\Fp$ there can be 0, 1 or 2 solutions depending on $\ps$ and $p$. 
It is a standard exercise in number theory to calculate the number of solutions and this
number is usually expressed in terms of the Legendre symbol. We find it interesting that
it is also possible to define periods and semiperiods for this situation and calculate  
the number of periods in terms of these. This leads to interesting relations between the
periods and the Legendre symbol.

In \SS8 we turn to the case of elliptic curves. For the case of a cubic in $\bb{P}_2$ one
may simply replace 5 by 3 in our expressions for the quintic to obtain expressions for
the number of rational points. For a nonsingular elliptic curve we do not have to consider
separately the case $3|q-1$ since there are no contributions from `other periods'. A fact
that is interesting is that the number of rational points for the manifold and for its
mirror are the same. 
 
In \SS9 we turn to a direct evaluation of $\n(\ps)$ in terms of Dwork's character and 
Gauss sums. This establishes the expressions that were given previously. It is
perhaps possible to omit the previous sections and begin the paper here. The evaluation
of $\n(\ps)$ in terms of Gauss sums, however, obscures the relation to the periods that we
have sought to stress. It also requires a knowledge of the Gross-Koblitz formula which,
though standard, is established by appeal to sophisticated cohomology theories. We derive
here the complete form of the contributions of the other 200 periods whose form we new
previously only mod~$p^2$. These contributions are in correspondence with monomials of
degree 5 and degree~10. The fact that monomials of degree contribute is surprising. The
monomials in question are those of the form $x_1^4\,x_2^3\,x_3^2\,x_4$ together
with their permutations, and these contribute precisely when $\psi^5=1$ which is the case
that the manifold singularises to a conifold.  

A companion paper~
\Ref{\CYII}{``\cy\ Manifolds Over Finite Fields, II'', by P.~Candelas, X.~de la Ossa and
F.~Rodriguez Villegas, in preparation.}\
deals with mirror symmetry and the application of our expressions to the zeta-function.
Mirror symmetry impinges on the discussion since, apart from other considerations, the
contribution of the degree 5 monomials to the count of rational points for the quintic has
an appealing interpretation in terms of the monomial divisor mirror map
whereby it is natural to associate these contributions with the rational points of the
exceptional divisors that are introduced as blow ups of the mirror manifold in the toric
construction. 
\newpage
\section{prelims}{Preliminaries}
\vskip-20pt
\subsection{Field theory for physicists}
A (number) field $\bb{F}$ is a set on which it is possible to perform the operations of
arithmetic. That is the operations of addition and multiplication are defined and
have the usual associative and distributive properties. Moreover $\bb{F}$ is an abelian
group with respect to addition and \hbox{$\bb{F}^*=\bb{F}\setminus\{0\}$} is an abelian
group with respect to multiplication.
A field can have a finite or an infinite number of elements. The rationals, $\bb{Q}$,
the reals, $\bb{R}$, and the complex numbers, $\bb{C}$, are fields but the
integers, $\bb{Z}$, do not form a field since apart from $\pm 1$ the inverse of an integer
is not an integer.

Finite fields are classified up to isomorphism by the number of elements which is $p^s$ for
some prime $p$ and integer $s$. There is therefore no ambiguity in writing $\bb{F}_{p^s}$
to denote such a field. Moreover these fields exist for each $p$ and $s$.

The simplest finite field is the set of integers mod $p$
$$
\Fp~=~\{0,1,\dots,p-1\}$$
and the field $\bb{F}_{p^s}$ is the field obtained by adjoining to $\Fp$ the root of an
irreducible monic polynomial that has coefficients in $\Fp$ and is of order $s$. If $\r$
denotes a root of such a polynomial then $\bb{F}_{p^s}$ is the set
 $$
\bb{F}_{p^s}~=~\left\{a_0+a_1\r+a_2\r^2+\ldots+a_{s-1}\r^{s-1}~|~a_i\in\Fp\right\}$$
and it does not matter which irreducible polynomial or which root is chosen.
  
It is important that $p$ should be prime. If we were to consider integers mod 6,
say, we would have $2\times 3\equiv 0$ so that neither 2 nor 3 could have an inverse. If
however $p$ is prime then every nonzero element of the field has an inverse as we
illustrate below for the case of~$\bb{F}_7$. A fact that is not obvious is that $\Fpstar$
and more generally $\bb{F}_{p^s}^*$ are cyclic as multiplicative groups. In other words
for each such group there exists a generator $g$ such that for every $x\in \bb{F}_{p^s}^*$
we can write $x=g^j$ for some $j$ in the range $0\leq j\leq p^s-2$. For the case of
$\bb{F}_7$ the generator can be taken to be 3 as we illustrate in the table.

We shall be concerned later with the existence of nontrivial fifth roots of unity. That is
when do there exist elements $\z\in\bb{F}_{p^s}^*$, $\z\neq 1$ such that $\z^5=1$. Since
$\bb{F}_{p^s}^*$ is a group of order $p^s-1$ and a nontrivial fifth root, when it
exists, generates a subgroup of order~5 a nontrivial fifth root cannot exist unless
$5|p^s-1$.  Since the group $\bb{F}_{p^s}^*$ is in fact cyclic the 
$\z$'s exist precisely when the order of the group is divisible by 5 and the nontrivial
fifth roots of unity are then $\z=g^{rk}$ with $r=1,\ldots,4$ and~$k=(p^s-1)/5$. 
\vskip5pt
 $$\vbox{\def\skip{\hskip7pt}
\offinterlineskip\halign{
\strut #\vrule height 12pt depth 6pt&\skip  #\skip \hfil\vrule
&\skip  # \skip \hfil&\skip  # \skip \hfil&\skip  # \skip \hfil&\skip  # \skip \hfil
&\skip  # \skip \hfil&\skip  # \skip \hfil&\skip  # \skip \hfil\vrule\cr
\noalign{\hrule}
&\multispan8{\hfil $\bb{F}_7$\hfil\vrule}\cr
\noalign{\hrule\vskip3pt\hrule}
&$x$&0&1&2&3&4&5&6\cr
\noalign{\hrule}
&$x^{-1}$&*&1&4&5&2&3&6\cr
\noalign{\hrule}
&$y~;~3^y=x$&*&0&2&1&4&5&3\cr
\noalign{\hrule}
}}$$

There are two elementary facts that will be very useful to us. The first, an old result
going back to Fermat, is that $a^p=a$, or equivalently $a(a^{p-1}-1)=0$.
Thus
 $$
a^{p-1}~=~\cases{1,~ \hbox{if}~ a\ne 0\cropen{10pt} 
0,~ \hbox{if}~a=0~.\cr}\eqlabel{factone}$$
We will also need to evaluate sums of powers of field elements. Consider therefore the sum
 $$
\sum_{a\in\sevenFp}\,a^n~=~\sum_{a\in\sevenFp}\,(ba)^n~=~b^n\sum_{a\in\sevenFp}\,a^n~~;
~~b\neq 0$$
Hence we have
 $$
\sum_{a\in\sevenFp}\,a^n~=~
\cases{\- 0,~ \hbox{if $p-1$ does not divide $n$}\cropen{10pt} 
-1,~ \hbox{if $p-1$ divides $n$~.}\cr}\eqlabel{facttwo}$$

For a given prime $p$ the number of rational points, that is solutions of the equation
$P(x,\ps)=0$, is a definite integer which is moreover less than $p^5$ since there are
$p^5$ points altogether in $\Fp^5$. 

It is natural to expand the number of solutions in base $p$ and write
 $$
\n~=~\n^{(0)} + \n^{(1)}\, p + \n^{(2)}\, p^2 + \n^{(3)}\, p^3 + \n^{(4)}\, p^4
\eqlabel{Nseries}$$  
with $0\leq \n^{(j)}\leq p-1$ and then to evaluate mod~$p^2$,
mod~$p^3$, and so on. One proceeds by expanding all quantities as power series in $p$ as
if performing perturbation theory. This leads naturally into p-adic
analysis since it is in this sense that it is legitimate to regard $p$ as a small parameter.
\subsection{p-adic numbers}
Given an $r\in \bb{Q}$ and a prime $p$ we can write
 $$
r~=~{m\over n}~=~{m_0\over n_0}\,p^\a $$
where $m_0,n_0$ and $p$ have no common factor.
The p-adic norm of $r$ is defined to be
 $$
\| r \|_p~=~p^{-\a}~,~~~\| 0 \|_p~=~0$$
and is a norm; that is it has the properties 
 $$\eqalign{
\|r\|_p~&\geq ~0,~~\hbox{with equality if and only if $r=0$}\cr  
\|r_1\, r_2\|_p~&=~\|r_1\|_p\,\|r_2\|_p\cr
\|r_1+r_2\|_p~&\leq~\|r_1\|_p+\|r_2\|_p~.\cr}$$
With respect to the p-adic norm a number is smaller the more it is divisible by $p$.
For the series \eqref{Nseries}, for example, the successive terms decrease with respect
to the p-adic norm while they increase with respect to the usual norm.
If this seems artificial then we refer the skeptical reader to Ostrowski's theorem~
\Ref{\Koblitz}{``p-adic Numbers, p-adic Analysis, and Zeta Functions'', Second Edition,\\ 
N.~Koblitz, Graduate Texts in Mathematics, Springer 1984.}\
which states that any nontrivial norm is equivalent to the (usual) Archimedean norm
or to the p-adic norm for some prime $p$. 
 
 Norms are important in analysis and enter into the way that we build up number systems.
In the beginning there are integers and from these we obtain the rationals. Then we
observe that there are sequences of rational numbers which converge, but to limits that are
not rational, and we complete the rationals to the reals. Finally we extend the reals by
adjoining $\sqrt{-1}$ to obtain the complex numbers; a field of the greatest utility for
describing the natural world. Could we have deviated from this path? The answer is that
the process of passing from $\bb{Q}$ to $\bb{R}$ required the notion of a limit and hence
a norm. If instead of the usual Archimedian norm we use the p-adic norm we pass from
$\bb{Q}$ to $\bb{Q}_p$, the field of p-adic numbers. By adjoining to $\bb{Q}_p$ the roots
of all polynomials whose coefficients are in $\bb{Q}_p$ but whose roots are not in
$\bb{Q}_p$ and then completing the result we arrive at $\bb{C}_p$ which is the p-adic
analogue of $\bb{C}$. This construction has many applications in Number Theory. It is an
intriguing question as to whether it has significant application in Physics.
\vskip10pt
 $$
\matrix{\bb{Z}&\longrightarrow &\bb{Q} &\longrightarrow &\bb{R} &\longrightarrow &\bb{C}\cr
          &&\downarrow &&&&\cr 
          &&\hskip5pt\bb{Q}_p &&&&\cr
     					&&\downarrow &&&&\cr
          &&\hskip5pt\bb{C}_p &&&&\cr}$$
\vskip10pt
The p-adic numbers can be thought of as series of the form
 $$
X~=~\sum_{n=n_0}^\infty \x_n\,p^n ~~;~~\x_n\in\bb{Z}~,~~0\leq \x_n\leq p-1~.$$
We shall make reference to the p-adic integers $\bb{Z}_p$ which is the ring of p-adic
numbers for which $n_0\geq 0$. Equivalently $\bb{Z}_p$ is the set of $X$'s for which
$\|X\|_p \leq 1$. If $n_0=0$ and $\x_0\neq 0$ then $X$ is a {\it p-adic unit}.
The p-adic units form a group, $\bb{Z}_p^*$. If $X\in \bb{Z}_p^*$ then both $X$ and
$X^{-1}$ are p-adic integers and $\|X\|_p = 1$.

Since there is a norm defined for p-adic numbers there is also a p-adic analysis, with 
p-adic notions of limits and convergence. We
will be concerned with the convergence of power series. Owing to the
properties of the p-adic norm a p-adic power series $\sum_{n=0}^\infty \a_n x^n$  
converges if and only if $\a_n x^n\to 0$ as $n\to\infty$. As in classical analysis power
series have a radius of convergence but in p-adic analysis there is no notion of
conditional convergence. If $\a_n x_0^n\to 0$ for some $x_0$ then $\a_n x^n\to 0$ for all
$x$ with $\|x\|_p\leq\|x_0\|_p$. In particular a power series either converges for all  
the points on the circumference of its disk of convergence or it diverges for all these
points. Moreover the circumference of a disk has a `thickness' in p-adic analysis. Consider
the disk
 $$
D(a,R)~=~\left\{x\in \bb{Q}_p~\big|~\| x-a\|_p\leq R\right\}$$
and let $b$ be a point such that $\| b-a\|_p = R$. Consider now the disk $D(b,r)$ for 
some $r<R$. If $x\in D(b,r)$ then 
 $$
\|x-a\|_p~=~\|(x-b)+(b-a)\|_p~\leq \hbox{max}\left(\|(x-b)\|_p,~R\right)~=~R~.$$
so in fact $D(b,r)\subset D(a,R)$.
\vskip0pt
\vbox{
\def\disks{\vbox{\vskip0pt\hbox{\hskip0pt\epsfxsize=2truein\epsfbox{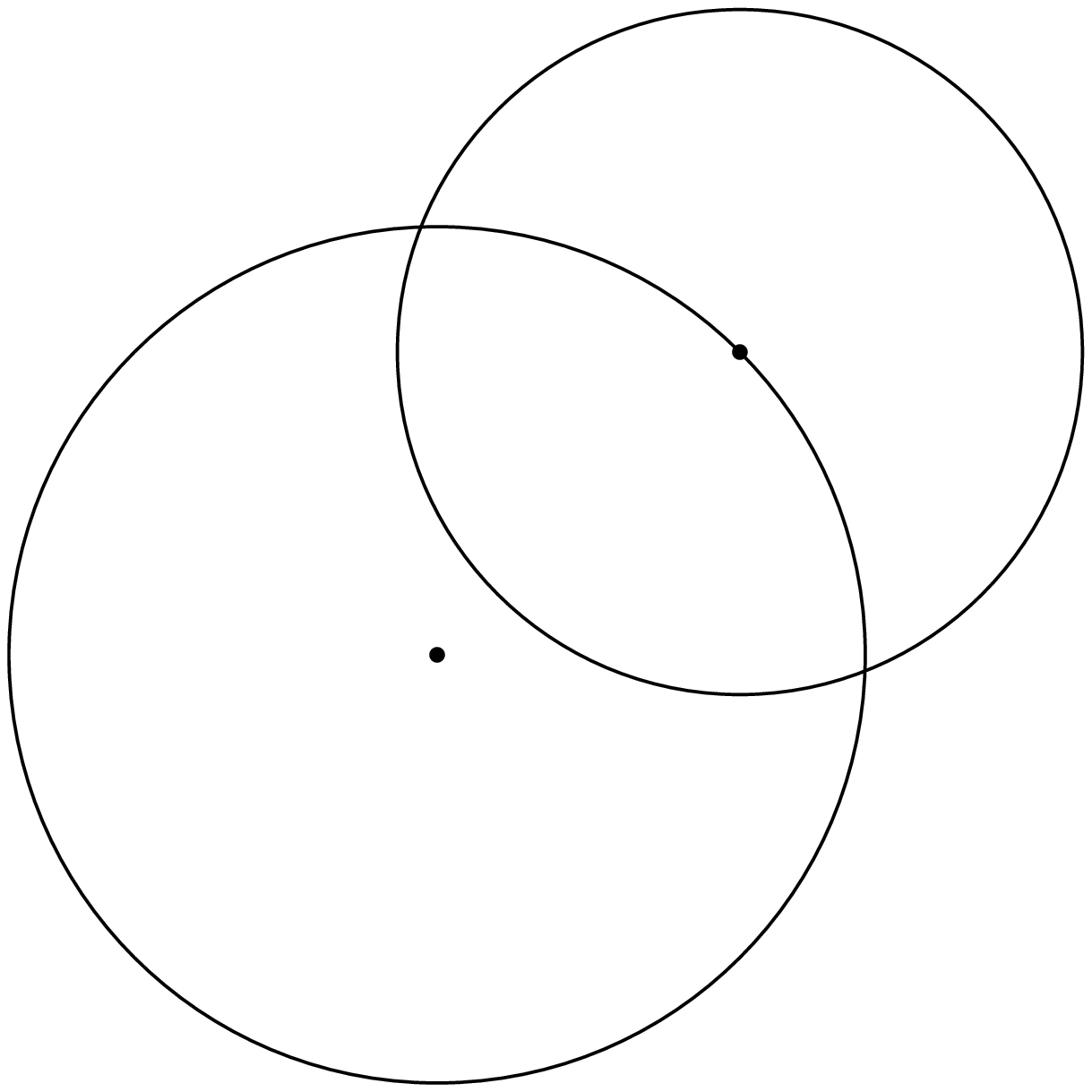}}}}
\figbox{\disks}{\figlabel{disk}}{A naive sketch of the disks $D(a,R)$ and $D(b,r)$. In 
reality $D(b,r)\subset D(a,R)$. }
\place{3}{1.6}{$a$}
\place{3.7}{2.35}{$b$}
}
\vskip10pt 
In p-adic analysis the series expansion of a composite function $f(g(x))$ is
more delicate than in classical analysis. Suppose $f$ and $g$ have series expansions that
converge in disks $D_f$ and $D_g$ and let $D\subset D_g$ be the largest disk such that
$x\in D\Rightarrow g(x)\in D_f$ then for $x\in D$ the series for $f(g(x))$ is obtained in
the usual way by expanding $f$ as a series in $g$ and then $g$ as a series in $x$. If the
resulting series is $\sum_{n=0}^\infty \a_n x^n$ then this series converges to $f(g(x))$
for $x\in D$ however for $x\not\in D$ the series may not converge or it may converge but
not to~$f(g(x))$.
We will meet a textbook example of this phenomenon in \SS\chapref{gauss}. If 
$f(Y)=\exp(\p Y)$ with $\exp$ denoting the usual exponential series and $\p$ a number in
$\bb{C}_p$ such that $\p^{p-1}=-p$ (this use of $\p$ has nothing to do with the classical
number that goes by that name) then it can be shown that the largest disk for which the
series expansion for $Y$ converges in the disk $\|Y\|_p < 1$. If we take
$g(X)=X-X^p$ then the $X$-disk that ensures $\| g(X)\|_p < 1$ is $\|X\|_p < 1$. For such 
$X$ 
 $$
f(g(X))~=~\sum_{n=0}^\infty c_n \p^n X^n$$
for certain coefficients $c_n$. It can be shown that the series on the right converges 
for $\|X\|_p < 1+\e$ for some fixed positive number $\e$ so that the series
$\sum_{n=0}^\infty c_n \p^n$, obtained by setting $X=1$, converges but it does not
converge to $f(g(1))=1$. 
\subsection{Some useful congruences}
We record here some useful congruences and relations which though elementary are perhaps
not generally known to physicists.

The first is Wilson's Theorem
 $$
(p-1)!~=~-1 + \ord{p}~.$$
To see this think of $(p-1)!$ as $\prod_{x\in \sevenFpstar} x$. Every $x$ has an inverse.
If $x$ and $x^{-1}$ are distinct then they cancel in the product. We are left with $x$'s
that satisfy $x^2-1=(x-1)(x+1)\equiv 0$ and these are 1 and $p-1$.

Next is a standard exercise~ 
\REF{\Schikhoff}{``Ultrametric Calculus: An Introduction to p-adic Analysis'', W. H.
Schikhoff,\\ Cambridge Studies in Advanced Mathematics, Cambridge University Press 1984.}
\cite{\Koblitz, \Schikhoff}: 
if $n\in \bb{Z}$ has the finite p-adic
expansion, that is expansion in base $p$, $n=c_0+c_1p+\ldots+c_\ell p^\ell$ then the p-adic
order of $n!$ is given by
 $$
\ordp{n!}~=~\sum_{j=1}^\ell \left[{n\over p^j}\right]~=~
{n-S(n)\over p-1} \eqlabel{p-order}$$
where $S(n)=\sum_{j=0}^\ell c_j$ denotes the sum of digits function and $[..]$ denotes the
integer part. The first equality states that between 1 and $n$ there are
$\left[{n\over p}\right]$ integers divisible by $p$, $\left[{n\over p^2}\right]$ integers
divisible by $p^2$ and so on. From this sum the result follows easily.

In this paper we shall need to make reference to the sums
 $$
\s_m~=~\sum_{j=1}^m\,{1\over j}~~~\hbox{and more generally to}~~~ 
\s_m^{(\ell)}~=~\sum_{j=1}^m\,{1\over j^\ell} \eqlabel{sigdef}$$
$\s_m$ being simply a shortened form of writing $\s_m^{(1)}$. Now it is follows from
\eqref{facttwo} that
 $$
\s_{p-1}^{(\ell)}~=~\ord{p}~~\hbox{if $(p-1)\notdiv\ell$}~.$$
a stronger result obtains for the important case $\ell=1$ which is that 
 $$
\s_{p-1}~=~\ord{p^2}~~\hbox{for $p>3$}~.$$
To see this we evaluate $(p-1)!$ in two different ways
 $$
(p-1)!~=~\prod_{j=1}^{p-1}(p-j)~=~\prod_{j=1}^{p-1}j\left( 1-{p\over j} \right)~=~
(p-1)!\left\{ 1 - p\s_{p-1} + \ord{p^3}\right\} $$
from which the result follows.

Consider now the ratio
 $$
{(rp)!\over ((r-1)p)!}~=~rp\prod_{j=1}^{p-1}(rp-j)~=~
r\,p!\prod_{j=1}^{p-1}\left(1-{rp\over j}\right)~=~r\,p!\,\left\{1+\ord{p^3}\right\}~.$$
By multiplying ratios of this form we find
 $$
(ap)!~=~a!(p!)^a\left\{1+\ord{p^3}\right\}\eqlabel{ap}$$
and from this it is easy to see that 
$$
(ap+b)!~=~(ap)!\prod_{j=1}^b (ap+j)~=~(ap)!b!\prod_{j=1}^b \left(1+{ap\over j}\right)
~=~a!b!(p!)^a\left\{1+ ap\s_b+\ord{p^2}\right\}~.$$
Our expression for $(ap)!$ is valid for arbitrary positive integers $a$ so we may replace 
$a$ by $ap$ to obtain
 $$
\left(ap^2\right)!~=~a!(p!)^{a(p+1)}\left\{1 + \ord{p^3}\right\}~.$$
If in our expression for $(ap+b)!$ we replace $a$ by $ap+b$ and $b$ by $c$ then we find
 $$
\left(ap^2 + bp + c\right)!
~=~ a!b!c!(p!)^{a(p+1)+b}\left\{1 + p(a\s_b + b\s_c) + \ord{p^2}\right\}~.$$ 
\subsection{The p-adic $\G$-Function}
We recall here a number of relations concerning the p-adic $\G$-function that we
will need in the following. Recall \cite{\Koblitz, \Schikhoff} that for $n\in \bb{Z}$,
$n\geq 0$ this function is defined by
 $$
\G_p(n+1)~=~(-1)^{n+1}{\prod_{j=1}^n}' j~~,~~~~\G_p(0)~=~1~,\eqlabel{gammadef}$$
where the prime denotes that $j$'s that are divisible by $p$ are omitted from the product.
The virtue of this omission as well as the improbable looking factor of $(-1)^{n+1}$ is
that thus defined $\G_p$ is continuous and even analytic in the p-adic sense. Continuity
in this sense means that $\G_p(n+mp^N)\to\G_p(n)$ as $N\to\infty$.
Since $\G_p$ is continuous and the integers are dense in $\bb{Z}_p$ (the partial sums of
the p-adic expansion of each $X\in \bb{Z}_p$ are integers that converge to $X$) the
domain of $\G_p$ can be extended to $\bb{Z}_p$ by continuity. For $X\in\bb{Z}_p$
it then satisfies the recurrence 
 $$
{\G_p(X+1)\over \G_p(X)}~=~
\cases{-X~;& $X\in \bb{Z}^*_p$\cropen{5pt}
       -1~~;& $X\in p\,\bb{Z}_p$~.\cr}$$

Now in computing the
product ${\prod_{j=1}^n}' j$ the factors that are omitted are
 $$
p\times 2p\times 3p\times\cdots\times \left[{n\over p}\right]\,p~=~
\left[{n\over p}\right]!\,p^{\left[{n\over p}\right]}~.$$
Hence
 $$
n!~=~(-1)^{n+1}\G_p(n+1) p^{\left[{n\over p}\right]}\left[{n\over p}\right]!~.$$
By writing $n=n_0+n_1 p+\ldots n_\ell p^\ell$ in base $p$ and iterating the above
expression by using a similar expression to replace $\left[n/p \right]!$ and
proceeding in this way we recover the useful textbook formula
 $$
n!~=~(-1)^{n+\ell+1} (-p)^{n-S(n)\over p-1}\prod_{j=0}^\ell 
\G_p\left( {\left[{n\over p^j}\right]} + 1 \right)~.\eqlabel{nfac}$$
 
Also useful is the reflection formula
 $$
\G_p(X)\G_p(1-X)~=~(-1)^{L(X)} \eqlabel{gammaformula}$$
where $L(X)$ denotes the mod $p$ reduction of the lowest digit of $X$ to the range
$\{1,2,\ldots,p\}$, that is $L(X)$ is the lowest digit unless the lowest digit is 0 in
which case $L(X)=p$. The previous formula is the analogue of the classical result
 $$
\G(\x)\G(1-\x)~=~{\p \over \sin(\p \x)}~.\eqlabel{classicalref}~. $$

We shall have need also of the multiplication formula for $\G_p$
 $$
{\prod_{i=0}^{n-1} \G_p\left({X+i\over n}\right) \over 
\G_p(X)\prod_{i=1}^{n-1} \G_p\left({i\over n}\right)}~=~
n^{1-L(X)}\left(n^{-(p-1)}\right)^{X_1} \eqlabel{premult}$$
where $n$ is an integer such that $p\notdiv n$ and $X_1$ is related to $X$ by
 $$
X~=~L(X) + pX_1~.$$
Note that the right hand side of \eqref{premult} has been carefully written so as to be
defined for p-adic
$X$. The quantity $n^{-(p-1)}$ is $1+\ord{p}$ and the result of elevating this to a p-adic
power is defined via the binomial series.
It is useful also to recall the classical multiplication formula for the $\G$-function.
This is usually presented in the form
 $$
\prod_{i=0}^{n-1}\G\left({\x+i\over n}\right)~=~
(2\p)^{n-1\over 2}n^{{1\over 2} -\x}\, \G(\x)~.\eqlabel{premultclass}$$
For our purposes it is better to eliminate the reference to the factors of $\p$ in
this formula. By taking $\x=1$ we find
 $$
\prod_{i=1}^n\G\left({i\over n}\right)~=~ (2\p)^{n-1\over 2}n^{-{1\over 2}}$$
and we use this expression to divide the previous one. This yields a formula remarkably
similar in form to the p-adic expression
$$
{\prod_{i=0}^{n-1}\G\left({\x+i\over n}\right)\over
\G(\x) \prod_{i=1}^{n-1}\G\left({i\over n}\right)}~=~
n^{1-\x}~.\eqlabel{multclass}$$
\subsection{The Teichm\"uller representative}
Given an $a\in \bb{Z}$ such that $p\notdiv a$ we know that $a^{p-1}=1+\ord{p}$ and by
raising this relation to the power $p$ we find that $a^{p(p-1)}=1+\ord{p^2}$ and by
continuing in this way we see that $a^{p^n(p-1)}=1+\ord{p^{n+1}}$. In virtue of this it is
natural to define the Teichm\"uller representative of $a$ as the limit
 $$
\teich(a)~=~\lim_{n\to\infty} a^{p^n} $$
which exists as a p-adic number. The Teichm\"uller representative has the property that 
 $$
\teich^p(a)~=~\teich(a) $$
exactly and so embeds $\Fp^*$ in $\bb{Q}_p^*$. 

Put differently: if we think of an $a\in\Fp$ abstractly then $a^p=a$ exactly. If however
we think of $a$ as an integer then $a^p = a + \ord{p}$. We will later be
performing perturbation theory in powers of $p$ and will want to work to $\ord{p^n}$ for
$n>1$. It is therefore useful to observe that $a^p$ also represents
$a$ in $\Fp$ (since $a^p = a + \ord{p}$) however by the above $(a^p)^p=a^p + \ord{p^2}$
so $a^p$ is a better representative in this sense and the sequence~$a^{p^n}$, which all
represent $a$ in $\Fp$, are successively better representatives. The Teichm\"uller
representative is the limit of this process. 

It is possible to define the Teichm\"uller representative also for fields $\Fq$ with
$q=p^s$ elements. In this case
 $$
\teich(a)~=~\lim_{n\to\infty} a^{q^n}~;~~~a\in\Fq$$
and $\teich^q(a)=\teich(a)$ which embeds $\Fqstar$ in $\bb{Q}_p^*$.
\subsection{The coefficients of the fundamental period}
It is of interest to consider the p-adic structure of the coefficients 
$a_m={(5m)!\over (m!)^5}$. Since these numbers are multinomial coefficients they are all
integers. Here are the first few:
\vskip0pt
 $$\eqalign{
        &1\cropen{-5pt} 
        &120\cropen{-5pt}
        &113400\cropen{-5pt} 
        &168168000\cropen{-5pt} 
        &305540235000\cropen{-5pt} 
        &623360743125120\cropen{-5pt} 
        &1370874167589326400\cropen{-5pt} 
        &3177459078523411968000\cropen{-5pt} 
        &7656714453153197981835000\cropen{-5pt}  
        &19010638202652030712978200000\cropen{-5pt} 
        &48334775757901219912115629238400\cropen{-5pt}
        &125285878026462826569986857692288000\cropen{-5pt} 
        &329981831728425465309559251123033960000\cropen{-5pt} 
        &880904182555008823696060440775388083200000\cropen{-5pt} 
        &2378829279642818668557063558238537401024000000\cr} $$ 
 
\noindent These coefficients have interesting structure if reduced mod $p$. For the case
$p=13$ the following is a list of the early coefficients thus reduced: 
\vskip15pt
\vbox{
$$
\vbox{
\halign{#\hfil \cr
  1,~3,~1,~0,~0,~0,~0,~0,~0,~0,~0,~0,~0, \cropen{-2pt}
  3,~9,~3,~0,~0,~0,~0,~0,~0,~0,~0,~0,~0, \cropen{-2pt}
  1,~3,~1,~0,~0,~0,~0,~0,~0,~0,~0,~0,~0, \cropen{-2pt}
  0,~0,~0,~0,~0,~0,~0,~0,~0,~0,~0,~0,~0, \cropen{-2pt}
  0,~0,~0,~0,~0,~0,~0,~0,~0,~0,~0,~0,~0, \cropen{-2pt}
  0,~0,~0,~0,~0,~0,~0,~0,~0,~0,~0,~0,~0, \cropen{-2pt}
  0,~0,~0,~0,~0,~0,~0,~0,~0,~0,~0,~0,~0, \cropen{-2pt}
  0,~0,~0,~0,~0,~0,~0,~0,~0,~0,~0,~0,~0, \cropen{-2pt}
  0,~0,~0,~0,~0,~0,~0,~0,~0,~0,~0,~0,~0, \cropen{-2pt}
  0,~0,~0,~0,~0,~0,~0,~0,~0,~0,~0,~0,~0, \cropen{-2pt}
  0,~0,~0,~0,~0,~0,~0,~0,~0,~0,~0,~0,~0, \cropen{-2pt}
  0,~0,~0,~0,~0,~0,~0,~0,~0,~0,~0,~0,~0, \cropen{-2pt}
  0,~0,~0,~0,~0,~0,~0,~0,~0,~0,~0,~0,~0, \cropen{-2pt} 
  3,~9,~3,~0,~0,~0,~0,~0,~0,~0,~0,~0,~0, \cropen{-2pt}
  9,~1,~9,~0,~0,~0,~0,~0,~0,~0,~0,~0,~0, \cropen{-2pt}
  3,~9,~3,~0,~0,~0,~0,~0,~0,~0,~0,~0,~0, \cr}} $$
\vskip-0.6truein
 $$
\hskip-0.1truein\underbrace{\underbrace{\hbox{\hphantom{3,~9,~3}}}%
\lower32pt\hbox{\hphantom{~0,~0,~0,~0,~0,~0,~0,~0,~0,~0}}}$$
\vskip-35pt\leftline{\hskip1.97truein$\left[ {p\over 5} \right]{+}1$}
\vskip20pt\centerline{\hskip-0.1truein$p$}
\place{1}{4.45}{\hbox{$\left\{\vrule height85pt width 0pt\right.$}}
\place{0.8}{3.1}{$p$}
\place{1.8}{4.48}{\hbox{$\left\{\vrule height20pt width 0pt\right.$}}
\place{1.2}{4.27}{$\left[ {p\over 5} \right]{+}1$}
\place{1.8}{1.67}{\hbox{$\left\{\vrule height20pt width 0pt\right.$}}
\place{1.2}{1.47}{$\left[ {p\over 5} \right]{+}1$}
}
\noindent
In order to understand the block structure of the preceding table let us compute the
p-adic order of the coefficients $a_m$. From \eqref{p-order} we see that this is given by
 $$
\ordp{a_m}~=~{5S(m) - S(5m) \over p-1}~.$$
Let the p-adic expansions of $m$ and $5m$ be
 $$\eqalign{
m~&=~c_0 + c_1 p + \ldots + c_r p^r\cr
5m~&=~d_0 + d_1 p + \ldots + d_s p^s~.\cr}$$
The digits of $5m$ are related to those of $m$ by
 $$\eqalign{
5c_0~&=~d_0 + \tilde{d}_0 p\cr
5c_1 + \tilde{d}_0~&=~d_1 + \tilde{d}_1 p\cr
5c_2 + \tilde{d}_1~&=~d_2 + \tilde{d}_2 p~~~\hbox{etc.}\cr}$$
where the $\tilde{d}_j$ are the `carries'. We see that 
$5S(m) - S(5m) = (p-1)\sum_j\tilde{d}_j$ so that
 $$
\ordp{a_m}~=~\sum_j \tilde{d}_j ~.$$
The right hand side of this expression is the total number of carries that are performed
when we multiply $m$ by 5. Consider again the list of the first $p^2$ coefficients
$a_m$. We~broke the list into lines of length $p$ and then stacked these lines to form a
column. It~would be better to stack the lines upwards since then $a_{\x+\eta p}$ would
correspond to Cartesian coordinates $(\x,\eta)$. The coefficients with both $\x$ and
$\eta$ less than $p/5$ are $\ord{1}$. If one of $(\x,\eta)$ is less than $p/5$ and the
other is in the range $(p/5,2p/5)$ then there is one carry and these coefficients are
$\ord{p}$. It is complicated to keep track of the precise number of carries but it is easy
to see that $a_{\x+\eta p}$ is at least as small as 
$\ord{p^{\left[\x/5\right]+\left[\eta/5\right]}}$.
Thus the first $p^2$ coefficients form a square made up of rectangles of
coefficients bounded the same order and the rectangles corresponding to the same order 
make up diagonal bands as in Figure~\figref{square}. If we consider the first $p^3$ terms
of the form $a_{\x + \eta p + \z p^2}$ then we first assemble the terms into squares and
then stack the squares into a cube. The small blocks then correspond to coefficients
bounded by the same order as in Figure~\figref{cube}. The next stage is to stack cubes 
into a four-dimensional cube and so on.
\midinsert
\def\square{\vbox{\vskip15pt\hbox{
\hskip0pt\epsfxsize=3truein\epsfbox{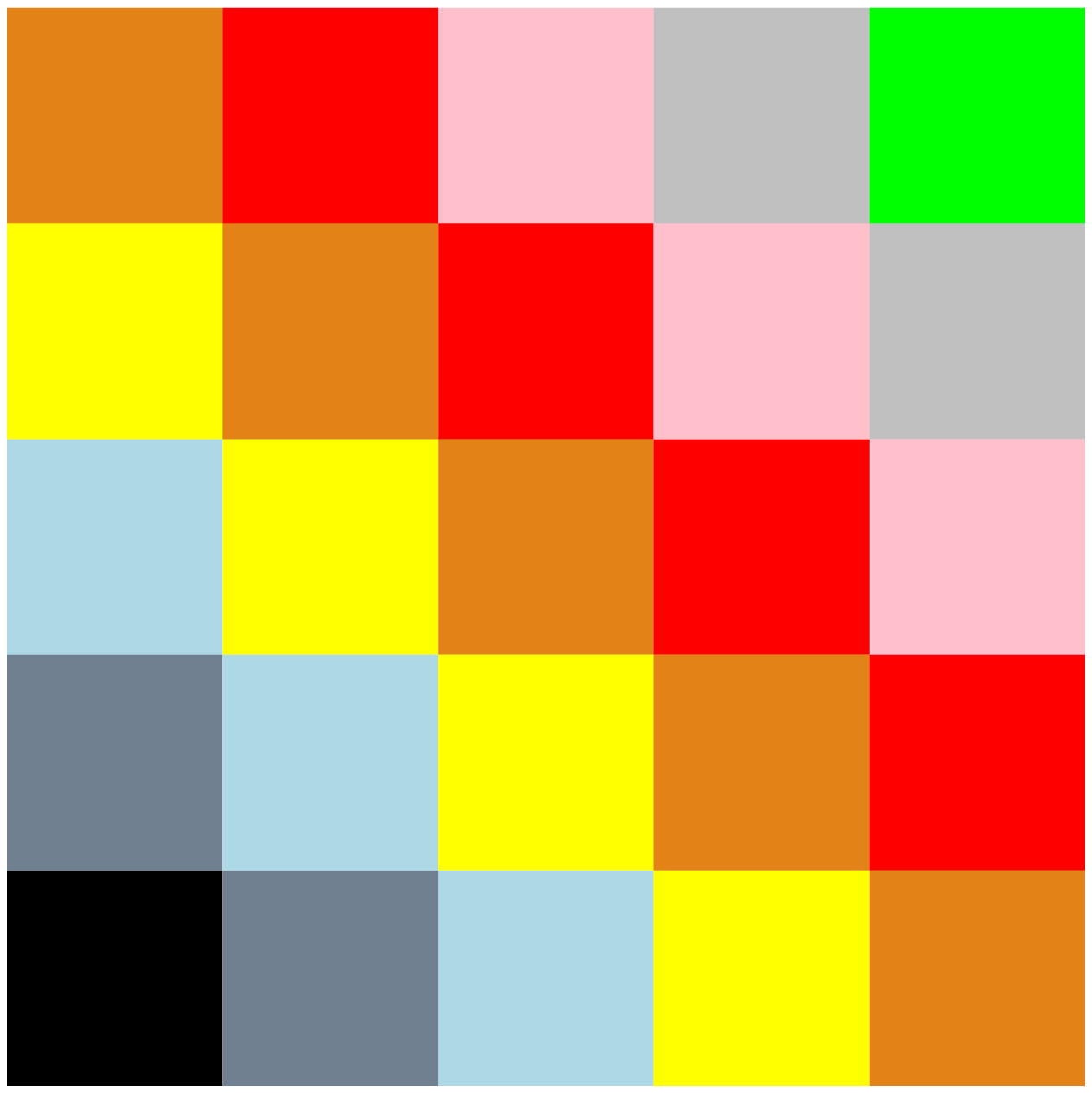}}}}
\figbox{\square}{\figlabel{square}}{Bounds on the orders of the first $p^2$
coefficients $a_m$.}
\vskip10pt
\def\cube{\vbox{\vskip15pt\hbox{
\hskip0pt\epsfxsize=3.5truein\epsfbox{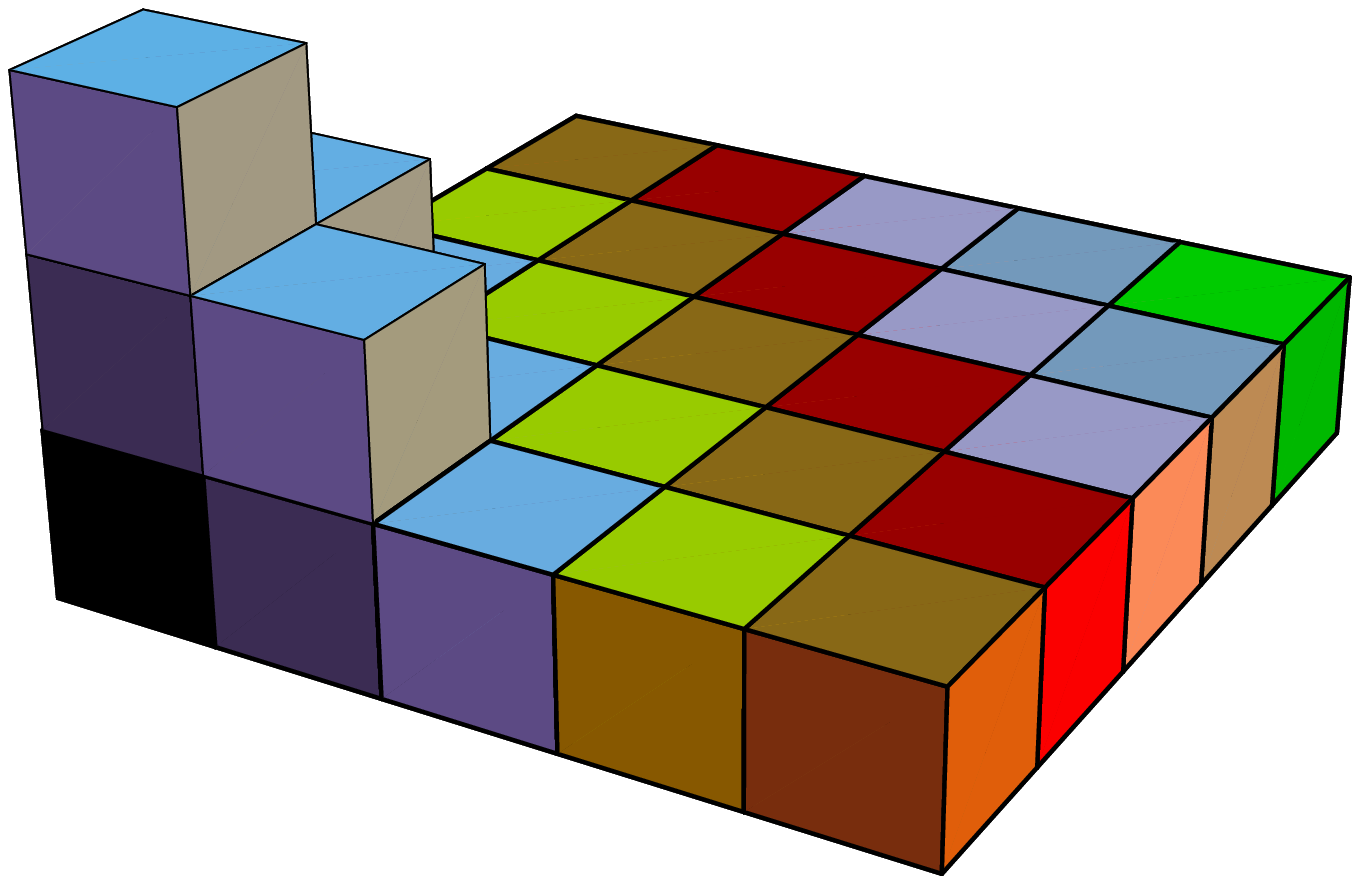}}}}
\figbox{\cube}{\figlabel{cube}}{Bounds on the orders of the first $p^3$
coefficients $a_m$.}
\endinsert
Although we shall have to obtain more detailed information about the periods there are
some very pretty results concerning the mod $p$ reduction of the fundamental period that
we pause to explain.
Consider the coefficient $a_m$ and write $m$ as a multiple of $p$ and a remainder
$m=rp+s$, $0\leq s\leq p-1$.
Now by means of \eqref{ap} and the expression for $(ap+b)!$ it is easy to see that
 $$
a_{rp+s}~=~a_r a_s + \ord{p}~.\eqlabel{acongruence}$$
We use this in the expansion for the fundamental period
 $$
f_0(\l)~\equiv~\sum_{s=0}^{p-1}\sum_{r=0}^\infty a_{rp+s}\l^{rp+s}~\equiv~
\sum_{s=0}^{\left[{p\over 5}\right]}a_s\l^s\,\sum_{r=0}^\infty a_r\l^{rp}~.$$ 
Hence
$$
\eqalign{f_0(\l)&~\equiv~\ftrunc{0}{\left[p/5\right]}(\l)\,f_0(\l^p)\cropen{5pt}
&~\equiv~\ftrunc{0}{\left[p/5\right]}(\l)\,\ftrunc{0}{\left[p/5\right]}(\l^p)\,
\ftrunc{0}{\left[p/5\right]}(\l^{p^2})\ldots\cr}\eqlabel{Frobzero}$$ 
The second congruence follows on iterating the first. The second congruence explains
the mod $p$ structure of $f_0(\l)$ and shows that to this order all the information is
contained in $\ftrunc{0}{\left[p/5\right]}$. We shall see later that to $\ord{p^2}$ it is
necessary to know only the first $\left[2p/5\right]$ terms and so on.

If we write
 $$
\ftrunc{0}{\left[p/5\right]}(\l)~\equiv~{f_0(\l)\over f_0(\l^p)} $$
then we see also that the truncated period satisfies the Picard-Fuchs
equation mod $p$
 $$
\ca{L}\left(\ftrunc{0}{\left[p/5\right]}(\l)\right)~\equiv~0~,~~~\hbox{since}~~~
{d\over d\l}\,\l^p~\equiv~0~.$$
\newpage
\section{allperiods}{The Periods and Semiperiods of the Quintic}
\vskip-20pt
\subsection{The periods}
In this section we will find all the periods and semiperiods of the one parameter family
of quintic threefolds, $M_{\psi}$, defined by the vanishing of the polynomials 
$$P(x,\psi) = \sum_{i=1}^5 x_i^5 - 5\psi\ x_1x_2x_3x_4x_5$$
in $\bb{P}^4$.  The nontrivial Hodge numbers of this manifold are $h^{1,1} = 1$ and
$h^{2,1}= 101$.  The dimension of the moduli space of the complex
structure is $b_3 = 2(1 + h^{2,1}) = 204$,  which means there are 204 periods of the
holomorphic $(3,0)$-form $\Omega$.  These periods satisfy a differential
equation known as the Picard Fuchs equation which is of order 204.  
Fortunately, a task
that seems quite impossible for a generic quintic three--fold, is greatly simplified by the
fact the one parameter family of
\cys\ we are considering has a large group of automorphisms
${\cal A} ={\cal Z}\times\cal{P}$, where $\cal{P}$ is the group of permutations of the
coordinates, and ${\cal Z} \cong \bb{Z}_5^3$ is the group of transformations of the form
$$ \ (x_1,x_2,x_3,x_4,x_5)\ \longrightarrow\ 
(\zeta^{n_1} x_1, \zeta^{n_2}x_2, \zeta^{n_3}x_3, \zeta^{n_4}x_4, \zeta^{n_5}x_5)\ ;
\quad \zeta^5 = 1\ ,\quad\sum_{i=1}^5 n_i = 0~,$$
generated by
$$\eqalign{
g_1&=(4\ 1\ 0\ 0\ 0)\cr
g_2&=(4\ 0\ 1\ 0\ 0)\cr
g_3&=(4\ 0\ 0\ 1\ 0)\cr
g_4&=(4\ 0\ 0\ 0\ 1)\ ,\cr
}\eqlabel{automorphisms}$$
where $g_1g_2g_3g_4 = 1$, and $g_1=(4\ 1\ 0\ 0\ 0)$, for example, means
$$g_1:(x_1,x_2,x_3,x_4,x_5)\ \longrightarrow\ (\zeta^4 x_1,\zeta x_2,x_3,x_4,x_5)~;\quad
\zeta^5 = 1\ .$$

There is a well defined prescription to find the Picard--Fuchs equation 
\REFS{\Dworkone}{B.~Dwork, ``On The Zeta function of a hypersurface, II'',\\
Ann.~of~Math.~(2) {\bf 80} (1964) 227--299.}
\REFSCON{\Griffiths}{P.~A.~Griffiths, ``On the Periods of Certain Rational Integrals'',\\
Ann.~of~Math.~(2) {\bf 90} (1969) 460--495.}\
\REFSCON{\Steenbrick}{J.~Steenbrick, ``Intersection Form for Quasi--Homogeneous
Singularities'',\\ Compositio Math.~{\bf 34} (1977) 211--223.}\
\REFSCON{\Dolgachev}{I.~Dolgachev, ``Weighted Projective Varieties'', in ``Group Actions
and Vector Fields'' (J.~B.~Carrell, ed.), Lecture Notes in Math., Vol. 956, Springer
Verlag, 1982, 34--71.}\
\REFSCON{\BatyrevCox}{V. V. Batyrev and D. Cox, ``On the Hodge Structure of Projective
Hypersurfaces in Toric Varieties'', Duke Math. J. {\bf 75} (1994) 293, alg-geom/9306011.}
\refsend~   
for a manifold defined as a hypersurface in a toric variety. There is a one to one correspondence 
between the elements in the polynomial ring
of the \cym\ and its periods, that is, for every monomial
$$ x^{\bf v} = x_1^{v_1}x_2^{v_2}x_3^{v_3}x_4^{v_4}x_5^{v_5}$$ 
of degree $\sum_{i=1}^5 v_i = 5w({\bf v})$, $0\le w({\bf v}) \le 3$, in the polynomial
ring of $M_{\psi}$, there exists a 3--cycle $\gamma_{\bf v}\in H_3(M_{\psi}, \bb{C})$ such
that the corresponding period $\varpi_{\bf v}(\psi)$ is given by
$$
\varpi_{\bf v}(\psi) = \int_{\gamma_{\bf v}} \Omega = 
{1\over (2\pi i)^5} \int_{\G}d^5x\ {x^{\bf v}\over  P^{w({\bf v})+1}}\ ,
\eqlabel{genericperiod}$$
where $\Gamma=C_1\times C_2\times\cdots\times C_5$ is a 5-torus that is the product of
loops $C_i$ that wind about the submanifolds $\partial_i P =0$ in $\bb{C}^5$.  
Another way to say this is that there is a one to one correspondence between the cohomology
classes in $H^3(M_{\psi}, \bb{C})$ and the equivalence classes of differential forms $d^5x\
x^{\bf v}/P^{w({\bf v})+1}$ modulo exact forms.
\midinsert
\def\loci{\vbox{\vskip0pt\hbox{\hskip-25pt\epsfxsize=5truein\epsfbox{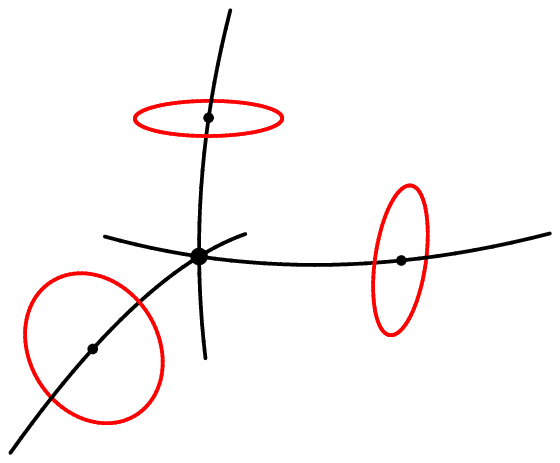}}}}
\figbox{\loci\vskip-60pt}{\figlabel{loci}}{The contour $\G$ is the product of five
$S^1$'s that wind about the loci $\partial_i P=0$. These loci are taken to have the
origin as the unique point in which they all meet.}
\endinsert
The Picard--Fuchs equation
in our example, which could be a very complicated coupled system of differential
equations, simplifies significantly because of the automorphisms of the manifold. The
elements of the polynomial ring can be classified according to their transformation
properties under the automorphisms, that is into representations of \ca{A}, and owing to
the correspondence with the periods, the periods can also be classified accordingly. 
Consider now derivatives with respect to $\psi$ of the periods \eqref{genericperiod}. 
These derivatives produce new periods which are in correspondence to monomials in the same
representation of \ca{A}.  For example, the period
$$
{{\rm d}\over {\rm d}\psi}\varpi_{\bf v}(\psi) =  5\big(w({\bf v})+1\big)
{1\over (2\pi i)^5}\int_{\G} d^5x\ 
{x^{\bf v+1}\over P^{w({\bf v})+2}}\ ,
$$
corresponds to the monomial
$x^{\bf v+1} = x^{\bf v} x_1x_2x_3x_4x_5$ of degree $5\big(w({\bf v})+1\big)$,
which is in the same representation as the monomial 
$x^{\bf v}$ because the monomial 
$x_1x_2x_3x_4x_5$ is invariant under the entire group of automorphisms.
As a consequence, the Picard--Fuchs system of equations breaks down into independent
blocks, one block for each representation of \ca{A} in the polynomial ring.  

The first step in finding the Picard--Fuchs equation is therefore to 
classify all the monomials in the polynomial ring according to their representation under
the group of automorphisms.  There are 126 monomials of degree 5, and they can be
classified according to their transformation under \ca{A} as follows:
$$
\eqalign{ 
{\bf case\ 1}\ :\qquad \hskip3pt 6\ \hbox{invariant monomials}\quad & 
\{x_1^5, x_2^5, x_3^5, x_4^5, x_5^5, x_1x_2x_3x_4x_5\}\cr 
{\bf case\ 2}\ :\qquad 20\ \hbox{sets of 2 monomials}\quad & 
\{x_1^4 x_2, x_2^2x_3x_4x_5\}, \hbox{and permutations}\cr 
{\bf case\ 3}\ :\qquad 20\ \hbox{sets of 1 monomial~}\quad & \{x_1^3x_2^2\}, \hbox{and
permutations} \cr 
{\bf case\ 4}\ :\qquad 30\ \hbox{sets of 1 monomial~}\quad & \{x_1^3x_2x_3\}, \hbox{and
permutations} \cr 
{\bf case\ 5}\ :\qquad 30\ \hbox{sets of 1 monomial~}\quad & \{x_1^2x_2^2x_3\},
\hbox{and permutations} \cr }$$
Each incomplete representation of $\cal A$ above has to be completed
to include also the monomials of degree zero, degree 10, and degree 15.  There are a total
of 101 representations and a differential operator corresponding to the periods
in each representation, the order of each differential equation being the dimension of the
representation.  Note that not all the monomials in each
representation are inequivalent. Two monomials are equivalent if they differ by an
element in the Jacobian ideal, that is the ideal $\ca{I}=(\partial_i P)$ generated by the
partial derivatives of $P$. For example, the monomials of degree 5 in the case 1 above,
are all equivalent because $x_i^5\simeq \psi Q$, and we choose
$Q = x_1x_2x_3x_4x_5$ to be the representative.  Similarly, the two monomials in each of
the sets in the second case above are also equivalent since, for example, 
$x_1^4x_2\simeq\psi\, x_2^2x_3x_4x_5$.  The representations are then
$$
\eqalign{ 
{\bf case\ 1}\ :\quad \,~\,~1\ \hbox{dimension 4 representation~}\quad & 
\{1\ ,\ Q ,\ Q^2,\ Q^3\}\ ,\cr 
{\bf case\ 2}\ :\quad \,~20\ \hbox{dimension 2 representations}\quad & 
\{x_1^4x_2,\ x_1^4x_2 Q\}, \hbox{and permutations}\ ,\cr 
{\bf case\ 3}\ :\quad \,~20\ \hbox{dimension 2 representations}\quad & 
\{x_1^3x_2^2,\ x_1^3x_2^2 Q\}, \hbox{and permutations} \ ,\cr 
{\bf case\ 4}\ :\quad \,~30\ \hbox{dimension 2 representations}\quad & 
\{x_1^3x_2x_3,\ x_1^3x_2x_3 Q\},\hbox{and permutations} \ ,\cr 
{\bf case\ 5}\ :\quad \,~30\ \hbox{dimension 2 representations}\quad & 
\{x_1^2x_2^2x_3,\ x_1^2x_2^2x_3 Q\}, \hbox{and permutations} \ ,\cr
{\bf case\ 6}\ :\quad 120\ \hbox{dimension 1 representations}\quad &
\{x_1^4x_2^3x_3^2x_4\}, \hbox{and permutations}\ .\cr }$$

The sixth case deserves some explanation.  The monomial
$x_1^4x_2^3x_3^2x_4$  is in the ideal {\it except when} $\psi^5 = 1$, that is, except
at the point in the moduli space where the manifold develops a conifold singularity.  
This can be seen by setting $\m_i=x_i^4x_{i+1}^3x_{i+2}^2x_{i+3}$, with the subscripts
on the $x_j$ understood mod 5. Since $x_i^4\simeq x_{i+1}x_{i+2}x_{i+3}x_{i+4}$, modulo 
\ca{I}, we have $\m_i\simeq \psi\, \m_{i+1}$ and by repeating this
relation we find
 $$
(1-\ps^5)\,\m_i~\simeq~0~. $$
Equivalently, the monomial $x_1^4x_2^3x_3^2x_4$ corresponds to a differential form which
is exact except when $\psi^5 = 1$
$$
\eqalign{
-10\, (1-\psi^5)\ {x_1^4x_2^3x_3^2x_4\over P^3}\ &=\cropen{5pt}
\partial_1\left({x_2^3x_3^2x_4\over P^2}\right) +
&\partial_2\left({x_3^3x_4^2x_5\over P^2}\right) +
\partial_3\left({x_1x_4^3x_5^2\over P^2}\right) +
\partial_4\left({x_1^2x_2x_5^3\over P^2}\right) +
\partial_5\left({x_1^3x_2^2x_3\over P^2}\right) \ .\cr}$$
Of course, as long as $\psi^5\ne 1$, the period associated to this monomial is zero.  The
reason we are including this case will not become clear until later when we calculate the
number of points over a finite field and we show that at
$\psi^5 = 1$ there is a ``period'' associated to this monomial which gives an
exceptional contribution to the number of points.  

We now proceed to find all the differential systems.  As described above, we
associate a period to each monomial as in \eqref{genericperiod}, and the Picard--Fuchs
equation for each representation can be found using repeatedly the operation
$$\eqalign{
D_i \left( {x^{\bf v}\over P^{w({\bf v})+1}}\right) &\define 
\partial_i \left( x_i\ {x^{\bf v}\over P^{w({\bf v})+1}}\right) \cropen{10pt}
&= (1 + v_i)\ {x^{\bf v}\over P^{w({\bf v})+1}} +  
5(w({\bf v})+1){x^{\bf v}\over P^{w({\bf v})+2}} (x_i^5 - \psi Q)\ ,\cr}
\eqlabel{exact}$$
where $x^{\bf v}$ is a monomial of degree $5 w({\bf v})$.  Since
 $$
\int_\G d^5x\,D_i \left( {x^{\bf v}\over P^{w({\bf v})+1}}\right)~=~0$$
the derivative $D_i \left( {x^{\bf v}\over P^{w({\bf v})+1}}\right)$ corresponds to an
exact form so \eqref{exact} establishes identities between the differential form associated
to the monomial $x^{\bf v}$ and those associated to  $x^{\bf v}x_i^5$ and $x^{\bf v}Q$, up
to an exact piece.  Note in particular that the period corresponding to the monomial
$x^{\bf v}Q$ corresponds to the derivative with respect to $\psi$ of the period for the
monomial $x^{\bf v}$.  It  will be very  useful to represent the operations
$D_i$ by diagrams. For $D_1$, for example, the diagram is as follows and the other cases
are similar
\vskip10pt
$$\vbox{
\halign{\strut # &\hfil $#$\hfil &\hfil $#$\hfil &\hfil $#$\hfil \cr
&(v_1,v_2,v_3,v_4,v_5)& \rightarrow~~~ &(v_1+1,v_2+1,v_3+1,v_4+1,v_5+1)\cr
& \qquad\downarrow D_1& & \cr
&(v_1+5,v_2,v_3,v_4,v_5)& & \cr
}}$$. 

\subsubsection{\bf Case 1} 
The monomials in this case are each invariant under
${\cal Z}$ and transform into each other under $\cal{P}$.  
The Picard--Fuchs equation for this
system is  well known\  
\Ref{\CDGP}{P.~Candelas, X.~de~la~Ossa, P.~Green, and L.~Parkes,\\
 ``A Pair of Calabi--Yau Manifolds as an Exactly Soluble Superconformal Theory'',
Nucl.~Phys.~B{\bf 359}(1991) 21--74.}\ 
(since by itself it is the Picard--Fuchs equation for the {\it
mirror} of the quintic \cym\ we are considering).  
Since we are writing for a mixed audience, and
to set the stage for generalizations to
\cys\ in other dimensions, we rederive this differential equation
here even though it is well known.  We are interested in finding an identity, up to an
exact form, between the differentials associated to the monomials $Q^n$. We obtain
this identity using the operations $D_i$ as above.  Using the schematic representation of
these operations, the desired relation can be found as shown below with the
the monomials in the given representation indicated by bold face.
\vskip7pt
$$\vbox{
\halign{\strut # &\hfil $#$\hfil &\hfil $\, #\,$\hfil 
&\hfil $#$\hfil &\hfil $\, #\,$\hfil&\hfil $#$\hfil &\hfil $\, #\,$\hfil
&\hfil $#$\hfil &\hfil $\, #\,$\hfil&\hfil $#$\hfil &\hfil $\, #\,$\hfil
&\hfil $#$\hfil &\hfil $\, #\,$\hfil\cr
& & & & & & & & & (4,4,4,4,-1)& \rightarrow &(5,5,5,5,0)\hskip-5pt\cr
& & & & & & & & &  \qquad\downarrow D_5\cr
&{\bf (0,0,0,0,0)}& \rightarrow &{\bf (1,1,1,1,1)}& \rightarrow &{\bf (2,2,2,2,2)}
& \rightarrow &\bf{(3,3,3,3,3)}& \rightarrow &{\bf (4,4,4,4,4)}\cr
& \qquad\downarrow D_1& & \qquad\downarrow D_1& & \qquad\downarrow D_1& &
 \qquad\downarrow D_1 \cr
&(5,0,0,0,0)& \rightarrow &(6,1,1,1,1)& \rightarrow &(7,2,2,2,2)
& \rightarrow &(8,3,3,3,3) \cr
& \qquad\downarrow D_2& & \qquad\downarrow D_2& & \qquad\downarrow D_2\cr
&(5,5,0,0,0)& \rightarrow &(6,6,1,1,1)& \rightarrow &(7,7,2,2,2)\cr
& \qquad\downarrow D_3& & \qquad\downarrow D_3\cr
&(5,5,5,0,0)& \rightarrow &(6,6,6,1,1)\cr
& \qquad\downarrow D_4 \cr
&(5,5,5,5,0)\cr
}}$$
Note that we are
allowing a (minimum) negative exponent of $x_i$ of $-1$ when operating with
$D_i$ to allow for equivalences of monomials in the monomial ring.  For example, the piece
of the diagram above
 $$\vbox{
\halign{\strut # &\hfil $#$\hfil &\hfil $#$\hfil &\hfil $#$\hfil \cr
&(4,4,4,4,-1)& \rightarrow &(5,5,5,5,0)\cr
& \qquad\downarrow D_5& & \cr
&(4,4,4,4,4)& & \cr
}}$$
means that the monomial $Q^4$ and $\psi\, x_1^5x_2^5x_3^5x_4^5$ are equivalent in \ca{I},
the identity being
 $$
D_5\left({x_1^4x_2^4x_3^4x_4^4\over x_5P^4}\right) =
\partial_5\left({x_1^4x_2^4x_3^4x_4^4\over P^4}\right) =
5 {Q^4\over P^5} - 5\psi {x_1^5x_2^5x_3^5x_4^5\over P^5}~.$$
Defining
 $$
Q_n = {Q^{n-1}\over P^n}\ , \quad n \ge 1\  ,$$
we see that the identity we are looking for is of the form
 $$\eqalign{
\psi Q_1 + A_2\psi^2 Q_2 + A_3\psi^3 Q_3 &+ A_4 \psi^4 Q_4 + A_5\psi^5 Q_5 - 
\tilde{A}_5 Q_5 =\cropen{3pt} 
&~~~\partial_1[ x_1(B_{11}\psi Q_1 + B_{12}\psi^2 Q_2 + B_{13}\psi^3 Q_3 
+ B_{14}\psi^4 Q_4)]\cropen{3pt} 
&+ \partial_2[ x_2 x_1^5 Q_1 (B_{21}\psi Q_1 + B_{22}\psi^2 Q_2 + B_{23}\psi^3
Q_3)]\cropen{3pt} 
&+ \partial_3[x_3 x_1^5 x_2^5 Q_1^2 (B_{31}\psi Q_1 + B_{32}\psi^2
Q_2)]\cropen{3pt} 
&+ \partial_4(x_4 x_1^5 x_2^5 x_3^5 B_{41}\psi Q_1^4)\cropen{3pt}
&+ \partial_5[(x_1x_2x_3x_4 B_5 Q_1)^4]~,\cr
}$$
with the $A_j,\tilde{A}_5, B_{jk}$ and $B_5$ denoting certain constants which may be
determined by evaluating the derivatives and comparing terms on each side of the equation.
In this way we find
 $$\eqalign{
\psi Q_1 + 75 \psi^2 Q_2 + 1250 \psi^3 Q_3 &+ 7500 \psi^4 Q_4 + 15000(\psi^5 - 1) Q_5 
=\cropen{3pt}
&~~~\psi\, \partial_1[ x_1(Q_1 + 35\psi Q_2 + 300\psi^2 Q_3 
+ 750 \psi^3 Q_4)]\cropen{3pt} 
&+ 5\psi\, \partial_2[ x_2 x_1^5 Q_1 (Q_1 + 30\psi Q_2 + 150\psi^2 Q_3)]\cropen{3pt}
&+ 50\psi\, \partial_3[x_3 x_1^5 x_2^5 Q_1^2 (Q_1 + 15\psi Q_2)]\cropen{3pt}
&+ 750\psi\, \partial_4(x_4 x_1^5 x_2^5 x_3^5 Q_1^4)\cropen{3pt}
& - 750\, \partial_5[(x_1x_2x_3x_4Q_1)^4]~.\cr
}\eqlabel{caseoneid}$$
Consider now the period integrals 
 $$
I_n = {1\over (2\pi i)^5}\int_{\G_0} d^5x\ Q_n\ ,$$
where $\G_0$ is the lifting to $\bb{C}^5$ of the three cycle in $M_{\psi}$ which is
topologically a three torus in the large complex structure limit $\psi\to\infty$ and which
intersects at a point with a three sphere which is shrinking to a point near 
$\psi^5 = 1$ (a conifold point); $\G_0$ is the five torus in $\bb{C}_5$ defined by   
$|x_i| = 1$.  These integrals are all related to $I_1$ by 
 $$
I_{n+1} = {1\over{5^n\, n!}}\ {{\rm d}^n\over {\rm d}\psi^n}\ I_1\ .$$ 
The Picard--Fuchs equation for this case is the differential equation for $I_1$ that
follows from~\eqref{caseoneid}
 $$
\psi\, I_1 + 15\, \psi^2\, {dI_1\over d\psi} + 25\,\psi^3\, {d^2I_1\over d\psi^2}
+10\,\psi^4\, {d^3I_1\over d\psi^3}  + (\psi^5 - 1)\,{d^4I_1\over d\psi^4} = 0\quad .$$
If we define $\varpi_0 = -5\,\psi\, I_1$, we obtain the usual form of the Picard--Fuchs
equation
 $$
\ca{L}\,\vp_0~=~0~~~\hbox{with}~~~
\ca{L}~=~\vth^4 - 5\l\,\prod_{i=1}^4(5\vth+i)\ , \eqlabel{PF}$$
where we have redefined the parameter to be $\l~=~(5\ps)^{-5}\ $, and have set 
$\vth = \l{{\rm d}\over{\rm d}\l}$.  Note that the form of the Picard--Fuchs equation
depends on the gauge chosen for the holomorphic $(3,0)$--form, $\Omega$.  This gauge
freedom corresponds to the fact that $\Omega$ is uniquely defined up to a holomorphic
function of the complex structure of the parameters (that is, it defines a holomorphic
line bundle over the moduli space).  In the canonical gauge of equation
\eqref{PF}, all four of its indices at $\l = 0$ are equal to zero, and the solutions are
asymptotically like $1$, log$\l$, log${}^2 \l$, and log${}^3 \l$ as $\l\to 0$.  The
solution of \eqref{PF} that is asymptotically $1$, called the fundamental period
$\varpi_0(\psi)$, is given by
$$\varpi_0(\psi) = 
{}_4F_3\left(\smallfrac{1}{5},\smallfrac{2}{5},\smallfrac{3}{5},\smallfrac{4}{5};
1,1,1;\psi^{-5}\right) =
\sum_{m=0}^{\infty} a_m \l^m\ ,\quad\quad a_m = {(5m)!\over m!^5}\ .$$

A basis for the space of solutions of \eqref{PF} can be found using the method of
Frobenius.  We review this procedure here to establish notation.  Consider the series
 $$ 
F(\l,\ve) = \sum_{m=0}^{\infty} A_m(\ve)\l^{m+\ve}\quad, \eqlabel{Ffrob}$$
satisfying the equation ${\cal L}F(\l,\ve) = \ve^4\l^{\ve}$, and $F(\l, 0) =
\varpi_0(\psi)$,  so that $A_m(0) = a_m$.  Then, the coefficients $A_m(\ve)$ must satisfy $
A_0(\ve) = 1$ and the recurrence relations
$$
A_m(\ve)~=~{(5m+5\ve)(5m+5\ve-1)\ldots(5m+5\ve-4)\over (m+\ve)^5}\,A_{m-1}(\ve)\ .$$
We choose
 $$
A_m(\ve)~=~{\G(5m+5\ve+1)\over \G^5(m+\ve+1)}{\G^5(\ve+1)\over
\G(5\ve+1)}~.\eqlabel{Am}$$   
The
logarithmic solutions are then obtained by derivatives of
$F(\l,\ve)$ with respect to $\ve$:
$$ 
\varpi_k(\psi) = {\partial^k\over \partial\ve^k} F(\l,\ve)\, \biggr|_{\ve=0} = 
\sum_{j=0}^k \left(k\atop j\right) f_j(\psi)\ \log^{k-j}\l\ ,\quad k= 0,\ldots, 3\ ,$$
where
$$f_j(\psi) 
= \sum_{m=0}^{\infty} {{\rm d}^j\over {\rm d}\ve^j}A_m(\ve)\biggr|_{\ve=0}\ \l^m \ .$$

\noindent{\bf Cases 2-5}:  In each of these cases we have a set of two dimensional
representations which are related to each other under $\cal P$, so that the periods
corresponding to each one of these representations will satisfy the same differential
equation.  Given a representation the
corresponding period integrals are
 $$I_{\bf v} = \psi \left({1\over 2\pi i}\right)^5 
\int_\G d^5x\ {x^{\bf v}\over P^2}\
\ ,\qquad I_{\bf v}' - I_{\bf v} =  10\psi^2 \left({1\over 2\pi i}\right)^5 
\int_\G d^5x\ {x^{\bf v} Q\over P^3}\ ,$$  
where we have chosen a convenient gauge.  We will also need second derivatives of   
$I_{\bf v}$ so we also define
 $$
Q_{{\bf v},n} = {x^{\bf v} Q^{n-1}\over P^{n+2}}\ , \quad n \ge 1\  .$$ 
Since all the cases at hand are similar, we now enumerate the results, which were obtained
following a procedure analogous to that in {\bf case 1}. For each case we give the diagrams
which provide the identities for the differentials
$Q_{{\bf v},n}d^5x$ associated to each monomial $x^{\bf v}$, the Picard--Fuchs equation
for the periods and the solutions.  As we will see, the Picard--Fuchs equations are, in all
four cases, second order hypergeometric differential equations
$$\{\vth(\vth -1 + c_{\bf v}) 
- 5^5\lambda(\vth + a_{\bf v})(\vth + b_{\bf v})\} I_{\bf v} = 0\ .$$ 
In all cases $c_{\bf v} = a_{\bf v} + b_{\bf v}$ which means that as $\psi^5\to 1$, there
is a regular solution and a logarithmic solution.  Also, in all cases there is a regular
solution asymptotic to $1$ as  $\l\to 0$ given by
$$I_{\bf v} = {}_2F_1\left(a_{\bf v},b_{\bf v};c_{\bf v};\psi^{-5}\right) =
\sum_{m=0}^{\infty} \lambda^m 
{(5m)!\over \prod_{i=1}^5 \left({{5-v_i}\over 5}\right)_m}~.$$
To find the second solution we need to distinguish between {\bf cases 2} and {\bf 3}
and {\bf cases 4} and {\bf 5}.  For {\bf cases 2} and {\bf 3}, we have
$c_{\bf v} = 1$, and thus as $\l\to 0$ the second solution is asymptotic to 
log$\l$.  We find this solution using the method of Frobenius, as in {\bf case 1}, that is
we consider the series
$$ F(\l,\ve) = 
\sum_{m=0}^{\infty} A_{{\bf v},m}(\ve)\l^{m+\ve}\quad, $$
satisfying
$$\eqalign{
{\bf case\ 2}\ :&\quad
\left\{\vth^2  - 5^5\lambda\left(\vth + \smallfrac{2}{5}\right)
   \left(\vth + \smallfrac{3}{5}\right)\right\} F(\l,\ve) = 
\ve^2 \l^{\ve}\ ,
\quad{\rm and}\quad F(\l, 0) = I_{(4,1,0,0,0)}\cr
{\bf case\ 3}\ :&\quad
\left\{\vth^2  - 5^5\lambda\left(\vth + \smallfrac{1}{5}\right)
    \left(\vth + \smallfrac{4}{5}\right)\right\} F(\l,\ve) = 
\ve^2 \l^{\ve}\ ,
\quad{\rm and}\quad F(\l, 0) = I_{(3,2,0,0,0)}\cr }$$
so that 
$$A_{{\bf v},m}(0) = {(5m)!\over \prod_{i=1}^5(1-\smallfrac{v_i}{5})_m}~.$$ 
Then, the coefficients $A_{{\bf v},m}(\ve)$ must satisfy 
$ A_{{\bf v},0}(\ve)= 1$ and the recurrence relations
$$
A_{{\bf v},m}(\ve)~=~5^3\, {(5m + 5 \ve- 5a_{\bf v})
(5m + 5 \ve- 5b_{\bf v})\over (m + \ve)^2}\,  
A_{{\bf v},m-1}(\ve)\ .$$
Hence
 $$
A_{{\bf v},m}(\ve)~=~{\G(5m+5\ve+1)\over \G(5\ve+1)}
\prod_{i=1}^5{\G\left( \ve+1-{v_i\over 5}\right)\over 
\G\left( m+\ve+1-{v_i\over 5}\right)}~.$$   
The logarithmic solution is then obtained by taking  the derivative of
$F(\l,\ve)$ with respect to $\ve$:
$$ 
J_{\bf v} (\psi) = {\partial\over \partial\ve} F(\l,\ve)\,
\biggr|_{\ve=0} = 
I_{\bf v}(\psi) \log\l + f_{1,{\bf v}}(\psi)\ ,$$
where
$$f_{1,{\bf v}}(\psi) 
= \sum_{m=0}^{\infty} {{\rm d}\over {\rm d}\ve}
A_{{\bf v},m}(\ve)\biggr|_{\ve=0}\ \l^m \ .$$
For {\bf cases 4} and {\bf 5}, $c_{\bf v}$ is {\it not} an integer, so the second solution,
which tends asymptotically to $\psi^{5(-1+c_{\bf v})}$ as $\l\to 0$, is given by
$$\eqalign{ J_{\bf v} ~=~ &\psi^{5(-1+c_{\bf v})}
{}_2F_1\left(1-b_{\bf v},1-a_{\bf v};2-c_{\bf v};\psi^{-5}\right) \cr
~=~ &\psi^{5(-1+c_{\bf v})} \sum_{m=0}^{\infty} \lambda^m 
{(5m)!\over (2-c_{\bf v})_m (b_{\bf v})_m (a_{\bf v})_m m!^2}~.\cr}$$
\newpage
$$\vbox{
\halign{\strut # &\hfil $#$\hfil &\hfil $\, #\,$\hfil 
&\hfil $#$\hfil &\hfil $\, #\,$\hfil&\hfil $#$\hfil &\hfil $\, #\,$\hfil
&\hfil $#$\hfil &\hfil $\, #\,$\hfil&\hfil $#$\hfil &\hfil $\, #\,$\hfil
&\hfil $#$\hfil &\hfil $\, #\,$\hfil\cr
\noalign{\noindent$\underline{{\bf Case\ 2}}\ :\quad {\bf v} = (4,1,0,0,0)$}\cr
\noalign{\noindent\bf Diagram:}\cr
\noalign{\vskip-40pt}\cr
& & & & & & & & & \hskip-5pt (3,0,4,4,-1)& \rightarrow &(4,1,5,5,0)\hskip-8pt\cr
& & & & & & & & &  \qquad\downarrow D_5\cr
& & & & & & &(2,-1,3,3,3)& \rightarrow &(3,0,4,4,4)\cr
& & & & & & & \qquad\downarrow D_2 \cr
&(-1,1,0,0,0)& \rightarrow &(0,2,1,1,1)& \rightarrow &(1,3,2,2,2)
& \rightarrow &(2,4,3,3,3) \cr
& \qquad\downarrow D_1& & \qquad\downarrow D_1& & \qquad\downarrow D_1\cr
&{\bf (4,1,0,0,0)}& \rightarrow &{\bf (5,2,1,1,1)}& \rightarrow &{\bf (6,3,2,2,2)}\cr
& \qquad\downarrow D_3& & \qquad\downarrow D_3\cr
&(4,1,5,0,0)& \rightarrow &(5,2,6,1,1)\cr
& \qquad\downarrow D_4 \cr
&(4,1,5,5,0)\cr
}}$$
\vskip10pt
{\noindent\bf Identity for differential forms:}
$$\eqalign{
5\,(\psi^5 - 2)\, &Q_{(4,1,0,0,0), 1} + 50\, \psi\, (3\psi^5 + 2)\, Q_{(4,1,0,0,0), 2} 
+ 750\, \psi^2 (\psi^5 - 1)\, Q_{(4,1,0,0,0), 3} \, =\cropen{5pt}  
& ~~~~5\,\psi^5 \partial_3(x_3(Q_{(4,1,0,0,0), 1} + 10\,\psi Q_{(4,1,0,0,0), 2})) 
+ 50\,\psi^3\partial_4\left(x_4\ {x_3^5\over P} Q_{(4,1,0,0,0), 1}\right)\cropen{5pt}
& + 50\, \psi^4\partial_5\left({x_1^3x_3^4x_4^4\over P^3}\right) 
+ 50\, \psi^3\partial_2\left({x_1^2x_3^3x_4^3x_5^3\over P^3}\right)\cropen{5pt}
&+ 2\,\partial_1\left( {x_2\over P} 
+ x_1\left( -5\psi\,{x_2^2x_3x_4x_5\over P^2} 
+ 25\,\psi^2\,{x_2^2x_3x_4x_5Q\over P^3}\right) \right)\cr}$$
\vskip10pt
{\noindent\bf Differential equation:}
$$\qquad\left\{\vth^2  - 5^5\lambda\left(\vth + \smallfrac{2}{5}\right)
         \left(\vth + \smallfrac{3}{5}\right)\right\} I_{(4,1,0,0,0)} = 0$$
{\noindent\bf Holomorphic solution at $\l = 0$:}
$$I_{(4,1,0,0,0)}(\psi) = 
{}_2F_1\left(\smallfrac{2}{5},\smallfrac{3}{5};1;\psi^{-5}\right) =
\sum_{m=0}^{\infty} \lambda^m 
{(5m)!\over (\smallfrac{1}{5})_m(\smallfrac{4}{5})_m m!^3}$$
\newpage
$$\vbox{
\halign{\strut # &\hfil $#$\hfil &\hfil $\, #\,$\hfil 
&\hfil $#$\hfil &\hfil $\, #\,$\hfil&\hfil $#$\hfil &\hfil $\, #\,$\hfil
&\hfil $#$\hfil &\hfil $\, #\,$\hfil&\hfil $#$\hfil &\hfil $\, #\,$\hfil
&\hfil $#$\hfil &\hfil $\, #\,$\hfil\cr
\noalign{\noindent$\underline{{\bf Case\ 3}}\ :\quad {\bf v} = (3,2,0,0,0)$}\cr
\noalign{\noindent\bf Diagram:}\cr
\noalign{\vskip-40pt}\cr
& & & & & & & & & \hskip-5pt(2,1,4,4,-1)& \rightarrow &(3,2,5,5,0)\hskip-4pt\cr
& & & & & & & & &  \qquad\downarrow D_5\cr
& & & & & (0,-1,2,2,2)& \rightarrow &(1,0,3,3,3)& \rightarrow &(2,1,4,4,4)\cr
& & & & & \qquad\downarrow D_2 & & \qquad\downarrow D_2\cr
& & &(-1,3,1,1,1)& \rightarrow &(0,4,2,2,2)& \rightarrow &(1,5,3,3,3) \cr
& & & \qquad\downarrow D_1& & \qquad\downarrow D_1\cr
&{\bf (3,2,0,0,0)}& \rightarrow &{\bf (4,3,1,1,1)}& \rightarrow &{\bf (5,4,2,2,2)}\cr
& \qquad\downarrow D_3& & \qquad\downarrow D_3\cr
&(3,2,5,0,0)& \rightarrow &(4,3,6,1,1)\cr
& \qquad\downarrow D_4 \cr
&(3,2,5,5,0)\cr
}}$$
\vskip10pt
{\noindent\bf Identity for differential forms:}
$$\eqalign{
\psi^4\, &Q_{(3,2,0,0,0), 1} + 10 \, (3\psi^5 + 2)\, Q_{(3,2,0,0,0), 2} 
+ 150\, \psi(\psi^5 - 1)\, Q_{(3,2,0,0,0), 3} \, =\cropen{5pt}  
& ~~~~\psi^4\, \partial_3(x_3(Q_{(3,2,0,0,0), 1} + 10\,\psi Q_{(3,2,0,0,0), 2})) 
+ 10\,\psi^4\partial_4\left(x_4\ {x_3^5\over P} Q_{(3,2,0,0,0), 1}\right)\cropen{5pt}
&+ 10\,\psi^3\partial_5\left({x_1^2x_2x_3^4x_4^4\over P^3}\right)\cropen{5pt}
&- 2\,\partial_1\left( {x_2^3x_3x_4x_5\over P^2} 
-5\psi x_1\,{x_2^4x_3^2x_4^2x_5^2\over P^3}\right)
-\psi\, \partial_2\left( {x_3^2x_4^2x_5^2\over P^2} 
-5\psi x_2\, {x_1x_3^3x_4^3x_5^3\over P^3}\right)\cr}$$
\vskip10pt
{\noindent\bf Differential equation:}
$$\qquad\left\{\vth^2  - 5^5\lambda\left(\vth + \smallfrac{1}{5}\right)
         \left(\vth + \smallfrac{4}{5}\right)\right\} I_{(3,2,0,0,0)} = 0$$
{\noindent\bf Holomorphic solution at $\l = 0$:}
$$I_{(3,2,0,0,0)}(\psi) = 
{}_2F_1\left(\smallfrac{1}{5},\smallfrac{4}{5};1;\psi^{-5}\right) =
\sum_{m=0}^{\infty} \lambda^m 
{(5m)!\over (\smallfrac{2}{5})_m(\smallfrac{3}{5})_m m!^3}$$
\newpage
$$\vbox{
\halign{\strut # &\hfil $#$\hfil &\hfil $\, #\,$\hfil 
&\hfil $#$\hfil &\hfil $\, #\,$\hfil&\hfil $#$\hfil &\hfil $\, #\,$\hfil
&\hfil $#$\hfil &\hfil $\, #\,$\hfil&\hfil $#$\hfil &\hfil $\, #\,$\hfil
&\hfil $#$\hfil &\hfil $\, #\,$\hfil\cr
\noalign{\noindent$\underline{{\bf Case\ 4}}\ :\quad {\bf v} = (3,1,1,0,0)$}\cr
\noalign{\noindent\bf Diagram:}\cr
\noalign{\vskip-40pt}\cr
& & & & & & & & & \hskip-5pt(1,4,-1,3,3)& \rightarrow &(2,5,0,4,4)\hskip-5pt\cr
& & & & & & & & & \qquad\downarrow D_3\cr
& & & & & & &(0,3,3,2,2)& \rightarrow &(1,4,4,3,3)\cr
& & & & & & & \qquad\downarrow D_1\cr
& & &{\bf (3,1,1,0,0)}& \rightarrow &{\bf (4,2,2,1,1)}& \rightarrow &{\bf (5,3,3,2,2)}\cr
& & & \qquad\downarrow D_4& & \qquad\downarrow D_4\cr
&(2,0,0,4,-1)& \rightarrow &(3,1,1,5,0)& \rightarrow &(4,2,2,6,1)\cr
& \qquad\downarrow D_5& & \qquad\downarrow D_5\cr
&(2,0,0,4,4)& \rightarrow &(3,1,1,5,5)\cr
& \qquad\downarrow D_2 \cr
&(2,5,0,4,4)\cr
}}$$
\vskip10pt
{\noindent\bf Identity for differential forms:}
$$\eqalign{
2\,\psi^4 &Q_{(3,1,1,0,0), 1} + 10 \, (4\psi^5 + 1) Q_{(3,1,1,0,0), 2} 
+ 150\, \psi (\psi^5 - 1)\, Q_{(3,1,1,0,0), 3} \, =\cropen{5pt}  
& ~~~~2\,\psi^4 \partial_4(x_4(Q_{(3,1,1,0,0), 1} + 5\psi Q_{(3,1,1,0,0), 2}))\cropen{5pt} 
&+ \psi^3\partial_5\left({x_1^2x_4^4\over P^2}
+ 10\, \psi x_5 {x_4^5\over P} Q_{(3,1,1,0,0), 1}\right)
+ 10\,\psi^3 \partial_2\left( x_2 \,{x_1^2x_4^4x_5^4\over P^3}\right)\cropen{5pt}
&+ 10\,\psi^2\, \partial_3\left({x_1x_2^4x_4^3x_5^3\over P^3}\right)
- \partial_1\left({x_2^2x_3^2x_4x_5\over P^2} 
- 10\,\psi x_1\, {x_1x_2^3x_3^3x_4^2x_5^2\over P^3}\right)\cr}$$
\vskip10pt
{\noindent\bf Differential equation:}
$$\qquad\left\{\vth\left(\vth - \smallfrac{1}{5}\right)  
       - 5\lambda^5\left(\vth + \smallfrac{1}{5}\right)
         \left(\vth + \smallfrac{3}{5}\right)\right\} I_{(3,1,1,0,0)} = 0$$
\vskip10pt
{\noindent\bf Holomorphic solution at $\l = 0$:}
$$I_{(3,1,1,0,0)}(\psi) = 
{}_2F_1\left(\smallfrac{1}{5},\smallfrac{3}{5};\smallfrac{4}{5};\psi^{-5}\right) =
\sum_{m=0}^{\infty} \lambda^m 
{(5m)!\over (\smallfrac{2}{5})_m(\smallfrac{4}{5})_m^2 m!^2}$$
\newpage
$$\vbox{
\halign{\strut # &\hfil $#$\hfil &\hfil $\, #\,$\hfil 
&\hfil $#$\hfil &\hfil $\, #\,$\hfil&\hfil $#$\hfil &\hfil $\, #\,$\hfil
&\hfil $#$\hfil &\hfil $\, #\,$\hfil&\hfil $#$\hfil &\hfil $\, #\,$\hfil
&\hfil $#$\hfil &\hfil $\, #\,$\hfil\cr
\noalign{\noindent$\underline{{\bf Case\ 5}}\ :\quad {\bf v} = (2,2,1,0,0)$}\cr
\noalign{\noindent\bf Diagram:}\cr
\noalign{\vskip-40pt}\cr
& & & & & & & & & \hskip-3pt(-1,4,3,2,2)& \rightarrow &(0,5,4,3,3)\hskip-10pt\cr
& & & & & & & & & \qquad\downarrow D_1\cr
& & & & & {\bf (2,2,1,0,0)}& \rightarrow &{\bf (3,3,2,1,1)}& \rightarrow
&{\bf (4,4,3,2,2)}\cr 
& & & & & \qquad\downarrow D_4& & \qquad\downarrow D_4\cr
& & &(1,1,0,4,-1)& \rightarrow &(2,2,1,5,0)& \rightarrow &(3,3,2,6,1) \cr
& & & \qquad\downarrow D_5& & \qquad\downarrow D_5\cr
&(0,0,-1,3,3)& \rightarrow &(1,1,0,4,4)& \rightarrow &(2,2,1,5,5)\cr
& \qquad\downarrow D_3& & \qquad\downarrow D_3\cr
&(0,0,4,3,3)& \rightarrow &(1,1,5,4,4)\cr
& \qquad\downarrow D_2 \cr
&(0,5,4,3,3)\cr
}}$$
\vskip10pt
{\noindent\bf Identity for differential forms:}
$$\eqalign{
3\,\psi^3 &Q_{(2,2,1,0,0), 1} + 50\, \psi^4 Q_{(2,2,1,0,0), 2} 
+ 150 \, (\psi^5 - 1) Q_{(2,2,1,0,0), 3} \, =\cropen{5pt}  
& ~~~~\psi^3\partial_4(x_4(3Q_{(2,2,1,0,0), 1} + 10\,\psi Q_{(2,2,1,0,0),2}))\cropen{5pt}  
&+ 2\,\psi^2\partial_5\left({x_1x_2x_4^4\over P^2}
+ 5 \psi\, x_5\, {x_4^5\over P} Q_{(2,2,1,0,0), 1}\right)
+\psi\,\partial_3\left({x_4^3x_5^3\over P^2}
+ 10\,\psi x_3\, {x_1x_2x_4^4x_5^4\over P^3}\right)\cropen{5pt}
&+ 10\,\psi\,\partial_2\left(x_2\,{x_3^4x_4^3x_5^3\over P^3}\right)
+10\, \partial_1\left({x_2^4x_3^3x_4^2x_5^2\over P^3}\right)
\cr}$$
\vskip10pt
{\noindent\bf Differential equation:}
$$\qquad\left\{\vth\left(\vth - \smallfrac{2}{5}\right)  
       - 5^5\lambda\left(\vth + \smallfrac{1}{5}\right)
         \left(\vth + \smallfrac{2}{5}\right)\right\} I_{(2,2,1,0,0)} = 0$$
\vskip10pt
{\noindent\bf Holomorphic solution at $\l = 0$:}
$$I_{(2,2,1,0,0)}(\psi) = 
{}_2F_1\left(\smallfrac{1}{5},\smallfrac{2}{5};\smallfrac{3}{5};\psi^{-5}\right) =
\sum_{m=0}^{\infty} \lambda^m 
{(5m)!\over (\smallfrac{3}{5})_m^2(\smallfrac{4}{5})_m m!^2}$$
\newpage
\subsection{The semiperiods}
We shall first review briefly the properties of what might be termed the
fundamental semiperiod integral pertaining to the quintic~
\REFS{\rperiods}{P. Berglund, P. Candelas, X. de la Ossa, A. Font, T. Hubsch, D.
Jan\v{c}i\'{c} and F.~Quevedo, ``Periods for \cy\ and Landau-Ginzburg Vacua'',\\
\npb{419} 352 (1994), hep-th/9308005.}
\REFSCON{\rsemiperiods}{A.C. Avram, E. Derrick and D. Jancic,\\ ``On Semiperiods'',
\npb{471} 293 (1996),  hep-th/9511152.}
\cite{\CDGP,\rperiods,\rsemiperiods}~
then we will extend the notion
of a semiperiod by associating one with each monomial in the variables $x_1,\ldots,x_5$.

As we have seen the fundamental period can be realised as a residue integral
 $$
\vp_0~=~{5\ps\over (2\p i)^4}\int_{\g_2\times\cdots\times\g_5}
{y_1dy_2dy_3dy_4dy_5\over P(y)}$$ 
with $y_1$ constant and the $\g_i,~i=2,\ldots,5$, circles that enclose the origin. For
$\ps$ sufficiently large one of the integrals, say the $y_5$-integral, has a pole close to
the origin and hence inside the contour. Consider the result of deforming the contours as
in Figure~\figref{loopandwedge}. The rays are chosen so that $\sum_{i=1}^5 y_i^5$ is real
and positive along them. We now use the representation
 $$
{1\over P}~=~\int_0^\infty ds\,e^{-sP}$$
under the integral and then change the variables of integration from
$\{s,\,y_2,\,y_3,\,y_4,\,y_5\}$ to $\{x_1,\,x_2,\,x_3,\,x_4,\,x_5\}$ by setting
$x_i=s^{1\over 5}y_i,~i=1,\ldots,5$, with $y_1$ treated as a constant. In~this way we see
that the fundamental period has the representation
 $$
\vp_0~=~{5\ps\over (2\p i)^4}\int_{\tilde{\g}_1\times\cdots\times\tilde{\g}_5}
\kern-30pt dx_1dx_2dx_3dx_4dx_5\,e^{-P(x,\ps)}$$
with the $\tilde{\g}_i$ as in the second of Figures~\figref{loopandwedge}. The integral
converges for $\ps$ sufficiently small.
\vskip0pt
\def\loopandwedge{\vbox{\vskip0pt\hbox{
\hskip0pt\epsfxsize=5truein\epsfbox{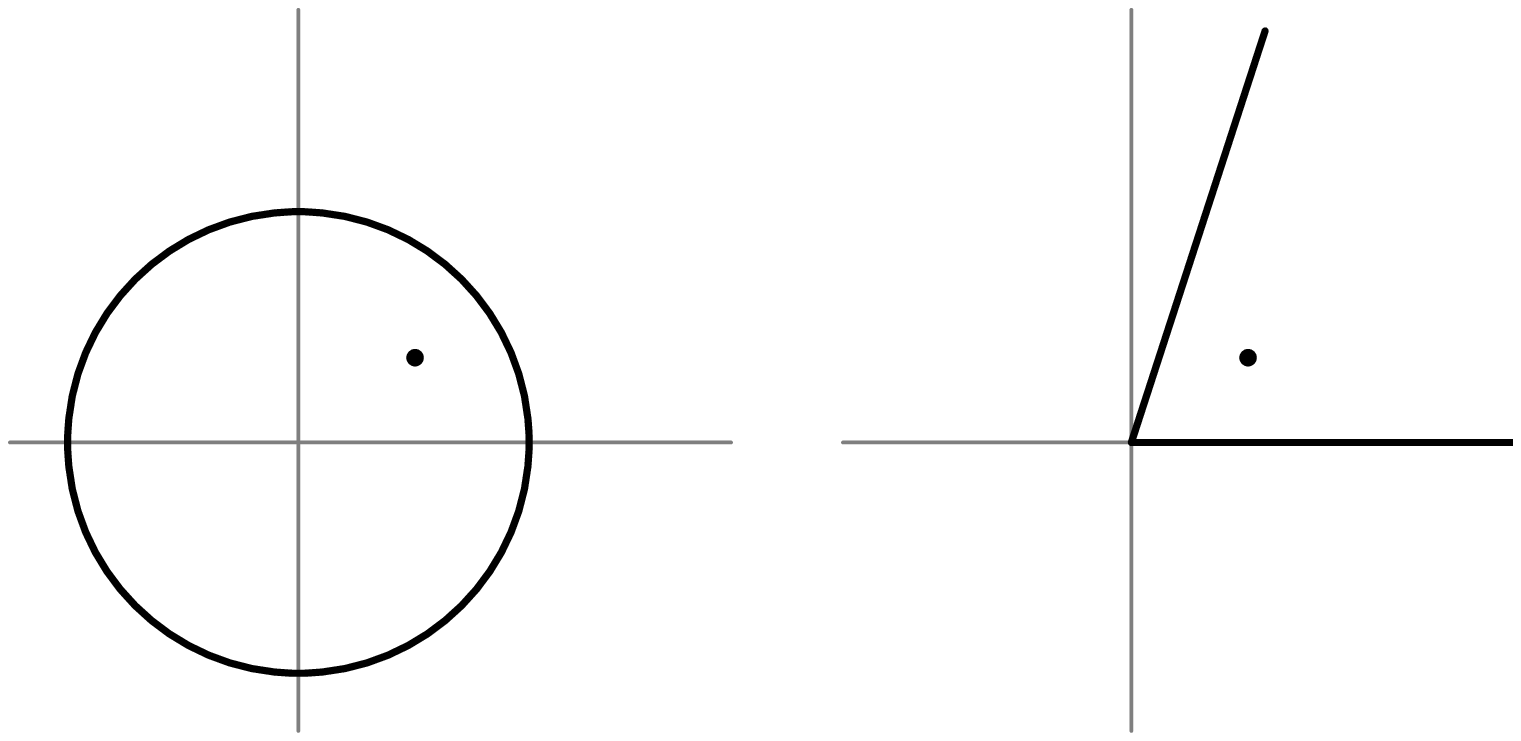}}}}
\figbox{\loopandwedge}{\figlabel{loopandwedge}}{The loops $\g_i$ may be deformed into
pairs of rays running to infinity in the directions of fifth roots of unity.}
It is therefore natural to consider integrals such as 
 $$
\U_0(\ps)~=~5\ps\,\int_0^\infty\kern-5pt d^5x\,e^{-P(x,\ps)} $$
where the integration runs from 0 to $\infty$ for each of the variables $x_i$. We wish to
show that $\U_0(\ps)$ is a semiperiod \ie that it satisfies the differential equation
$\ca{L}\U_0=\hbox{constant}$.
To~this end take $\ps$ to be small and expand the term $e^{-5\ps\prod x}$ in the
integrand as a series:
 $$
\U_0(\ps)~=~5\ps\sum_{n=0}^\infty {(5\ps)^n\over n!}
\left(\int_0^\infty\kern-5pt d\x\, e^{-\x^5} \x^n\right)^5
~=~{1\over 5^5}\sum_{n=0}^\infty\G^5\left({n+1\over 5}\right)\,
{\l^{-\left({n+1\over 5}\right)} \over n!}~.\eqlabel{upsilonzero}$$
From this series it is immediate that
 $$
\ca{L}\U_0~=~-{1\over 5^5} $$
and hence that $\vth\ca{L}\U_0=0$.

Now it is clear that all the functions $\U_0(\z^j\ps)$, with $\z$ a nontrivial
fifth root of unity and $j=0,\ldots, 4$, satisfy the same equation. This gives a set of
five semiperiods. The difference of any two satisfies the homogeneous Picard-Fuchs equation
and hence is a period. We obtain in this way a full set of solutions to the Picard-Fuchs
equation. The logarithms are not apparent in the series \eqref{upsilonzero} since this
series converges for small $\ps$. The logarithms appear when we analytically continue
$\U_0$ to large $\ps$ and expand around $\ps=\infty$.

The function $\U_0$ is a semiperiod associated with the fundamental
period. We wish to discuss also semiperiods associate with the other periods. These other
periods are associated with monomials $x^{\bf v}$ so let us consider integrals of the form
 $$
\U_{\bf v}(\ps)~=~5\ps\,\int_0^\infty\kern-5pt d^5x\, x^{\bf v} e^{-P(x,\ps)} 
~=~{1\over 5^5}\sum_{n=0}^\infty {\prod_{i=1}^5 \G\left({n+v_i+1\over 5}\right) \over n!}
\,\l^{-\left({n+1\over 5}\right)}
$$
These series converge for $|\ps|\leq 1$. To obtain explicit expressions valid for large
$\ps$ one may continue the $\U_{\bf v}$ analytically. For our present purpose however it
suffices to note that we can see from the series that the $\U_{\bf v}$ satisfy the
differential equations $\ca{L}_{\bf v}\U_{\bf v}=0$ with operators
 $$
\ca{L}_{\bf v}~=~\prod_{i=1}^5\left(\vth -{v_i\over 5}\right) - 
\l\prod_{i=1}^5 (5\vth +i)~.\eqlabel{Lv}$$
We can seek solutions to these equations that converge for large $\ps$ \ie small $\l$.
These will contain logarithms so we initially seek series solutions of the form
 $$
\sum_{n=0}^\infty A_{{\bf v},n}(\ve)\l^{n+\ve}$$
with coefficients that can be taken to be
 $$
A_{{\bf v},n}(\ve)~=~{\G(5n+5\ve+1)\over \G(5\ve+1)}
\prod_{i=1}^5{\G\left( \ve+1-{v_i\over 5}\right)\over 
\G\left( n+\ve+1-{v_i\over 5}\right) }~.\eqlabel{gencoeffs}$$

We saw previously that the periods associated with the monomials of cases 2 -- 5 are in
fact hypergeometric functions of second order. It is of interest to observe that the
operator $\ca{L}_{\bf v}$ accommodates this by factorizing into the product of a third
order operator and the second order operator corresponding to the hypergeometric equation.
For the case ${\bf v}=(4,1,0,0,0)$ for example we have
 $$
\ca{L}_{(4,1,0,0,0)}~=~\vth \left(\vth - \smallfrac{1}{5}\right) 
\left(\vth - \smallfrac{4}{5}\right)
\left\{\vth^2 - 5^5\l \left(\vth + \smallfrac{2}{5}\right)
\left(\vth + \smallfrac{3}{5}\right)\right\}
$$
and the operators corresponding to the other cases factor similarly.
\newpage
\section{calczero}{\bignupsi\ in Zeroth Order}
\vskip-20pt
\subsection{Generalities}
The first thing to note is that we work over $\F$ rather than the projective
space $\Fp\bb{P}^4$ so the cardinality of the \cym\ is given by
 $$
N(\ps)~=~{\n(\ps) - 1\over p-1}~.$$
It is also often easier to work over $\Fstar$, the set of points for which no $x_i$ is
zero. We therefore distinguish between
 $$
X_\ps~=~\left\{x\in\F~|~P(x,\ps)=0\right\}~~~\hbox{and}~~~
X_\ps^*~=~\left\{x\in\Fstar~|~P(x,\ps)=0\right\}$$
and denote by $\n(\ps)$ and $\n^*(\ps)$ the cardinality of these two sets.

Note that $\Fstar$ is the disjoint union of the $X_\ps^*$ since if $x\in\Fstar$ then 
$\prod x_i\neq 0$ so 
 $$
{\sum_{i=1}^5 x_i^5\over \prod_{i=1}^5 x_i}~=~5\,\ps $$
for some $\ps$ and clearly no $x$ can correspond to more than one $\ps$. Thus we have
 $$
\sum_{\ps\in\sevenFp}\n^*(\ps) ~=~ (p-1)^5~.$$
Consider now $x\in \F\setminus\Fstar$, that is the set with at least one $x_i$ zero. On
this set the quintic reduces to $\sum_{i=1}^5 x_i^5=0$. Thus the set
 $$
Y~=~X_\ps\cap(\F\setminus\Fstar)$$
is independent of $\ps$ and
 $$
X_\ps~=~X_\ps^*\cup Y~~\hbox{and}~~\n(\ps)~=~\n^*(\ps) + \n_Y$$
with $\n_Y=\hbox{Card}(Y)$. It is easy to evaluate $\n$ for the case that $5\notdiv (p-1)$ by
calculating the number of points of $Y$ with precisely $j$ coordinates nonzero for 
$0\leq j\leq 4$. We find
 $$
\n_Y~=~ - 4 + 10p - 10p^2 + 5p^3~~;~~5\notdiv (p-1)~.$$
It is also easy to calculate $\n(0)$ for the case $5\notdiv (p-1)$ by simply observing that
the equation $\sum_{i=1}^5x_i^5=0$ can be solved by assigning arbitrary values to the
variables $x_1,x_2,x_3,x_4$ and then solving uniquely for $x_5$ since there are no
nontrivial fifth roots of unity when $5\notdiv (p-1)$.
Hence
 $$
\n(0)~=~p^4~~;~~5\notdiv (p-1)~.$$
When $5|(p-1)$ expressions for $\n_Y$ and $\n(0)$ may be found in terms of Gauss sums or by
expanding these quantities p-adically as in the following.
\subsection{The calculation to zeroth order}
We begin with the zeroth order calculation for the case that $5\notdiv (p-1)$. This is
elementary but a good introduction to the nature of the computation.

The number of points mod $p$ is, as we see from \eqref{facttwo}, given by
 $$
\n(\ps)~=~\sum_{x\in\sevenF}\left( 1 - P(x,\ps)^{p-1}\right) + \ord{p}~.$$
In this expression we expand $P^{p-1}$ as binomial series and the powers of 
$\sum x_i^5$ using the multinomial series
 $$\eqalign{
\n(\ps)&= p^5 - \sum_{x\in\sevenF}\sum_{n=0}^{p-1}\sum_{\sum n_i = n}
{(p-1)!\over (p-1-n)!\prod_{i=1}^5 n_i!}\,(-5\ps)^{p-1-n}\, 
\prod_{i=1}^5 x_i^{5n_i-n+p-1}+\ord{p}\cropen{5pt}
&= - \sum_{x\in\sevenFstar}\sum_{n=0}^{p-2}\sum_{\sum n_i = n}
{n!\over \prod_{i=1}^5 n_i!}\,(5\ps)^{-n}\,\prod_{i=1}^5 x_i^{5n_i-n}+\ord{p}\cr}~.$$
Where, in passing to the second equality, we have recognised that
 $$
{(p-1)!\over (p-1-n)!}~=~(p-1)(p-2)\ldots(p-n)~=~(-1)^n n! + \ord{p}$$
as well as the fact that $a^{p-1}=1+\ord{p}$ for $a\neq 0$. It is useful to recognise also
that the $n=p-1$ term in the $n$ sum corresponds to $\n(0)=p^4$ and so may be omitted; also
owing to the powers of $\prod_{i=1}^5 x_i$, in the remaining terms the $x$-sum can be
restricted to $\Fstar$. Now we use the fact that
 $$
\sum_{x_i\in\sevenFp}\,x_i^{5n_i-n}~=~
\cases{\ord{p} &,~ \hbox{if $5n_i-n\neq 0$}\cropen{10pt} 
-1+\ord{p} &,~ \hbox{if $5n_i-n = 0$~.}\cr}\eqlabel{facttwo}$$
So the $\ord1$ term is obtained by setting
 $$
5n_i~=~n~,~~i=1,\ldots,5~;$$
so the $n_i$ are all equal, $n_i=m$ say, and $n=5m$. Thus we find
 $$
\n(\ps)~=~\sum_{m=0}^{\left[{p\over 5}\right]} {(5m)!\over (m!)^5}\,\l^m+\ord{p}~=~
\ftrunc{0}{\left[p/5\right]}(\l) + \ord{p}~.$$
The case $5|(p-1)$ is requires a more careful analysis but the result is the same to this
order.
\subsection{The zeroth order expression for a general CY hypersurface in a toric variety}
With only a little more work an analogous result may be established for the case that $M$
is a \cy\ hypersurface in a toric variety. Such a hypersurface is described by a reflexive
polyhedron~$\D$ 
 $$
P(x, \a)~=~\sum_{{\bf m}\in\D}\,\a_{\bf m}\,x^{\bf m}~. $$
In general there will be $N+4$ coordinates $x_i$ and $P$ is homogeneous with respect to $N$
scaling relations. Thus there are $N$ degree vectors ${\bf h}_j$ such that the $N$
multidegrees 
 $$
\deg_j(\bm)~=~{\bf h}_j\cdot\bm~=~{\bf h}_j\cdot\bone$$
are independent of $\bm$. The polyhedron $\D$ is therefore a four-dimensional polyhedron
embedded in an $N+4$ dimensional space.
There are now many parameters $\a_{\bf m}$ but among them is a parameter
$\a_\bone$ analogous to $\ps$ that multiplies the fundamental monomial 
$Q=x_1\ldots x_{N+4}$. First we shall write an expression for the fundamental period
$\vp_0(\a)$ as a function of the coefficients $\a_\bm$ and then we shall show that, mod
$p$, the number of rational points, $\n(\a)$, is given by a truncation of the
fundamental~period. We derive the relation 
$\vp_0(\a)\equiv \vptrunc{p-1}(\a) \vp_0(\a^p)$ which is the generalization of
\eqref{Frobzero}.  

First we write $\ph=-\a_\bone$ and separate out the fundamental monomial 
 $$
P(x,\a)~=~\widetilde{P}(x,\a) - \ph\, Q\qquad\hbox{with}\qquad
\widetilde{P}(x,\a)~=~\sum_{{\bf m}\in\D \atop \bm\neq \bone} \a_{\bf m} x^{\bf m}$$
then following \cite{\rperiods} and our discussion of \SS\chapref{allperiods} we have
 $$
\vp_0(\a)
~=~{\ph\over (2\p i)^{N+4}}\oint {d^{N+4}x\over \ph\,Q - \widetilde{P}}
~=~{1\over (2\p i)^{N+4}}\oint {d^{N+4}x\over Q}
\left(1 - {\widetilde{P}\over \ph\, Q}\right)^{-1}~.$$
In the integral on the right we expand the bracket which yields the expansion in inverse
powers of $\ph$:
 $$\eqalign{
\vp_0(\a)
~&=~{1\over (2\p i)^{N+4}}\sum_{n=0}^\infty {1\over \ph^n}
\oint{d^{N+4}x\over Q}{\widetilde{P}^n\over Q^n}\cropen{10pt}
~&=~{1\over (2\p i)^{N+4}}\sum_{n=0}^\infty {1\over \ph^n}
\sum_{n_\bm\atop \sum\limits_\bm n_\bm =n}
\left(\prod_\bm\a_\bm^{n_\bm}\right)\,{n!\over \prod_\bm n_\bm!}
\oint {d^{N+4}x\over Q}\,x^{\sum_\bm n_\bm \bm - n {\bf 1}}\cr}$$
where in this and the following equations the label $\bm$ is understood to run over all
monomials of $\D$ apart from $\bone$. The effect of the contour integral is to pick out
coefficients $n_\bm$ such~that
 $$
\sum_\bm n_\bm\, \bm~=~n{\bf 1}~.\eqlabel{nmcondition}$$ 
Note that this condition implies that $\sum_\bm n_\bm = n$, since the monomials all have
the same multidegrees, so this condition need not be imposed separately.
We are left with the following series for $\vp_0$
 $$
\vp_0(\a)~=~
\sum_{n=0}^\infty\, {n!\over \ph^n}\kern-7pt
\sum_{n_\bm\atop \lower5pt\hbox{$\scriptscriptstyle\sum\limits_\bm n_\bm \bm =n {\bf 1}$}}
\kern-7pt\prod_\bm\,{\a_\bm^{n_\bm}\over n_\bm!}~.\eqlabel{piofalpha}$$
There are now many sums in this series; however there is a principal sum over $n$ and
subordinate sums such that, if we truncate the principal sum, the subordinate sums are
over a finite number of terms.

In order to compute $\n(\a)$ mod $p$ we proceed by summing $\left(1-P^{p-1}\right)$ as in
the previous subsection. After a few lines of calculation we find the expression
 $$
\n(\a)~=~-\sum_{n=0}^{p-1}\,{n!\over \ph^n}
\sum_{n_\bm\atop\lower5pt\hbox{$\scriptscriptstyle\sum\limits_\bm n_\bm = n$}}
\prod_\bm\, {\a_\bm^{n_\bm}\over n_\bm!}
\sum_{x\in\sevenFp^{N+4}} x^{\sum_\bm n_\bm \bm + (p-1-n){\bf 1}} + \ord{p}~.$$
The $x$-sum further restricts the $n_\bm$ to those that satisfy
 $$
\sum_\bm n_\bm \bm ~=~ n{\bf 1} + (p-1){\bf u}$$
with $\bf u$ an integral vector with nonnegative components. By taking degrees and
comparing with the restriction $\sum_\bm n_\bm = n$ we see that $\hbox{deg}({\bf u}) = 0$
and hence that $\bf u$ vanishes. In this way we see that the $n_\bm$ are restricted by
\eqref{nmcondition}. On taking into account the factor of $(p-1)^{N+4}$ that arises from
the $x$-sum we see that
 $$
\n(\a)~=~(-1)^{N-1}\,\vptrunc{p-1}(\a) + \ord{p}~.$$
This result holds irrespective of whether or not $p-1$ is divisible by the multidegrees
of~$P$.
\subsection{A congruence involving the Frobenius map}
Consider again the expression \eqref{piofalpha} for the fundamental period and break up
the sums by writing 
 $$
n_\bm~=~r_\bm\, p + s_\bm\qquad\hbox{and}\qquad n~=~rp + s $$
with $0\leq s_\bm\leq p-1$ and $0\leq s\leq p-1$. Now it will not, in general, be the case
that 
 $$
\sum r_\bm = r \qquad\hbox{and}\qquad \sum s_\bm = s \eqlabel{simplecase}$$
owing to the carries associated with the sum $\sum n_\bm$. However the ratio
${n!/\prod_\bm\, n_\bm!}$ is $\ord{1}$ precisely when there are no carries so to this order
we may impose \eqref{simplecase}. Thus 
 $$\eqalign{
\vp_0(\a)~&=~\sum_{\sum r_\bm\,\bm = r}\; \sum_{\sum s_\bm\,\bm = s}\,
{r!\, s!\over \ph^{rp+s}}\> \prod_\bm {\a_\bm^{r_\bm p}\, \a_\bm^{s_\bm}\over r_\bm!\,
s_\bm!} + \ord{p}\cropen{7pt}
~&=~\vptrunc{p-1}(\a)\,\vp_0(\a^p) + \ord{p}~.\cr}$$
\newpage
\section{calcone}{\bignupsi\ in First Order}
\vskip-20pt
\subsection{The first order calculation}
To calculate $\n(\ps)$ mod $p^2$ we now use the formula
 $$
\n(\ps)~=~\sum_{x\in\sevenFp^5} \left( 1 - P(x,\ps)^{p(p-1)}\right) + \ord{p^2}~~,$$
since if $P(x,\ps)\neq 0~~\hbox{mod $p$}$, then
$P(x,\ps)^{(p-1)}~= 1 + \ord{p}$, and
$P(x,\ps)^{p(p-1)}= 1 + \ord{p^2}$.

A lengthy calculation which we do not give here yields $\n(\ps)$ correct through
$\ca{O}(p)$
 $$\eqalign{
\n(\ps)~&=~\ftrunc{0}{\left[{2p/5}\right]}(\l^p) +
p\,\ftrunc{1}{\left[{2p/5}\right]}'(\l^p)\cropen{10pt}  
&\hskip20pt -\d_p\, p\left\{ \sum_{m=0}^k\l^{mp}\hskip-5pt
\sum_{{{\bf v}\in \th\atop 1\leq\hbox{\eightrm dim}\th\leq 4}} 
{(5m)!\over \prod_{i=1}^5 (m{+}v_i k)!} - {20\over (2k)!(k!)^3}
\right\} +\ord{p^2}\cr} \eqlabel{newnu}$$
where $f_1'(\l^p)$ means $(\vth f_1)(\l^p)$ with $\vth=\l{d\over d\l}$, and 
where $\d_p$ distinguishes the cases that \hbox{$5|(p-1)$}
 $$
\d_p~=~\cases{0~, &if $5\notdiv (p-1)$\cropen{5pt}              
              1~, &if $5\hskip2.6pt|\;(p-1)$\cr}$$ 
and where one of the sums is taken over faces, $\th$, of $\D$ of dimension greater than
one; that is the sum is taken over all quintic monomials apart from the interior point and
the vertices of $\D$.  
 
There are perhaps two puzzles that present themselves with regard to the case that
$5|p-1$. The first is that hypergeometric functions emerge that appear to be of fifth
order rather than the second order hypergeometric functions that we might have expected
from our discussion of the periods. The second puzzle is that the sum over the points of
$\D$ includes points interior to codimension one faces which do not contribute to the
count of the periods. We shall see that both of these difficulties disappear on closer
inspection.
\subsection{The sum over monomials}
For the case $5|p-1$ let us consider the sum over monomials ${\bf v}\in \D$. It is
sufficient to list these up to permutation of the components which we do in the table
below. In the table it is the monomial $(2,1,1,1,0)$ together with its permutations that
lies in the interior of the codimension one faces.
\vskip5pt
$$\vbox{\def\skip{\hskip7pt}
\offinterlineskip\halign{
\strut #\vrule height 12pt depth 6pt&\hfil\skip #\skip\hfil\vrule
&\hfil\skip #\skip\hfil\vrule&\hfil\skip $\{#\}$\skip\hfil\vrule\cr
\noalign{\hrule}
&monomial &permutations &a_{\bf v},\,b_{\bf v};\,c_{\bf v} \cr
\noalign{\hrule\vskip3pt\hrule}
&(4,1,0,0,0) &20 &{2\over 5},\,{3\over 5};\,1\cr
&(3,2,0,0,0) &20 &{1\over 5},\,{4\over 5};\,1\cr
&(3,1,1,0,0) &30 &{1\over 5},\,{3\over 5};\,{4\over 5}\cr
&(2,2,1,0,0) &30 &{1\over 5},\,{2\over 5};\,{3\over 5}\cr
&(2,1,1,1,0) &20 &\omit{\hfil\vrule}\cr
\noalign{\hrule}
}}$$
\vskip3pt
\centerline{Table~\tablabel{vtableone}}
\vskip10pt

Consider now the new terms that appear when $5|p-1$. These can be written as above which
emphasizes the simple dependence of these terms on the monomials of $\D$. On the other
hand we may rewrite these terms to show that they correspond to the periods that we met in
\SS\chapref{allperiods}. In order to simplify the expression further we note that
 $$
(m+kv_i)!~=~\poch{1+kv_i}{m}\, (kv_i)!~=~\poch{1-{v_i\over 5}}{m}\,(kv_i)! + \ord{p}
\eqlabel{poch}$$
where $\poch{a}{m}=\G(a+m)/\G(a)=a(a+1)\ldots(a+m-1)$ denotes the Pochhammer symbol.
Before proceeding note that this shows that to the required order the term corresponds to
a fifth order differential equation with operator $\ca{L}_{\bf v}$ as in \eqref{Lv}.

By using \eqref{poch} and also the multiplication formula \eqref{multclass} we may rewrite
the terms enclosed in braces in \eqref{newnu} in the form
 $$
\sum_{\bf v} {\g_{\bf v}\over \prod_{i=1}^5 (v_i k)!}\sum_{m=0}^k 
{\prod_{i=1}^5 \poch{i\over 5}{m}\over \prod_{i=1}^5 \poch{1 - {v_i\over 5}}{m}}\,\ps^{-5m}
 - {20\over (2k)!(k!)^3} +\ord{p} \eqlabel{vcontribs}$$
where the coefficients $\g_{\bf v}$ account for the permutation factors of the table. For
a given $\bf v$ there is cancellation between the Pochhammer symbols in the numerator and
denominator. Consider the contribution of the monomial $(4,1,0,0,0)$. The ratio of
Pochhammer symbols reduces to $\poch{2/5}{m}\poch{3/5}{m}/(m!)^2$. So the contribution of
this monomial reduces to the truncation of a second order hypergeometric function
 $$
{20\over (4k)! k!}\sum_{m=0}^k {\poch{2\over 5}{m}\poch{3\over 5}{m}\over (m!)^2}\,
\ps^{-5m}~=~{20\over(4k)!k!}\,
{}^k_2F_1^{}\kern-3pt\left(\smallfrac{2}{5},\smallfrac{3}{5};1;\,\ps^{-5}\right)~.$$   
In this way the contribution of each $\bf v$, apart from $(2,1,1,1,0)$, reduces to the form
 $$
{\g_{\bf v}\over \prod_{i=1}^5 (v_ik)!}\,
{}^k_2F_1^{}\kern-3pt\left(a_{\bf v},b_{\bf v};c_{\bf v};\,\ps^{-5}\right)$$
and we list the parameters of the hypergeometric functions in Table~\tabref{vtableone}.

We turn now to discuss the contribution of the discrepant monomial $(2,1,1,1,0)$ which
appears to be related to a ${}_3F_2$ but turns out to be a prolongation of the
contribution of $(4,1,0,0,0)$. As a preliminary we note some elementary congruences. First
by writing
 $$
(p-1)!~=~(p-1)(p-2)\ldots(p-k)\times (4k)!~=~(p-1)(p-2)\ldots(p-2k)\times (3k)!$$
we see that 
 $$
(4k)!\,k!~=~-1+\ord{p}~~~\hbox{and}~~~(3k)!\,(2k)!~=~-1+\ord{p}~.$$
We may also write 
 $$
\G(a+\m)~=~{\G(a+\m)\over \G(a)}\,\G(a)~=~\poch{a}{\m}\G(a) $$
and applying this to $(ak)!$ we find
 $$
(ak)!~=~\poch{(a-1)k+1}{k}\,\big((a-1)k\big)!~=~
\poch{1-{(a-1)\over 5}}{k}\,\big((a-1)k\big)! + \ord{p}~;~~2\leq a\leq5 $$
from which we see that
 $$
(ak)!~=~k!\,\prod_{i=1}^{a-1}\poch{1-{i\over 5}}{k} + \ord{p}~.$$

Turn now to the contribution of the monomial $(2,1,1,1,0)$. First note that the extra
constant in \eqref{vcontribs} is just such as to remove the $m=0$ term. Thus we are
concerned with the sum
 $$
{20\over (2k)!\,(k!)^3}\sum_{m=1}^k 
{\poch{1\over 5}{m}\poch{2\over 5}{m}\over \poch{4\over 5}{m}^2}\,\ps^{-5m}~.
\eqlabel{discrepant}$$
Consider now the prolongation of the contribution of the monomial $(4,1,0,0,0)$ to $2k$
terms.
 $$\eqalign{
{20\over (4k)!\,k!}\sum_{m=1}^{k}
{\poch{2\over 5}{m+k}\poch{3\over 5}{m+k}\over \poch{1}{m+k}^2}\,\ps^{-5m}~&=~
-{\poch{2\over 5}{k}\poch{3\over 5}{k}\over (k!)^2}\sum_{m=1}^{k}
{\poch{k+{2\over 5}}{m}\poch{k+{3\over 5}}{m}\over \poch{k+1}{m}^2}\,\ps^{-5m}\cropen{5pt}
~&=~{20\over (2k)!\,(k!)^3}\sum_{m=1}^k 
{\poch{1\over 5}{m}\poch{2\over 5}{m}\over \poch{4\over 5}{m}^2}\,\ps^{-5m} 
+ \ord{p}\cr}$$
which is the same as \eqref{discrepant} to the required order and where in passing to the
last equality we have used the congruences above. 
To write the  contributions uniformly we may now extend the summation as far as $m=p-1$
since the additional terms are $\ord{p^2}$
 $$\eqalign{
\n(\ps)&=\ftrunc{0}{(p-1)}(\l^p) + p\,\ftrunc{1}{(p-1)}'(\l^p)\cropen{3pt}  
&\hskip20pt -\d_p\, p\,\sum_{\bf v}{\g_{\bf v}\over \prod_{i=1}^5 (v_ik)!}\kern5pt
{}^{(p-1)}\kern-5pt{}^{}_2F_1^{}\kern-3pt\left(a_{\bf v},b_{\bf v};
c_{\bf v};\,\ps^{-5}\right) +\ord{p^2}\cr} \eqlabel{orderone}$$
where the monomial $(2,1,1,1,0)$ is now omitted from the $\bf v$-sum. 
\newpage
\section{higher}{\bignupsi\ in Higher Order and for Finer Fields}
In this section we will make an ansatz for $\n(\ps)$ for the case that $5\notdiv p-1$.
Which we are then able to compare with the numbers obtained by counting solutions to the
equation $P=0$. It turns out that this process is considerably simpler for the case 
$5\notdiv p-1$ since then we do not have to take into account the contributions of the
other periods. It will be natural then to extend the ansatz to the case of $\bb{F}_q$
with $q=p^s$ providing $5\notdiv q-1$. Note that 5 will divide $p^s-1$ for some values of
$s$ even if $5\notdiv p-1$. In fact it is easy to see that $5|p^4-1$ for all $p\neq 5$ so
this restriction is quite serious. Fortunately we will be able to compute $\n(\ps)$ in
terms of Gauss sums in \SS\chapref{gauss} for the case that $5|q-1$ as well as the case
$5\notdiv q-1$. This will in fact establish the expressions that we propose here. We
prefer however to proceed in the following way since this emphasizes the role of the
periods and this is somewhat harder to see when one works with the Gauss sums.
\subsection{An ansatz for the case $5\notdiv p-1$}
For the case $5\notdiv (p-1)$ we can guess the form of
$\n(\ps)$ to next order
 $$
\n(\ps)~=~(1+A\,p^2)\ftrunc{0}{\left[{3p/5}\right]}(\l^{p^2}) +
p(1+B\,p)\ftrunc{1}{\left[{3p/5}\right]}'(\l^{p^2}) 
+ C\,p^2\ftrunc{2}{\left[{3p/5}\right]}''(\l^{p^2}) +\ca{O}(p^3)$$
and fix the constants by comparing with the value of $\n(\ps)$ found by counting points
with a computer for various values of $p$. Having fixed these constants we make a similar
ansatz at next order and so on. In this way we find what appears to be the exact
expression for~$\n(\ps)$
$$\eqalign{
\n(\ps)~=~&\ftrunc{0}{(p-1)}(\l^{p^4}) +
\left({p\over 1-p}\right)\ftrunc{1}{(p-1)}'(\l^{p^4}) 
+ {1\over 2!}\left({p\over 1-p}\right)^2\ftrunc{2}{(p-1)}''(\l^{p^4})\cropen{7pt}
&+ {1\over 3!}\left({p\over 1-p}\right)^3\ftrunc{3}{(p-1)}'''(\l^{p^4}) 
+ {1\over 4!}\left({p\over 1-p}\right)^4\ftrunc{4}{(p-1)}''''(\l^{p^4})\hskip20pt
(\hbox{mod}\,p^5)\cr}$$
The most interesting aspect is the appearance of the semiperiod.
There is a procedure for finding the differential equations satisfied by the periods
directly from the reflexive polyhedron $\nabla$ corresponding to the manifold; this yields
the GKZ system. This does not quite give the PF equation. For our case the PF operator~is
 $$
\ca{L}~=~\vth^4 - 5\l\,\prod_{i=1}^4(5\vth+i)~~,~~~~\hbox{with}~~~\vth=\l{d\over d\l}$$
while the GKZ operator is 
 $$
\ca{L^\nabla}~=~\vth\,\ca{L}~=~\vth^5 - \l\,\prod_{i=1}^5(5\vth+i)~.$$
All the solutions to the PF equation satisfy the higher order equation, but there is now an extra solution
 $$
\vp_4(\l)~=~f_0(\l)\,\log^4\l + 4\,f_1(\l)\,\log^3\l + 6\,f_2(\l)\,\log^2\l  
+ 4\,f_3(\l)\,\log\l + f_4(\l)$$
The semiperiod is also an integral of $\O$
 $$
\vp_4(\l)~=~\int_C \O~,~~~\hbox{but}~~~\partial C\neq 0~.$$
\subsection{The Method of Frobenius}
We have to be clear now about the basis for the periods. Saying that they have the form
 $$\eqalign{
\vp_0~&=~f_0\cr
\vp_1~&=~f_0\,\log\l + f_1\quad\hbox{etc.}\cr}$$
does not specify the $f_k$ uniquely. We can pick a particular basis using the method of 
Frobenius. We seek a series solution to the GKZ equation $\ca{L}^\nabla F(\l,\ve)=\ve^5$
of the form
 $$
F(\l,\ve)~=~\sum_{m=0}^\infty A_m(\ve)\,\l^{m+\ve}$$
and our periods are obtained as the derivatives of $F(\l,\ve)$ with respect to $\ve$. We
work mod~$\ve^5$ and expand
 $$
F(\l,\ve)~=~\sum_{k=0}^4 {\ve^k\over k!}\,\vp_k(\l)~.$$
The coefficients satisfy the recurrence
 $$
A_m(\ve)~=~{(5m+5\ve)(5m+5\ve-1)\ldots(5m+5\ve-4)\over (m+\ve)^5}\,A_{m-1}(\ve)$$
so we choose
 $$
A_m(\ve)~=~{\G(5m+5\ve+1)\over \G^5(m+\ve+1)}{\G^5(\ve+1)\over \G(5\ve+1)}\eqlabel{Am}~.$$
Differentiating $A_m(\ve)$ with respect to $\ve$ we have
 $$
A'_m(\ve)~=~5\Ph_m(\ve)A_m(\ve)~~,~~~
A''_m(\ve)~=~5\left(\Ph'_m(\ve)+5\Ph^2(\ve)\right)A_m(\ve)~~,~~~\hbox{etc.}$$
where $\Ph_m$ can be written in terms of the digamma function
 $$\eqalign{
\Ph_m(\ve)~&=~\big( \Ps(5m + 5\ve + 1) - \Ps(5\ve + 1)\big) -
\big( \Ps(m + \ve + 1) - \Ps(\ve + 1)\big)\cropen{3pt}
&=~\sum_{j=1}^{5m}{1\over j+5\ve} - \sum_{j=1}^{m}{1\over j+\ve} \cr}$$
Thus the $A_m^{(\ell)}(0)$ are of the form $b_{m\ell}a_{m}$ with $b_{m\ell}$ a
polynomial in the
 $$
\Ph_m^{(k)}(0)~=~(-1)^k k!\left( 5^k \s_{5m}^{(k+1)} - \s_{m}^{(k+1)}\right) $$
with $0\leq k\leq \ell$ and with the $\s_m^{(\ell)}$ as in \eqref{sigdef}.

 We work mod $\ve^5$ and expand
 $$
A_m(\ve)~=~\sum_{k=0}^4{1\over k!}\,\ve^k a_m b_{mk} \eqlabel{Amfour}$$
then the $f_k$ are simply
 $$
f_k~=~\sum_{m=0}^\infty a_m b_{mk} \l^m~.\eqlabel{fk}$$
The extra $\G$-functions that appear in the $A_m$ serve to remove transcendental quantities
$\Psi'(1),~\Psi''(1)$ etc.~that would otherwise appear in the coefficients $b_{mk}$.

Before proceeding note a useful identity. In the definition \eqref{Am}
set $m = s$ and $\ve = rp$. We see that
 $$
a_{rp+s}~=~a_{rp}\,A_s(rp)$$
which we may regard as the exact form of the congruence \eqref{acongruence}.
(Note that we work mod $p^5$ in order to calculate $\n(\ps)$.)
\vskip20pt
Returning to our discussion of the choice of basis note that we may replace the $A_m$ by
another choice. We can parametrize this freedom by means of a function $h(\ve)$ with
$h(0)=1$
 $$
\widetilde{A}_m(\ve)~=~h(\ve)\,A_m(\ve)~~\hbox{with}
~~h(\ve)~=~\sum_{k=0}^4 {1\over k!}\,h_k\, \ve^k~.$$
This yields another basis:

 $$
\pmatrix{\tilde{f}_0\cropen{2pt}\tilde{f}_1\cropen{2pt}\tilde{f}_2\cropen{2pt}\tilde{f}_3\cropen{2pt}
\tilde{f}_4\cr}~=~
\pmatrix{1   &0   &0   &0   &0  \cropen{2pt}
         h_1 &1   &0   &0   &0  \cropen{2pt}
         h_2 &2h_1&1   &0   &0  \cropen{2pt}
         h_3 &3h_2&3h_1&1   &0  \cropen{2pt}
         h_4 &4h_3&6h_2&4h_1&1  \cr}
\pmatrix{f_0\cropen{2pt} f_1\cropen{2pt} f_2\cropen{2pt} f_3\cropen{2pt} f_4\cr}$$

In the $f_k$ basis our result for $\n(\ps)$ takes the form
 $$\eqalign{
\n(\ps)~&=~\ftrunc{0}{(p-1)}(\l^{p^4}) +\ldots + 
 {1\over 4!}\left({p\over 1{-}p}\right)^4\ftrunc{4}{(p-1)}^{(4)}(\l^{p^4})\cropen{7pt}
&\hskip20pt + C_3\left({p\over 1{-}p}\right)^3
\left\{\hskip-3pt\ftrunc{0}{(p-1)}^{(3)}(\l^{p^4}) + \left({p\over 1{-}p}\right)
\ftrunc{1}{(p-1)}^{(4)}(\l^{p^4})\right\}+\ord{p^5}\cr}\eqlabel{Npsi}$$
where $C_3$ is a constant (depending on $p$) that, on the basis of numerical
experimentation, we identify as
$$
C_3~=~240\,{\s_{p-1}\over p^2} + \ord{p^2}~.$$ 
The `ugly' terms in \eqref{Npsi} may be removed by a change of basis
corresponding to the function
 $$ 
h(\ve)~=~1 + {C_3\over 3!}\,\ve^3 + \ord{\ve^5}~.$$
This $h$ is interesting because it is the same function that appears in the expansion
 $$
{a_{rp}\over a_r}~=~ h(rp) ~.\eqlabel{hratio}$$
We can therefore rewrite our expression for the number of points
 $$\eqalign{
\n(\ps)~&=~\sum_{s=0}^{p-1}
\sum_{k=0}^4{1\over k!}\left({ps\over 1-p}\right)^k\,b_{sk}\,a_s\,
h\left({ps\over 1-p}\right)\,\l^{sp^4}\cropen{5pt}
&=~\sum_{s=0}^{p-1}
A_s\left({ps\over 1-p}\right)
\,{a_{sp(1+p+p^2+p^3)}\over a_{s(1+p+p^2+p^3)}}\,\l^{sp^4}\cropen{5pt}
&=~\sum_{s=0}^{p-1}{a_{s(1+p+p^2+p^3+p^4)}\over a_{s(1+p+p^2+p^3)}}\,\l^{sp^4}\cr}
\eqlabel{aovera}$$
the right hand side being understood as evaluated mod $p^5$.

We have succeeded in finding an expression for the exact number by evaluating the
expressions in \eqref{Npsi} and \eqref{aovera} mod $p^5$. We can however do
better by seeking an exact p-adic expression. In virtue of our discussion of
\SS\chapref{prelims}.5 we recognise
$\l^{p^4}$ as the appropriate approximation to $\teich(\l)$. It is therefore natural to
suppose that this is given by
 $$
\n(\ps)~=~\sum_{m=0}^{p-1}\b_m\,\teich^m(\l) \eqlabel{padicN}$$
with 
 $$
\b_m~=~\lim_{n\to\infty} 
{a_{m(1+p+p^2+\ldots+p^{n+1})}\over a_{m(1+p+p^2+\ldots+p^{n})}}~. \eqlabel{betam}$$
We shall shortly evaluate this limit. We can
however also retrace the steps that led to \eqref{padicN} to write expressions for
$\n(\ps)$ in terms of an extended set of semiperiods that will obtain mod $p^6$, mod $p^7$
and so on. To do this we simply forget the restriction $\ve^5=0$ and expand $A_m(\ve)$ as
in
\eqref{Am} but now to all orders in $\ve$
 $$
A_m(\ve)~=~\sum_{k=0}^\infty{1\over k!}\,\ve^k a_m b_{mk} \eqlabel{Aminfty}$$
and through \eqref{fk} this defines semiperiods $f_j$ for all $j$. These semiperiods
satisfy the equations 
 $$
\vth^{j-3}\ca{L}\,f_j~=~0~~,~~~\hbox{for $j \geq 3$.}$$

Returning to the evaluation of the coefficients $\b_m$ it is useful also to write the 
ratio $a_{rp}/a_r$ as in
\eqref{hratio} in terms of the p-adic $\G$-function. 
From the relation \eqref{nfac} it is easy to see that, for $r\in\bb{Z}$, $r\geq 0$,
 $$
{a_{rp}\over a_r}~=~{\G_p(5rp+1)\over \G_p^5(rp+1)}~$$
and the right hand member is analytic in the p-adic sense and so tends to the limit
 $$
{\G_p\left({5mp\over 1-p}+1\right)\over \G_p^5\left({mp\over 1-p}+1\right)}
\eqlabel{limit}$$ 
as $r\to {m\over 1-p}$. Dwork,
writing under the pseudonym of Boyarski~
\Ref{\Boyarski}{M. Boyarski,``$p$-adic Gamma Functions and Dwork Cohomology'',\\
Transactions of the American Mathematical Society {\bf 257}, number 2 (1980).}, 
has given the following useful series for
$\G_p(1+pZ)$, $Z\in\bb{Z}_p$
 $$
\G_p(1+pZ)~=~ - \G_p(pZ) ~=~ - \sum_{n=0}^\infty c_{np}\,p^n\, (Z)_n \eqlabel{Boyarski}$$
where $(Z)_n$ is again the Pochhammer symbol, and the coefficients are
defined via the function
 $$
F(X)~=~\exp\left( X+{X^p\over p}\right)~=~\sum_{n=0}^\infty c_n X^n~. \eqlabel{Dworkfn}$$
By differentiating $F(X)$ it is easy to see that the $c_n$ satisfy the recurrence
 $$
nc_n~=~c_{n-1} +
c_{n-p}~~,~~~c_0~=~1~~,~~~c_n~=~0~~\hbox{for}~~n<0~.\eqlabel{Dworkcoeffs}$$ 
These
relations permit rapid numerical evaluation of the limit \eqref{limit}. We will return in
\SS\chapref{gauss} to discuss the remarkable properties of this function.
 
Now the remaining factor that we need to consider to calculate $\b_m$ is
 $$
A_m\left({mp\over 1-p}\right)~=~
{\G\left( 5m+{5mp\over 1-p}+1 \right) \G\left({mp\over 1-p}+1 \right)^5 \over
\G\left({5mp\over 1-p}+1 \right) \G\left(m + {mp\over 1-p}+1 \right)^5}~=~
{\prod_{j=1}^{5m}\left({5mp\over 1-p}+j \right) \over 
\prod_{j=1}^{m}\left({mp\over 1-p}+j \right)^5} \eqlabel{Acoeffs}$$
and the right member is seen to be a rational number. Thus
 $$
\b_m~=~{\G_p\left({5mp\over 1-p}+1\right)\over \G_p^5\left({mp\over 1-p}+1\right)}
{\prod_{j=1}^{5m}\left({5mp\over 1-p}+j \right) \over 
\prod_{j=1}^{m}\left({mp\over 1-p}+j \right)^5}~.
\eqlabel{numericalbeta}$$

The series \eqref{padicN} for $\n(\ps)$ has two constant terms corresponding to
$m=0$ and \hbox{$m=p{-}1$}. For these values of $m$ we have
 $$
\b_0~=~1~~~,~~~\b_{p-1}~=~p^4 \eqlabel{ends}$$
while for $1\leq m\leq p-2$ we may write
 $$
\b_m~=~(-p)^{\left[{5m\over p-1}\right]}\,
{\G_p\left( 1-\langle{5m\over p-1}\rangle\right)\over
\G_p^5\left( 1-\langle{m\over p-1}\rangle\right)}\eqlabel{shortbeta}$$
where $\langle..\rangle$ denotes the fractional part.
This last expression may be derived from \eqref{gammadef}. To see this take $p>5$ and
$0 < t< 5(p-1)$ an integer such that $p-1\notdiv t$. Set therefore $t=r(p-1)+\m$ with
$1\leq\m\leq p-2$. Then, by separating the terms in the product
 $$
\prod_{j=1}^t\left(j+{pt\over 1-p}\right)$$
into those that are divisible by $p$ and those that are not, we see that
 $$\eqalign{
\G_p\left(1+{pt\over 1-p}\right)\,\prod_{j=1}^t\left(j+{pt\over 1-p}\right)
&=(-1)^t\,
p^{\left[t\over p \right]}\,
\G_p\left(1+{t\over 1-p}\right)\,\prod_{j=1}^{\left[{t\over p}\right]}
\left(j+{t\over 1-p}\right)\cropen{5pt} 
&=(-1)^{\m+r}\, p^{\left[t\over p\right]}\,\G_p\left(1+{\m\over 1-p}\right)
{\prod_{j=1}^{\left[{t\over p}\right]} \left(j-r+{\m\over 1-p}\right)\over
{\prod_{j=1}^r}'\left(j-r+{\m\over 1-p}\right)}
~.\cr}
$$
Now 
 $$
\left[{t\over p}\right]~=~\left[r+{\m-r\over p}\right]~=~
\cases{r & if $\m\geq r$,\cropen{5pt}
      r-1 & if $\m< r$,\cr}$$
and we also have
 $$
{\prod_{j=1}^{\left[{t\over p}\right]} \left(j-r+{\m\over 1-p}\right)\over
{\prod_{j=1}^r}'\left(j-r+{\m\over 1-p}\right)}~=~
\cases{1 & if $\m\geq r$,\cropen{5pt}
       p & if $\m< r$,\cr}$$
since, when $\m<r$, the omission of terms divisible by $p$ causes the term with $j=r-\m$
to be omitted from the denominator. In this way we see that
 $$
\G_p\left(1+{pt\over 1-p}\right)\,\prod_{j=1}^t\left(j+{pt\over 1-p}\right)~=~
(-1)^t\,(-p)^{\left[{t\over p-1}\right]}\, 
\G_p\left(1-\left\langle{t\over p-1}\right\rangle\right)
\eqlabel{forcalc}$$  
and from this \eqref{shortbeta} follows immediately. As a matter of practical computation
it is useful to note that \eq\eqref{shortbeta} shows that it is necessary only to
evaluate $\G_p\left(1-{t\over p-1}\right)$ for \hbox{$0\leq t\leq p-2$} and for such
$t$'s \eq\eqref{forcalc}, read from right to left, together with Boyarski's series
\eqref{Boyarski} provide an efficient means of computation. 

We note from \eqref{shortbeta} the interesting relation
 $$
\b_{p-1-m}~=~(-p)^{4-\left[{5m\over p-1}\right]}\,
{\G_p\left(\left\langle{5m\over p-1}\right\rangle\right)\over 
\G_p^5\left(\left\langle{m\over p-1}\right\rangle\right)}~=~
{p^4\over \b_m}~.$$
The first equality is true for $0\leq m\leq p-2$  but owing to \eqref{ends} the
relation $\b_{p-1-m}=p^4/\b_m$ is true also for $m = p-1$. In passing to the second
equality in the equation above we have used the reflection formula \eqref{gammaformula}.

It is easy to see also that $\b_{p-1\over 2}=p^2$ for $p\geq 3$
so we can write $\n(\ps)$ in the form
 $$\eqalign{
\n(\ps)&=1+p^4 +\sum_{m=1}^{p-2}\b_m\teich^m(\l)\cr
&=1 + p^4 + p^2\teich^{p-1\over 2}(\l)+
\sum_{m=1}^{p-3\over 2}\left(\b_m\teich^m(\l) +
{p^4\over\b_m\teich^m(\l)}\right)~.\cr}
\eqlabel{finalN}$$
Curiously, as a consequence of these expressions, there are very simple
expressions for the mean and variance of $\n(\ps)$ as $\ps$ ranges over $\Fpstar$
 $$
{1\over p-1}\sum_{\psi\in \sevenFpstar}\n(\ps)~=~p^4 + 1$$
and
 $$
{1\over p-1}\sum_{\psi\in \sevenFpstar}\n^2(\ps) - 
\left({1\over p-1}\sum_{\psi\in \sevenFpstar}\n(\ps)\right)^2 ~=~p^4(p-4)~.$$ 
\subsection{The number of rational points over $\bb{F}_{p^s}^5$ when $5\notdiv p^s-1$}
Given \eqref{padicN} and \eqref{betam} it is natural to conjecture that
the number of rational points for $x\in\bb{F}_{p^s}^5$ will be given by replacing $p$ by
$q=p^s$ in these expressions. That is
 $$
\n_s(\ps)~=~1 + q^4 + \sum_{m=1}^{q-2}\b_{s,m}\,\teich^m(\l) \eqlabel{nus}$$
with 
 $$
\b_{s,m}~=~\lim_{n\to\infty} 
{a_{m(1+q+q^2+\ldots+q^{n+1})}\over a_{m(1+q+q^2+\ldots+q^{n})}}~. $$
In order to evaluate the p-adic limit we proceed as before. Now we meet the ratio
 $$
{a_{rq}\over a_r}~=~{a_{rp^s}\over a_{rp^{s-1}}}\,{a_{rp^{s-1}}\over a_{rp^{s-2}}}\ldots
{a_{rp}\over a_{r}}~=~
\prod_{\ell=1}^s {\G_p(5rp^\ell + 1)\over \G_p^5(rp^\ell + 1)} $$
with $r=m/(1-q)$. The quantities $A_m\left(mq/(1-q)\right)$ have the same form as in
\eqref{Acoeffs} but with $p$ replaced by $q$.
Thus we have
 $$
\b_{s,m}~=~\prod_{\ell=1}^s {\G_p\left({5mp^\ell\over 1-q} + 1\right)\over 
\G_p^5\left({mp^\ell\over 1-q} + 1\right)}\,
{\prod_{j=1}^{5m}\left({5mq\over 1-q}+j \right) \over 
\prod_{j=1}^{m}\left({mq\over 1-q}+j \right)^5} \eqlabel{betas}$$
One may check this result numerically for low values of $p$ and $s$. The computation is
easy however the `experimental data' is difficult to obtain since one is dealing
with spaces that have $p^{5s}$ points that defeat the computer already for low values of
$p$ and $s$.
\newpage
\section{zerodim}{\cy\ Manifolds of Zero Dimension}
\vskip-20pt
\subsection{The rational points of a quadric}
In this section we will repeat our analysis for a \cym\ of dimension zero.
We shall understand by this the points that  solve the quadratic equation
$$P(x) = x_1^2 + x_2^2 - 2\psi x_1 x_2$$
in $\Fp\bb{P}^1$.  Note also that the three monomials correspond to the points
of the unique
reflexive polyhedron in one dimension
\vskip-35pt
$$\epsfxsize=2truein\epsfbox{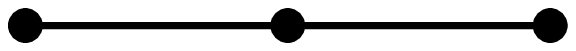}$$
\vskip-30pt
Over $\bb{P}^1$ the situation is, of course, that for $\psi^2\ne 1$ there are two
solutions and these coincide precisely when $\psi^2=1$.  Over $\Fp\bb{P}^1$ the situation
is more involved since the solutions that exist over $\bb{P}^1$ need not be rational.  So
the number of solutions over $\Fp\bb{P}^1$ may be 0, 1, or~2, depending on $p$ and $\psi$.

Let us first recall some terminology. For $a\in\Fp^*$ we say that $a$ is a
{\it square residue} in $\Fp$ if it has a square root,  that is  $a=y^2$
for some $y\in\Fp$. The number $a$ is a {\it square non-residue} otherwise. For
example, in $\bb{F}_7$, the numbers $1=(\pm 1)^2\, ,\ 2=(\pm 3)^2\, ,\ {\rm and}\  4=(\pm
2)^2 $ are square residues, whereas 3, 5 and 6 are square non-residues.

The {\it Legendre symbol} $\left({a\over p}\right)$ is defined as
 $$
\left({a\over p}\right) = \cases{
~\- 1\, ,\ \hbox{if $a$ is a square residue}\ ,\cropen{10pt}
~- 1\, ,\ \hbox{if $a$ is a non-square residue}\ ,\cropen{10pt}
~\- 0\, ,\ \hbox{if}\ p|a\ .\cr}
$$
Let now $M_{\psi}$ be a \cym\ of zero dimensions as defined above.  We denote by $N(\ps)$
the number of solutions of the equation $P=0$ over $\Fp\bb{P}^1$. It is easy to show, by
considering the formula for the solution of a quadratic equation, that this number is
given~by
$$ 
N(\psi) = 1 + \left( {{\psi^2 - 1}\over p}\right) =
\cases{~~1,\ {\rm if}\ \psi^2 - 1\equiv 0\ {\rm mod}\ p\cropen{10pt}
~~2,\ {\rm if}\ \psi^2 - 1\ \hbox{is a non-zero square residue of}\ p\cropen{10pt}
~~0,\ {\rm if}\ \psi^2 - 1\ \hbox{is a non-square residue of}\ p~.\cr}
\eqlabel{reclaw}$$
Note that this means that $N(\psi)$ depends only on $\psi^2 -1$.  For $p=7$, for example,
\eqref{reclaw}~yields
 $$\eqalign{ 
N(\pm 3) &= 2 \quad\hbox{since}\quad (\pm 3)^2-1=1,\quad\hbox{a square residue,}\cr
N(\pm 2) &= 0 \quad\hbox{since}\quad (\pm 2)^2-1=3,\quad\hbox{a non-square residue,}\cr
N(\pm 1) &= 1 \quad\hbox{since}\quad (\pm 1)^2-1=0,\cr
N(0) &= 0\quad\hbox{since}~~~~~\quad 0^2-1=6,\quad\hbox{a non-square residue.}\cr
}$$ 

One of the objects of this section is to obtain the simple result
\eqref{reclaw} using the methods of the previous sections, \ie in terms of the `periods'
and `semiperiods' of $M_\ps$.  Because of the simplicity of this example,
we will be able to see explicitly how the arithmetic structure is finer than the
complex structure.
\subsection{Solution in terms of periods}
Consider first the periods for $M_{\psi}$ over the complex numbers and set 
$\l=(2\psi)^{-2}$.   The Picard--Fuchs equation may be derived by the methods of
\SS\chapref{allperiods}. We find that
$$
\ca{L}\,\vp_0~=~0~~~\hbox{with}~~~
\ca{L}~=~\vth - 2\l\, (2\vth+1)\quad , $$
is a first order equation that can be solved in closed form.  Its solution,
corresponding to the fundamental period, is
$$\vp_0(\l) = f_0(\l) = (1 - 4\l)^{-\half} = \sum_{n=0}^{\infty} {(2n)!\over n!^2}
\l^n\quad .\eqlabel{period}$$
The semiperiods $\vp_j\, ,\ j>1$ are solutions of the equations
$$\vth^j\ca{L} \vp_j = 0\quad.$$
and in particular 
 $$
\vp_1~=~f_0(\l)\log\l + f_1(\l)~~~\hbox{with}~~~
f_1 (\l) = 2\ \sum_{n=1}^{\infty} {(2n)!\over n!^2} (\sigma_{2n} -\sigma_n)\ \l^n $$

Over the complex numbers, the equation $P(x) = 0$, has two different solutions
except when $\psi^2=1$, in which case there is only
one solution $x_1 = \psi x_2$.  If $\ps\neq 1$ it may be eliminated from $P$ by a
coordinate transformation.  The case $\ps^2=1$ is the analog of the conifold singularity
for the quintic in the sense that it corresponds to a point for which both $P = 0$ and
${\rm d}P = 0$.

To compute $N(\psi)$ we recall the relation $N(\psi) = {{\nu(\psi) - 1}\over {p-1}}$
with 
 $$ 
\nu(\psi) ~=~ \sum_{\sevenFp^2}\left(1 - P(x)^{p(p-1)}\right) + \ord{p^2}\quad.
\eqlabel{prenpsi}$$
These formulas give $\n(\psi) + \ord{p^2}$ which is
actually the exact result since $\nu(\psi) < p^2$.  We set $k={(p-1)\over 2}$ and after
some straightforward algebra we find:
 $$\eqalign{
{\rm For}\ \psi=0\ :\quad
\nu(0) &~=~ (-1)^{k+1} + p \left( 1 + (-1)^k\right)
 = \cases{ 2p-1,\ &if $p\equiv 1\ {\rm mod}\, 4$\cropen{10pt}
              1,  &if $p\equiv 3\ {\rm mod}\, 4$\cr}\cropen{20pt}
{\rm For}\ \psi\ne 0\ :\quad
\nu(\psi) &~=~ 
-\ {}^{(p-1)}f_0(\l^p) - p\ {}^{(p-1)}f_1'(\l^p) + 2p + \ord{p^2}\cr} 
\eqlabel{npsi}$$
and therefore
 $$\eqalign{
{\rm For}\ \psi=0\ :\quad N(0) ~&=~ 1 + (-1)^k = 
\cases{
2,\ \ {\rm if}\ p\equiv 1\ {\rm mod}\, 4\cropen{10pt}
0, \ \ {\rm if}\ p\equiv 3\ {\rm mod}\, 4\cr}\cropen{20pt}
\hskip4pt{\rm For}\ \psi\ne 0\ :\quad
N(\psi) ~&=~ (1 + p) \left(1 +  {}^{(p-1)}f_0(\l^p)\right) +
p\ {}^{(p-1)}f_1'(\l^p) - 2p +\ord{p^2}.\cr}\eqlabel{mpsi}$$
As before, ${}^{(p-1)}f_j(\l^p)$ is the truncation of the series
$f_j$ to $p$ terms.
Note the difference in signs in \eqref{npsi} as compared with the similar result for the
quintic three-fold. This difference comes from the factor of $(p-1)^2=1+\ord{p}$, as
opposed to a factor of $(p-1)^5=-1+\ord{p}$ for the quintic threefold, obtained after
summing \eqref{prenpsi} over
$\Fp^2$.  Comparing with \eqref{reclaw},  we obtain interesting identities that relate the
periods to the Legendre symbol. For $\psi = 0$, we have the well known result
 $$ 
~~\left({{p-1}\over p}\right) = (-1)^{\half (p-1)}\ ,$$
and for $\psi\ne 0$ we have
 $$ 
\left({{\psi^2-1}\over p}\right) \equiv \ {}^{\half(p-1)}f_0(\l)
\rlap{\quad{\rm mod}\ p\ .}$$

We would like to be able to write down exact p-adic expressions for our results for
$\psi\ne 0$. To this end let
 $$
\beta(\l)~=~\sum_{m=0}^{p-1} \beta_m {\rm Teich}^m(\l)\ ,\qquad\hbox{with}\qquad
\b_m~=~{\G_p\left({2mp\over 1-p}+1\right)\over \G_p^2\left({mp\over 1-p}+1\right)}
{\prod_{j=1}^{2m}\left({2mp\over 1-p}+j \right) \over
\prod_{j=1}^{m}\left({mp\over 1-p}+j \right)^2}\ .$$
By means of Gauss sums, one obtains the exact p-adic expression we are looking for
 $$ 
N(\psi) + {{\beta(\l) - 1}\over {p-1}} = 2\,\qquad\hbox{or equivalently,}\qquad
\nu(\psi) = 2p -\beta(\l)\ .$$
It is also easy to check these relations numerically.

Comparing with our results for the quintic 3-fold, we note again the
difference in the sign in front of $\beta(\l)$ and  the extra term $2p$.  The latter is
the analog of the extra contributions in $\nu(\psi)$ whenever $5|(p-1)$; for the zero
dimensional \cym, this term is always present because $2|(p-1)$ for all odd primes.

Comparing with \eqref{reclaw}, we have the following exact relation:
 $$
\left({{\psi^2-1}\over p}\right) = {{p - \beta(\l)}\over {p-1}}\ ,$$
which implies very intriguing relations between the period and the
semiperiods:
$$\vbox{\def\skip{\hskip10pt}
\offinterlineskip\halign{
\strut#\vrule height 12pt depth 8pt&\skip $#$\skip\hfil\vrule
&\hfil\skip $#$&$#$\skip\hfil\vrule\cr
\noalign{\hrule}
&\hfil\hbox{Condition on $\ps$} &\multispan{2}\hfil\hbox{Implication}\hfil\vrule\cr
\noalign{\hrule\vskip3pt\hrule}
&\ps^2=1 &\b(1/4)& ~=~p\cr
&\psi^2 - 1\ \hbox{is a square residue}& \beta(\l)& ~=~1\cr
&\psi^2 - 1\ \hbox{is a non-square residue}& \beta(\l)& ~=~ 2p-1\cr
\noalign{\hrule}
}}$$
These relations imply, in turn, congruence relations for the periods and semiperiods at
each order in a $p$ expansion. Examples of these congruences for the truncation of the
fundamental period are:
$$\eqalign{
\sum_{n=0}^{\half (p-1)} {(2n)!\over n!^2}\, 2^{-2n} &\equiv 0\quad {\rm mod}\ p\ ,
\ {\rm for}\ \psi^2 = 1 \ ,\cropen{10pt}
\sum_{n=0}^{\half (p-1)} {(2n)!\over n!^2}\, \l^{-n} &\equiv
\cases{\- 1\quad {\rm mod}\ p\ ,\ {\rm if}\ \psi^2 -1\ \hbox{is a square residue}\ ,
\cropen{10pt}
- 1\quad {\rm mod}\ p\ ,\ {\rm if}\ \psi^2 -1 \ \hbox{is a non-square residue}\ .\cr}
}$$
\newpage
\section{cubics}{CY Manifolds of Dimension One: Elliptic Curves}
In this section we consider the case of nonsingular cubics
$$
A_\psi: \qquad x^3+y^3+z^3-3\psi\, xyz=0\eqlabel{Acubic}
$$
for which $\ps^3\neq 1$.
In order to count the number of points of $A_\psi$ over a finite field
we will work instead with its mirror (we will be interested later in comparing the number
of rational points of the quintic with its mirror), which, in affine coordinates, can be
given~by
 $$
B_\l: \qquad  xy(1-x-y)=\l, \qquad \lambda=1/(3\psi)^3\;.
$$
The monomial change of variables
$$
x\mapsto x^2/y, \qquad y \mapsto y^2/x
$$
on the torus $x\neq 0, y\neq 0, z=1$ gives rise to a  degree three map
between these families. It is not hard to check that $B_\l$ is the quotient of $A_\psi$ by
the automorphisms 
$(x,y)\mapsto (\zeta x, \zeta^{-1} y)$, where $\zeta^3=1$.

We may obtain Weierstrass models for our families, for
example, using the algorithm in~
\Ref{\RT}{F. Rodriguez-Villegas and J. Tate, ``On the Jacobian of plane cubics'',\\
  http://www.ma.utexas.edu/users/villegas/cnt/jac\_cubic.gp~.}. 
We find
$$
\eqalign{
A_\psi:&\qquad y^2-3\psi xy+9y=x^3-27(\psi^3+1)\cr
B_\lambda:&\qquad y^2+xy+\lambda=x^3~.}
$$
An elliptic curve over $\bb{C}$ has a natural group structure and this structure is
manifest also when we consider the elliptic curve over a finite field~
\Ref{\SilvermanTate}{``Rational Points on Elliptic Curves'', by J. H. Silverman and J.
Tate,\\ Undergraduate Texts in Mathematics, Springer-Verlag 1992.}.
With respect to this group structure we verify that our previous map is a degree $3$
isogeny (an isogeny is a map that preserves the group structure of the elliptic curve);
more precisely,
$B_\lambda$ is isomorphic to the quotient of
$A_\psi$ by the subgroup of order $3$ generated by
\hbox{$\big(3(\psi+1),9\psi(\psi+1)\big)$}.

It is known that isogenous elliptic curves have the same number of
points over a given finite field and we will hence restrict ourselves
to counting points on $B_\lambda$. It is not difficult to check that
for $\lambda\neq 0$ there are exactly three points of $B_\lambda$ not
included among those satisfying
 $$
xy(1-x-y)=\lambda,\qquad x\neq0,\qquad y\neq0~.
$$
(We may note, for example, that each  side of the Newton polygon of
this equation  has integral length 1, corresponding to a single
rational point on the appropriate non-singular toric surface.)

For $\lambda \in \Fqstar$ let $N^*(\lambda)$ be the number of projective
solutions of \eqref{Acubic} with $x,y \in \Fqstar$. By the above discussion
the number of points $N(\lambda)$ of $B_\lambda$ over $\Fq$, and also
of $A_\psi$ if there is a $\psi \in \Fq$ with $\lambda=1/(3\psi)^3$,
is $N^*(\lambda) + 3$.

For any $x\in\Fqstar$ we have 
 $$
\sum_\chi \chi(x) =\cases{q-1 & $x=1$\cropen{3pt}
                             0& otherwise~,\cr}
$$
where the sum is over all multiplicative characters of $\Fqstar$.
Therefore,  for $\lambda \in \Fqstar$
 $$
N^*(\lambda)=
\frac1{q-1} \sum_\chi \chi(\lambda^{-1}) 
\hskip-10pt\sum_{\scriptstyle x_i\in\sevenFqstar \atop x_1+x_2+x_3=1}\hskip-10pt
\chi(x_1x_2x_3)~.
$$
Let us recall the definition of the {\it Jacobi sum} associated to
characters $\chi_1,\ldots, \chi_d$ of $\Fqstar$
 $$
J(\chi_1,\ldots,\chi_d)~=
\hskip-10pt\sum_{\scriptstyle x_i\in \sevenFq \atop x_1+\cdots+x_d=1}\hskip-10pt
\chi_1(x_1)\cdots \chi_d(x_d)~,
$$
where we follow the standard convention
 $$
\chi(0)=
\cases{1&if $\chi={\bf 1}$\cropen{3pt} 
       0& otherwise~,\cr}
$$
where, in this context, ${\bf 1}$ denotes the trivial character. For standard facts
relating to Jacobi sums see, for example,~
\Ref{\IrelandRosen}{``A Classical Introduction to Modern Number Theory'', by K. Ireland
and M. Rosen,\\ Graduate Texts in Mathematics, Number 84, Springer-Verlag 1972, 1982}.

The inner sum in (1.2) is almost the Jacobi sum $J(\chi,\chi,\chi)$
except for the terms where one of the $x_i$'s is zero. Such a term
contributes to the Jacobi sum only if $\chi$ is trivial, in which case
it contributes a $1$. By the principle of inclusion/exclusion the
neglected terms amount to $3q-3$ ($q$ for each of the three cases
where one $x_i$ is zero, $1$ for each of the three cases where two of
them are zero). Hence,
$$
N^*(\lambda)= \frac1{q-1} \sum_\chi \chi(\lambda^{-1})
J(\chi,\chi,\chi)-3\,,
$$
and 
$$
N(\lambda)= \frac1{q-1} \sum_\chi \chi(\lambda^{-1})
J(\chi,\chi,\chi)\,.
\eqlabel{}
$$
Using the Gross-Koblitz formula (which we will come to in the following section) we may
write $J(\chi,\chi,\chi)$ in terms of $\Gamma_p$ and is not hard to verify then that we
recover the analogue of our previous formula for $N(\lambda)$.
\newpage
\section{gauss}{The Relation to Gauss Sums}
\vskip-20pt
\subsection{Dwork's character}
\REF{\DGS}{``An Introduction to G-Functions'', by B. Dwork, G. Gerotto and F. J. Sullivan,
Annals of Mathematics Studies {\bf 133}, Princeton University Press 1994.} 
We record in this section expressions for the $\n(\ps)$ in terms of Gauss sums. We begin 
by reviewing the properties of Dwork's character $\Th:\,\Fp\rightarrow \bb{C}_p$; a full
account may be found in
\cite{\DGS,\Koblitz}. The character is defined by
 $$
\Th(x)~=~F(\p\,\teich(x))$$
with $F$ the function given in \eqref{Dworkfn} and $\p$ a number in $\bb{C}_p$ such that
$\p^{p-1}=-p$. 

The introduction of $\p$ merits some comment. Since the p-adic order of
$\p$ is $1/(p-1)$, which is not an integer, $\p$ is not a p-adic number but lies in an
extension of $\bb{Q}_p$.
$\p$ arises in connection with finding nontrivial $q$'th roots of unity which is clearly
closely allied to the question of finding multiplicative characters. It can be shown that
all the roots of the equation $X^q-1=0$ lie in $\bb{Q}_p(\p)$, the field of elements of
the form
$\sum_{r=0}^{p-2}a_r\,\p^r~,~a_j\in\bb{Q}_p$. A~second fact is that the series $\exp(X)$
is less convergent in the p-adic case than in classical analysis, owing to the presence of
the factorials in the denominators. It is straightforward to see that it converges in
the disk $|X|_p < |\p|_p$. It is perhaps natural therefore to consider the series
$\exp(\p Y)$ which converges in the disk $|Y|_p < 1$.   

Now
 $$
F(\p X)~=~\exp\bigl(\p(X-X^p)\bigr) ~=~\sum_{n=0}^\infty c_n\,(\p X)^n \eqlabel{series}$$
and it is an essential fact that the series on the right converges for $|X|_p < 1+\e$ for
a fixed positive number $\e$. 

There is a trap
here for the unwary which is to assume that since a Teichm\"uller representative satisfies
the equation $X^p=X$ that $\Th(x)=\exp(0)=1$. This turns out to be false owing to the
fact that the $X$-disk that ensures that $|X-X^p|_p <1$ is $|X|_p <1$. There is thus an
issue of analytic continuation as discussed in \SS\chapref{prelims}.2. We are to regard
$F(\p X)$ and hence
$\Th$ as defined by the series on the right of~\eqref{series}. $F(\p X)$ then equals
$\exp\bigl(\p(X-X^p)\bigr)$ for $|X|_p < 1$ but not for $|X|_p \geq 1$. 
From the series expansion we have
 $$
\Th(1)~=~1 + \p + \ord{\p^2}$$
from which it follows that $\Th(1)\neq 1$. On the other hand if $X^p=X$ then
 $$
F(\p X)^p~=~\exp(p\p X - p\p X^p)~=~1$$
since the presence of the $p$ in the exponent ensures the convergence of the series.
In particular $\Th(1)$ is a $p$'th root of unity and since 
$\Th(x)=1+\p\,\teich(x) +\ord{\p^2}$ 
we see also that $\Th(x)=\Th(1)^{\teich(x)}=\Th(1)^x$. 

Since for $x,y\in\Fp$
 $$
\teich(x+y) ~=~ \teich(x) + \teich(y) + pZ $$
for some $Z\in \bb{Z}_p$ it follows that $\Th$ is a nontrivial additive character, that is
 $$
\Th(x+y)~=~\Th(x)\,\Th(y)~.$$

Dwork also adapted this construction to give a additive character
$\Th_s:\,\bb{F}_q\rightarrow \bb{C}_p$ for~$q=p^s$
 $$
\Th_s(x)~=~\Th(\tr(x))$$
where $\tr:\,\bb{F}_q\rightarrow\Fp$ is the trace map
 $$
\tr(x)~=~x + x^p + x^{p^2} + \cdots + x^{p^{s-1}} $$
Now the limit $\teich(x)=\lim_{n\to\infty}x^{q^n}$ exists also for $x\in\bb{F}_q$ though
$\teich(x)$, for $x\neq 0$, is now a unit (has unit norm) of $\bb{C}_p$ but is not in
general in $\bb{Z}_p$. 
 We have
 $$ \teich\left(\tr(x)\right)~=~\sum_{\ell=0}^{s-1}\teich^{p^\ell}\!(x) + pZ $$
with $Z$ an integer of $\bb{C}_p$.
It follows that 
 $$
\Th_s(x)~=~\Th(1)^{\sum_{\ell=0}^{s-1}\teich^{p^\ell}\!(x)}~.$$
Note however that we cannot, in general, write the right hand side of this relation as
the product
 $$
\prod_{\ell=0}^{s-1} \Th(1)^{\teich^{p^\ell}\!(x)}$$
since the $\teich^{p^\ell}(x)$ are not in general in $\bb{Z}_p$. Dwork showed nevertheless
that $\Th_s$ has the remarkable splitting property
 $$
\Th_s(x)~=~\prod_{\ell=0}^{s-1} \Th\!\left(x^{p^\ell}\right)~.$$
\subsection{Evaluation of $\n(\ps)$ in terms of Gauss sums}
It is useful to define `Gauss sums'
 $$ 
g_n(y)~=~\sum_{x\in\sevenFpstar} \Th(yx^5)\, \teich^n(x)~~~\hbox{and}~~~
G_n~=~\sum_{x\in\sevenFpstar}\Th(x)\,\teich^n(x)~.\eqlabel{gausssums}$$
$G_n$ is a genuine Gauss sum, that is a sum of an additive character, $\Th$, times a
multiplicative character, $\teich^n(x)$. Note however that our convention is nonstandard
with respect to~$G_0$. The standard convention for the Gauss sum for the case that the
multiplicative character is the identity takes the sum to be over $\Fp$ instead of $\Fp^*$
with the result that the Gauss sum is zero in this case. We find it convenient here to
define $G_0$ as above with the result that $G_0=\sum_{x\in\sevenFpstar}\Th(x)=-1$. 

Gauss sums are the analogues for finite fields of
the classical $\G$-function
 $$
\G(s)~=~\int_{0}^\infty {dt\over t}\,t^s e^{-t} $$
which is an integral, with multiplicative Haar measure $dt/t$, of a product of the
additive character $e^{-t}$ and the multiplicative character $t^s$. In our case it is a
consequence of standard properties of Gauss sums~
\REF{\Lang}{``Cyclotomic Fields, vol II'', by S. Lang,\\ Graduate Texts in Mathematics,
Number 69, Springer-Verlag 1980}
\cite{\IrelandRosen,\Lang}
that
 $$
G_n\,G_{-n}~=~(-1)^n\, p ~~~\hbox{for}~~~p-1\notdiv n\eqlabel{inversion}$$
which is an analogue for finite fields of the reflection formulae \eqref{classicalref} and
\eqref{gammaformula}. An explicit expression for $G_n$ for the the case $p-1\notdiv n$ is
provided by the Gross-Koblitz formula (see, for example, \cite{\Lang};
the original reference is~
\Ref{\GrossKoblitz}{B. H. Gross and N. Koblitz, 
``Gauss Sums and p-adic $\G$-function'',\\
Annals of Math. {\bf 109} 569 (1979).})
 $$
G_n~=~p\, (-p)^{-\left\langle{n\over p-1}\right\rangle}\, 
\G_p\left(1-\left\langle{n\over p-1}\right\rangle\right)~;~~ p-1\notdiv n~.
\eqlabel{GKformula}$$ 

We will make repeated use in the following of an identity. Let $t$ be an integer and let
$t'$ be the reduction of $t~\hbox{mod}~p-1$ to the range $0\leq t'\leq p-2$ then by
considering the definition
\eqref{gausssums} for
$G_{-t}$ and expanding $\Th(x)$ as a series we find
 $$
 (p-1)\sum_{r=0}^\infty c_{r(p-1)+t'}\p^{t'} (-p)^r~=~G_{-t} $$
and by returning with this identity to the series expansion of $\Th$ we see that
 $$
\Th(x)~=~{1\over p-1}\,\sum_{m=0}^{p-2} G_{-m}\,\teich^m(x)~.$$
Now we write
 $$\eqalign{
\Th(yP(x))~&=~
\left(\prod_{i=1}^5 \Th(yx_i^5)\right)\Th\left(-5\ps y
\prod_{i=1}^5 x_i\right)\cropen{5pt}   
&=~{1\over p-1}\,\sum_{m=0}^{p-2} G_{-m}\,\teich^m(-5\ps y) \prod_{i=1}^5
\Th(yx_i^5)\,\teich^m(x_i)~.\cr}\eqlabel{yPeqn}$$   
Since $\Th$ is a character we have
 $$
\sum_{y\in \sevenFp}\Th(yP(x))~=~\cases{p~,&~~~if $P(x)=0$\cropen{5pt}
       		                               0~,&~~~if $P(x)\ne 0$~.\cr}\eqlabel{charcases}$$
so if we sum this relation for $x\in \Fstar$ we have
 $$
p\n^*(\ps) - (p-1)^5 ~=~{1\over p-1}\,\sum_{m=0}^{p-2} G_{-m}\,\teich^m(-5\ps) 
\sum_{y\in\sevenFpstar}g_m^5(y) \,\teich^m(y)~. \eqlabel{intermediategauss}$$
Consider now the case that $5\notdiv (p-1)$ and choose $\ell$ such that $5\ell = 1$ 
mod $p-1$. We~have
 $$
\sum_{y\in\sevenFpstar}g_m^5(y) \,\teich^m(y)~=~
\sum_{y\in\sevenFpstar}\left(\sum_{x\in\sevenFpstar} \Th(yx^5)\,\teich^{lm}(yx^5)\right)^5
~=~(p-1)G_{lm}^5~.$$
If we use this identity in \eqref{intermediategauss}, choose $\ell m$ as the variable of
summation and also use expressions from \SS\chapref{calczero}.1 to relate $\n^*(\ps)$ to
$\n(\ps)$ we find
 $$
\n(\ps)~=~ 1 + p^4 + \sum_{m=1}^{p-2}{G_m^5\over G_{5m}}\,\teich^{-m}(\l)~.
\eqlabel{nuexpr}$$ 
Comparing this to our expression for $\n(\ps)$ in \SS\chapref{higher} we see that 
$\b_{p-1-m}={G_m^5\over G_{5m}}$ 
and  the identity that relates $\b_{p-1-m}$ to $\b_m$ is now a simple consequence
of\eqref{inversion}:
 $$
\b_{p-1-m}~=~{G_m^5\over G_{5m}}~=~p^4\,{G_{5m}\over G_m^5}~=~{p^4\over \b_m}~.$$
By using the Gross-Koblitz formula \eqref{GKformula} in \eqref{nuexpr} we recover (and
establish) \eqref{shortbeta}.
\subsection{The case $5|p-1$}
We return now to \eqref{intermediategauss} and note that when $5|(p-1)$ there are
nontrivial fifth roots of unity. If $\z$ is such a root then by replacing $x$ by $x\z$ in
the first of \eqref{gausssums} we see that $g_m(y)=0$ unless
$5|m$. Thus for $\ps\neq 0$ 
 $$\eqalign{
p\n^*(\ps)-(p-1)^5~&=~{1\over p-1}\sum_{m=0}^{k-1}(-1)^m G_{-5m} \teich^{-m}(\l)
\sum_{y\in \sevenFpstar} g_{5m}^5(y) \,\teich^{5m}(y)\cr
&=~{1\over p-1}\sum_{m=0}^{k-1}(-1)^m G_{5m} \teich^{m}(\l)
\sum_{y\in \sevenFpstar} g_{-5m}^5(y) \,\teich^{-5m}(y)\cr~.}$$
Now by again expanding the character as a series we have
 $$
 g_{-5m}(y)\, \teich^{-m}(y)~=~{1\over p-1}\sum_{n=0}^{p-2}G_{-n}
\sum_{x\in \sevenFpstar}\teich^{n-m}(yx^5)~=~\sum_{s=0}^4 G_{-(m+ks)}\,\teich^{ks}(y)
\eqlabel{yg}$$ 
In this way we find
  $$
p\n^*(\ps) - (p-1)^5~=~ 
\sum_{s_i=0\atop 5\left|{\scriptstyle\sum} s_i\right.}^4 \sum_{m=0}^{k-1}
(-1)^m \teich^m(\l)\, G_{5m}\prod_{j=1}^5 G_{-(m+ks_j)}~.\eqlabel{ssum}$$ 
In this expression it is natural to think of the vectors ${\bf s}=(s_1,\ldots,s_5)$ as
corresponding to monomials $x^{\bf s}$ of degree $5w$, $0\leq w\leq 4$ that are subject
to the constraint $s_i\leq 4$. Let us call the set of such monomials $\ca{S}$ and consider
the operation, $\vph:~\ca{S}\to\ca{S}$, of adding $\bf 1$ to
${\bf s}$ and then reducing the result mod 5. There is no ${\bf s}$ invariant
under $\vph$, and $\vph^5$ is the identity so the monomials fall into orbits of $\vph$ of
length 5. Consider now the contribution of such an orbit to the sum \eqref{ssum}. The
contribution of $\vph({\bf s})$ is obtained by making the replacement $m\to m+k$ in the
term corresponding to $\bf s$ so we may sum over the orbit by choosing a representative
and increasing the range of $m$ to $0\leq m\leq p-2$. Since two monomials which differ
merely by a permutation of their components contribute equally to the sum we may as well
identify such monomials. If we do this then it is no longer true that no monomial is
invariant since now $\vph(4,3,2,1,0)\simeq (4,3,2,1,0)$. It is simple to enumerate all the
orbits and assign representative monomials which we choose to be of lowest degree. This
process is unambiguous apart from the orbit of the monomial $(4,1,0,0,0)$ which contains
also $(2,1,1,1,0)$. In this case we choose $(4,1,0,0,0)$ to represent the orbit. As a
consequence, and as was to be anticipated from our discussion of \SS\chapref{calcone} the
monomial $(2,1,1,1,0)$, which is interior to a codimension one face of
$\D$, does not occur explicitly in our list. Which we present in Table~\tabref{vtabletwo}
below
\vskip5pt
$$\vbox{\def\skip{\hskip7pt}
\offinterlineskip\halign{
\strut #\vrule height 12pt depth 6pt&\hfil\skip #\skip\hfil\vrule
&\hfil\skip #\skip\hfil\vrule&\hfil\skip #\skip\hfil\vrule&\hfil\skip #\skip\hfil\vrule\cr
\noalign{\hrule}
&monomial &degree &$\g_{\bf v}$ &length of orbit\cr
\noalign{\hrule\vskip3pt\hrule}
&(0,0,0,0,0) &0 &1  & 5\cr
\noalign{\hrule}
&(4,1,0,0,0) &5 &20 & 5\cr
&(3,2,0,0,0) &5 &20 & 5\cr
&(3,1,1,0,0) &5 &30 & 5\cr
&(2,2,1,0,0) &5 &30 & 5\cr
\noalign{\hrule}
&(4,3,2,1,0) &10 &24 & 1\cr
\noalign{\hrule}
}}$$
\vskip3pt
\centerline{Table~\tablabel{vtabletwo}}
\vskip10pt
Incorporating these considerations brings us to the expression
 $$
p\n^*(\ps) - (p-1)^5~=~ 
\sum_{\bf v}\g_{\bf v} \sum_{m=0}^{p-2}
(-1)^m \teich^m(\l)\, G_{5m}\prod_{j=1}^5 G_{-(m+kv_j)}\eqlabel{lsum}$$
with coefficients $\g_{\bf v}$, given in the table, that account for the permutations of
the components of the monomials ${\bf v}$. 

The quantity $\n^*(\ps)$ differs from $\n(\ps)$ by the constant $\n_Y$ but this is given by
a more complicated expression than was the case for $5\notdiv p-1$. In order to calculate
$\n_Y$ as the difference $\n(0)-\n^*(0)$ we return to the first of equations \eqref{yPeqn}
from which it follows that
 $$
p\n(0) ~=~ p^5 + \sum_{y\in \sevenFpstar}\left( g_0(y)+1 \right)^5 \quad\hbox{and}\quad
p\n^*(0)~=~ (p-1)^5 + \sum_{y\in \sevenFpstar} g_0^5(y)~.$$ 
now by specialising \eqref{yg} to the case $m=0$ we have
 $$
g_0(y)~=~\sum_{s=0}^4 G_{-ks}\,y^{ks} \quad\hbox{and}\quad 
g_0(y) + 1~=~\sum_{s=1}^4 G_{-ks}\,y^{ks}~.$$
and from this we obtain the useful expression
 $$
p\big(\n(\ps)-\n^*(\ps)\big)~=~p^5 - (p-1)^5 -(p-1)\hskip-5pt 
\sum_{s_i\in\ca{S}\atop \prod_i s_i=0} \hskip-3pt\prod_{i=1}^5 G_{-ks_i}\eqlabel{nuY}$$ 
with the sum on the right being over vectors of $\ca{S}$ which have at least one
component that is zero. This expression will guide us in choosing how best to write the
coefficients that appear in the expansion
 $$
\n(\ps)~=~p^4 + \sum_{\bf v}\g_{\bf v} \sum_{m=0}^{p-2} \b_{{\bf v},m}\teich^m(\l)~.
\eqlabel{newnewnu}$$
This question is more involved for the present case that $5|p-1$ 
since there are now five constant terms on the right hand side, those with 
$m=ks,~0\leq s\leq 4$, and the question arises as to how best to distribute the constant
that we have to transfer to the right hand side when we pass from $\n^*(\ps)$ to $\n(\ps)$.
For $m\neq ks$ the coefficients are not affected by this constant and a little algebra,
making use of \eqref{forcalc} and the Gross-Koblitz formula yields {\sl for the case
$k\notdiv m$\/} the relations
 $$\eqalign{
\b_{{\bf v},m}~
&=~{(-1)^m\over p}\,G_{5m}\prod_{j=1}^5 G_{-(m+kv_j)}\cropen{7pt}
&=~ p^4\,{G_{5m}\over \prod_{i=1}^5 G_{m+kv_i}}\cropen{7pt}
&=~ (-p)^{w({\bf v})}\, {\goth_{5m}\over \prod_{i=1}^5 \goth_{m+kv_i}}\cropen{7pt}
&=~ (-p)^{w_m({\bf v})}\, {\G_p\left(1-\left\langle{5m\over p-1}\right\rangle\right) \over
\prod_{i=1}^5 \G_p\left(1-\left\langle{m+kv_i\over p-1}\right\rangle\right)}\cr} 
\eqlabel{coeffrels}$$ 
where 
$$
w_m({\bf v})~=~\sum_{i=1}^5\left\langle{\left[{m\over k}\right] +v_i\over 5}\right\rangle
~~\hbox{and}~~~w({\bf v})=w_0({\bf v})={1\over 5}\sum_{i=1}^5 v_i$$ 
are the weights of ${\bf v} + \left[m/k\right]{\bf 1}$ and of ${\bf v}$ and we have defined
 $$
\goth_t~=~\G_p\left(1 + {tp\over 1-p}\right) 
\prod_{j=1}^t \left(j + {tp\over 1-p}\right)~. $$
The point to be made by \eqref{coeffrels} is that these four expressions for 
$\b_{{\bf v},m}$ are all the same when $k\notdiv m$ but are all different when $k|m$. If
however we compare the desired expression \eqref{newnewnu} with \eqref{lsum} and
\eqref{nuY} we see that we can choose the $\b_{{\bf v},ak}$ such that
 $$
p\sum_{\bf v}\g_{\bf v}\sum_{a=0}^4 \b_{{\bf v},ak} ~=~ - \sum_{\bf v}\g_{\bf v}\left\{
(p-1)\hskip-8pt\sum_{a=0\atop z({\bf v} + a\bone) > 0}^4 \prod_{i=1}^5 G_{-(v_i+a)k} + 
\sum_{a=0}^4 \;\prod_{i=1}^5 G_{-(v_i+a)k} \right\}$$
where $z({\bf u})$ is the function that counts the number of components of a vector $\bf u$
that are zero mod 5. We therefore choose 
 $$
\b_{{\bf v}, ak} 
~=~ - {p^{5-z-\d(z)}\over \prod_{i=1}^5 G_{(v_i+a)k}} 
~=~ - {p^{-\d(z)}\,(-p)^{w_{ak}({\bf v})}\over 
\prod_{i=1}^5 \G_p\left( 1 - \left\langle{v_i+a\over 5}\right\rangle\right)}$$
with $z$ denoting $z({\bf v} + a\bone)$.

We recognise, in \eqref{newnewnu}, the terms corresponding to $(0,0,0,0,0)$ as
corresponding to the analog of our previous expression for the case $5\notdiv p-1$. The
terms corresponding to monomials of degree 5 yield the prolongations of the `extra terms'
that we found in
\eqref{newnu}. The degree 10 monomial is however new and we pause to examine its
contribution. For reasons that will become apparent we shall refer to this as the conifold
term and denote it by $\cnf(\ps)$. A little simplification quickly reveals that
 $$
\cnf(\ps)~=~120\, p^2 
 \sum_{m=0}^{k-1} \;{\prod_{i=0}^4 \G_p\left({{m\over k}+i\over 5}\right)\over
\G_p\left({m\over k}\right)}\;\teich^m(\l) ~.\eqlabel{precnf}$$ 
We recognise the combination of $\G_p$-functions that appears in the sum as corresponding
to the multiplication formula \eqref{premult} and we come to some bookwork~
\Ref{\Koblitztwo}{``p-adic Analysis: a Short Course on Current Work'', by N.~Koblitz,\\  
London Mathematical Society Lecture Notes Series, Number 46,\\ 
Cambridge University Press 1980.}\
that we cannot resist recalling. We need the multiplication formula for
$n=5$ and when in addition $5|p-1$ the product 
\hbox{$\prod_{i=1}^4 \G_p\left({i\over 5}\right)=1$} as we see by using
\eqref{gammaformula} and noting that $L(1/5)=p-k$ and $L(2/5)=p-2k$. Thus we have
 $$   
{\prod_{i=0}^4 \G_p\left({Z+i\over 5}\right) \over \G_p(Z)}~=~
5^{1-L(Z)}\left(5^{-(p-1)}\right)^{Z_1}~. \eqlabel{gammamult}$$
In this formula we take 
  $$
Z~=~{m\over k}~=~{5m\over p-1}$$
so we have $L(Z)=p-5m$ and 
 $$
Z_1~=~{1\over p}\left(Z - L(Z)\right)~=~{5m-p+1\over p-1}~=~{1-L(Z)\over p-1}~.$$
Let us denote the right hand side of the multiplication formula by $t_m$. As remarked
previously $5^{-(p-1)}=1+\ord{p}$ so $\left(5^{-(p-1)}\right)^{Z_1}$ is also $1+\ord{p}$.
Thus we see that
 $$
t_m~=~5^{5m} + \ord{p}~.$$
Now following from our observation concerning $Z_1$ we have
 $$
t_m = 5^{1-L(Z)}\left(5^{-(p-1)}\right)^{1-L(Z)\over p-1}$$
It is perhaps tempting to rearrange the exponents on the right and conclude, falsely, that
\hbox{$t_m=1$}. This is another trap for the unwary since the term
$\left(5^{-(p-1)}\right)^{Z_1}$ has to be understood as the binomial series of a quantity
that is $1+\ord{p}$. However if we consider $t_m^{p-1}$ we have
 $$
t_m^{p-1}~=~5^{(p-1)(1-L(Z))}\left(5^{-(p-1)}\right)^{1-L(Z)}~=~1$$
and the cancellation is now valid. 
Thus it has been shown that $t_m~=~5^{5m} + \ord{p}$ and that $t_m^{p-1}=1$. In other
words 
 $$
t_m~=~\teich(5^{5m})~.$$
If we now use this result to simplify \eqref{precnf} we find, in an obvious notation, that
 $$
\cnf(\ps)~=~24p^2(p-1)\,\d\!\left(\teich^5(\ps) - 1\right)~.$$
\subsection{Relation to the periods}
In this subsection we rewrite the expression for the number of points \eqref{newnewnu} in
terms of the periods of the holomorphic
$(3,0)$ form $\Omega$.  To do this we begin by defining
$$
\tilde\b_{{\bf v},m}~
=~ (-p)^{w({\bf v})}\, 
{{\G_p\left(1 + {5mp\over 1-p}\right) 
\prod_{j=1}^{5m} \left(j + {5mp\over 1-p}\right)}\over 
\prod_{i=1}^5 \G_p\left(1 + {(m+kv_i)p\over 1-p}\right) 
\prod_{j=1}^{m+kv_i} \left(j + {(m+kv_i)p\over 1-p}\right)}~~.\eqlabel{newcoefrels}$$
These coincide with $\b_{{\bf v},m}$ in \eqref{coeffrels} when $k\notdiv m$.  In terms of
$\tilde\b_{{\bf v},m}$ the number of points can be rewritten as
$$
\n(\ps)~=~p^4 + {\cal I} +
\sum_{\bf v}\g_{\bf v} \sum_{m=0}^{p-2} \tilde\b_{{\bf v},m}\teich^m(\l)~,
$$ 
where the term ${\cal I}$ is independent of $\lambda$
$$
{\cal I} ~=~ 30 (p-1) p^2 \left( 
{1\over{\G_p\left({1\over 5}\right)^2 \G_p\left({3\over 5}\right)}}
+ {1\over{\G_p\left({1\over 5}\right) \G_p\left({2\over 5}\right)^2}}\right)~.$$
We will discuss the interpretation of this constant in \cite{\CYII}.
By writing 
$$
{(m+kv_i)p\over 1-p} = {mp\over 1-p} - {v_i\over 5} - kv_i ~,$$
it is easy to see that $\tilde\b_{{\bf v},m}$ can be written as
$$
\tilde\b_{{\bf v},m}~
=~ (-p)^{w({\bf v})}\, 
{{\G_p\left(1 + {5mp\over 1-p}\right) 
\prod_{j=1}^{5m} \left(j + {5mp\over 1-p}\right)}\over 
\prod_{i=1}^5 \G_p\left(1 + {mp\over 1-p} - {v_i\over 5}\right) 
\prod_{j=1}^m \left(j + {mp\over 1-p} - {v_i\over 5}\right)}~~.$$
Recalling \eqref{gencoeffs}
 $$
A_{{\bf v},n}(\ve)~=~{\G(5n+5\ve+1)\over \G(5\ve+1)}
\prod_{i=1}^5{\G\left( \ve+1-{v_i\over 5}\right)\over 
\G\left( n+\ve+1-{v_i\over 5}\right) }~,$$
and writing $a_{{\bf v},m} = A_{{\bf v},m}(0)$, we have
$$
\tilde\b_{{\bf v},m}~
=~ - {(-p)^{w({\bf v})}\over \prod_{i=1}^5\G_p(1 - {v_i\over 5})}\,  
\lim_{l\to\infty} {a_{{\bf v},mp(1+p+\ldots +p^l)}\over
a_{{\bf v},m(1+p+\ldots +p^l)}}
\  A_{{\bf v},m}(mp(1+p+\ldots +p^l))~.$$
Therefore, by using the identity 
 $$
a_{{\bf v},rp+s} = a_{{\bf v},rp}\, A_{{\bf v},s}(rp)~,$$
we see that
 $$
\tilde\b_{{\bf v},m}~
=~ - {(-p)^{w({\bf v})}\over \prod_{i=1}^5\G_p(1 - {v_i\over 5})}\,  
\lim_{l\to\infty} {a_{{\bf v},m(1+p+\ldots +p^{l+1})}\over
a_{{\bf v},m(1+p+\ldots +p^l)}}~.$$
This expression for the coefficients $\tilde\b_{{\bf v},m}$ generalizes the one we had
obtained for the case ${\bf v}=(0,0,0,0,0)$ (see \eqref{betam}).

We now follow the method of Frobenius as in Section (6.2).  It will be convenient to
define, in analogy with \eqref{hratio}, the quantity
$$
h_{\bf v}(rp) ~=~ {a_{{\bf v},rp} \over a_{{\bf v},r}}
   ~=~ -\, \G_p(1 + 5rp)\, 
   \prod_{i=1}^5{\G_p(1 - {v_i\over 5})\over \G_p(1 - {v_i\over 5} +rp)}~,$$
and to rewrite $\tilde\beta_{{\bf v},m}$ in terms of these
$$   
\tilde\b_{{\bf v},m}~
=~ - {(-p)^{w({\bf v})}\over \prod_{i=1}^5\G_p(1 - {v_i\over 5})}\  
h_{\bf v}\left({mp\over{1-p}}\right)\, A_{{\bf v},m}\left({mp\over{1-p}}\right)~~.$$
For each ${\bf v}$, a basis of periods and semiperiods is obtained from
$$ F_{\bf v}(\lambda,\epsilon) 
~=~ \sum_{m=0}^{\infty} A_{{\bf v},m}(\epsilon) \,\lambda^{m+\epsilon}~,$$
as
$$
\varpi_{{\bf v}, k} (\lambda) ~=~ 
\left. {d^{(k)}F_{\bf v}(\lambda,\epsilon)\over d\epsilon^k}\right|_{\epsilon = 0}
~=~ \sum_{i=0}^k {k\choose i} f_{{\bf v},i}(\l) \log^{k-i}\l$$
where the
$$
f_{{\bf v},i}(\l) ~=~ \sum_{m=0}^{\infty} \ 
\left.{d^i A_{{\bf v},m}(\epsilon)\over d\epsilon^i}\right|_{\epsilon = 0}\ \l^m ~,$$
are regular at $\l=0$.  We work mod $\e^5$ as before and expand
$$\eqalign{
A_{{\bf v},m}(\epsilon) &= \sum_{k=0}^4 {1\over k!}\, 
a_{{\bf v},m}\, b_{{\bf v},m,k}\ \e^k~,\cr
h_{{\bf v}}(\e) &= \sum_{k=0}^4 {1\over k!}\,
h_{{\bf v}}(0)\, g_{{\bf v},k}\ \e^k~.\cr  
}$$
Noting that $h_{{\bf v}}(0)=1$ and that $g_{{\bf v},0} = 1$
we obtain for the number of points in terms of the periods
$$
\n(\ps)=p^4 + {\cal I} +
\sum_{\bf v}\g_{\bf v} (-p)^{w({\bf v})}\ 
\sum_{k=0}^4 {1\over k!} \left({p\over{1-p}}\right)^k g_{{\bf v},k}
\sum_{j=0}^4 {1\over j!} \left({p\over{1-p}}\right)^j 
{}^{(p-2)}f_{{\bf v},j}^{(k)}({\rm Teich}(\l))~~,$$
where, as before, $ {}^{(p-2)}f_{{\bf v},j}^{(k)}$ refers to the truncated series up to
the $(p-2)$-th term.  
\subsection{The calculation for finer fields}
The calculation for the field with $q=p^s$ points proceeds in complete analogy to the
calculation presented above. It suffices to replace $\Th$ by $\Th_s$, $p$ by $q$, and
$G_n$ and $g_n(y)$ by $G_{s,n}$ and $g_{s,n}(y)$. Where
 $$ 
g_{s,n}(y)~=~\sum_{x\in\sevenFqstar} \Th_s(yx^5)\, \teich^n(x)~~~\hbox{and}~~~
G_{s,n}~=~\sum_{x\in\sevenFqstar}\Th_s(x)\,\teich^n(x)~.\eqlabel{gaussqsums}$$

The only expressions that change their form are the explicit forms of the coefficients in
the expansion of the character which are no longer given by \eqref{Dworkcoeffs} and the
Gross-Koblitz formula which is now
 $$
G_{s,n}~=~(-1)^{s+1} q\,\p^{-S(n)}
\prod_{\ell=0}^{s-1}\G_p\left(1-\left\langle{p^\ell n\over q-1}\right\rangle\right)~;~~
1\leq n\leq q-2~.$$
The important point is that apart from the change to the Gross-Koblitz formula the direct
analogs of our expressions from \eqref{inversion} on are true. For the case $5\notdiv q-1$
we check from these expressions that $\n_s(\ps)$ is as we proposed in \eqref{nus} and
\eqref{betas}. For the case $5|q-1$ we have an expression analogous to \eqref{newnewnu}
$$
\n_s(\ps)~=~q^4 + \sum_{\bf v}\g_{\bf v} \sum_{m=0}^{q-2} \b_{s,{\bf v},m}\,\teich^m(\l)~.
\eqlabel{newnewnuq}$$ 
with
 $$
\b_{s,{\bf v},m}~=~\cases{\displaystyle
q^4\,{G_{s,5m}\over \prod_{i=1}^5 G_{s,(m+v_i k)}}~,\quad &$k\notdiv m$ \cropen{15pt}
\displaystyle - {q^{5-z-\d(z)}\over \prod_{i=1}^5
G_{s,(v_i+a)k}}~,\quad &$k|m$~.\cr}$$
\subsection{A CY hypersurface in a toric variety}
For a general \cy\ hypersurface in a toric variety we return to the notation and
conventions of \SS\chapref{gauss}.3. We have
 $$
p\n^*(\a) - (p-1)^{N+4}~=~\sum_{y,x_i\in\sevenFpstar}\,
\Th\kern-3pt\left(\sum_\bm\,\a_\bm x^\bm - \ph\,Q\right)
~=~\sum_{y,x_i\in\sevenFpstar}\,\Th(-\ph\,Q)\prod_\bm \Th(\a_\bm x^\bm)~.$$
By expanding the characters as a series we find
 $$\eqalign{
p\n^*(\a) - (p-1)^{N+4}~&=~{1\over (p-1)^{\smallpts(\D)}}\sum_{s=0}^{p-2}\,(-\ph)^{-s}\,G_s
\prod_\bm\sum_{s_\bm=0}^{p-2}\,\a_\bm^{s_\bm}\,G_{-s_\bm}\,\times\cropen{5pt}
&\hskip20pt\times\sum_{y\in\sevenFpstar}\,
y^{\sum_\bm s_\bm - s}\,\sum_{x_i\in\sevenFpstar}\,x^{\sum_\bm s_\bm\,\bm - s\bone}~,\cr}$$
where $\pts(\D)$ denotes the number of monomials of $\D$. The $x$ and $y$ sums impose the
conditions
 $$
\sum_\bm s_\bm\,\bm ~\equiv~ s\bone \qquad\hbox{and}\qquad \sum_\bm s_\bm ~\equiv~ s
\qquad\hbox{mod}~(p-1)~.$$
The second of these is redundant unless all the multidegrees of $P$ divide $p-1$. In
either event we have
 $$ 
p\n^*(\a) - (p-1)^{N+4}~=~{1\over (p-1)^{\smallpts(\D)-N-5}}
\sum_{{s_\bm=0}\atop \sum\limits_\bm s_\bm \bm = \s\bone}^{p-2}
(-\ph)^{-\s}\,G_\s \prod_\bm\,\a_\bm^{s_\bm}\,G_{-s_\bm}~,$$
where, in this last expression, $\s=\sum_\bm s_\bm$. The summation is over all
configurations of `masses' $s_\bm$ attached to points $\bm$ of $\D$ such that the centroid
of the masses is the unique internal point $\bone$.
\vskip1in
\leftline{\bf Acknowledgements}
It is a pleasure to acknowledge fruitful conversations with Sir Michael Atiyah,
John Coates, Bogdan Florea, Roger Heath-Brown, Michael McQuillan, Gregory Moore, Jan
Nekovar, Albert Schwarz, John Tate and Felipe Voloch. The work of FR-V is supported by a
grant of the NSF and TARP.
\newpage
\frenchspacing
\immediate\closeout\referencewrite\referenceopenfalse
\line{\fourteenbold\hfil References\hfil}\bigskip\parindent=0pt\input referenc.texauxil
\bye